\documentclass[12pt]{article}
\usepackage[top=2cm,bottom=3cm,left=2cm,right=2cm]{geometry}
\usepackage{graphicx}
\usepackage{amsmath}
\usepackage{amssymb} 
\usepackage{hyperref}
\usepackage[table]{xcolor}
\usepackage[mathscr]{euscript}

\numberwithin{equation}{section}
\numberwithin{figure}{section}

\def\eq#1{(\ref{eq:#1})}
\def\lineup{\!\!\!\!\!\!\!\! &&}

\def\d{\partial}
\def\eps{\epsilon}

\def\smallpile#1#2{\genfrac{}{}{0pt}{2}{#1}{#2}}

\definecolor{verylight}{rgb}{.95,.95,.95}

\def\Qlc{Q_\mathrm{lc}}

\def\S{S}

\def\Lpar{L_\parallel}
\def\Lperp{L_\perp}
\def\Llong{L_\mathrm{long}}
\def\LDDF{L_\mathrm{DDF}}

\def\bpar{b_\parallel}
\def\bperp{b_\perp}
\def\blong{b_\mathrm{long}}
\def\bDDF{b_\mathrm{DDF}}

\def\deltalc{\delta_\mathrm{lc}}

\def\H{\mathcal{H}}
\def\Hcov{\mathcal{H}_\mathrm{cov}}
\def\Hlc{\mathcal{H}_\mathrm{lc}}
\def\Hperp{\mathcal{H}_\perp}
\def\Hpar{\mathcal{H}_\parallel}
\def\HDDF{\mathcal{H}_\mathrm{DDF}}
\def\Hlong{\mathcal{H}_\mathrm{long}}
\def\Hgrass{\mathcal{H}_\mathrm{grass}}
\def\Hdeg{\mathcal{H}_\mathrm{deg}}

\def\Deltapar{\Delta_\parallel}
\def\Deltaperp{\Delta_\perp}
\def\DeltaDDF{\Delta_\mathrm{DDF}}
\def\Deltalong{\Delta_\mathrm{long}}

\def\kpar{k_\parallel}
\def\ppar{p_\parallel}

\def\Psilc{\Psi_\mathrm{lc}}
\def\Psipar{\Psi_\parallel}
\def\Psiperp{\Psi_\perp}

\def\P{\mathcal{P}}
\def\Plc{\mathcal{P}^\mathrm{lc}}

\def\C{\mathcal{C}}
\def\Clc{\mathcal{C}^\mathrm{lc}}
\def\M{\mathcal{M}}
\def\Mlc{\mathcal{M}^\mathrm{lc}}
\def\V{\mathcal{V}}
\def\Vlc{\mathcal{V}^\mathrm{lc}}
\def\R{\mathcal{R}}
\def\Rlc{\mathcal{R}^\mathrm{lc}}
\def\Slc{\mathcal{S}^\mathrm{lc}}

\def\pilc{\pi_\mathrm{lc}}
\def\sigmalc{\sigma_\mathrm{lc}}
\def\varphilc{\varphi_\mathrm{lc}}

\def\Omegalc{\Omega^\mathrm{lc}}
\def\vlc{v^\mathrm{lc}}
\def\Alc{A^\mathrm{lc}}

\def\flc{f^\mathrm{lc}}
\def\rlc{r^\mathrm{lc}}

\def\puncture{\mathrm{puncture}}
\def\propagator{\mathrm{propagator}}
\def\quartic{\mathrm{quartic}}
\def\cubic{\mathrm{cubic}}
\def\llink#1{\mathrm{link}(#1)}

\def\BCFT{\mathrm{BCFT}}
\def\UHP{\mathrm{UHP}}
\def\bpz{\mathrm{bpz}}

\def\F{\mathscr{F}}
\def\Flc{\mathscr{F}_\mathrm{lc}}

\begin{document}

\hypersetup{pageanchor=false}
\begin{titlepage}
\rightline\today

\begin{center}

\vspace{3.5cm}

{\large \bf{Open string field theory in lightcone gauge}}

\vspace{1cm}

{\large Theodore Erler}

\vspace{1cm}

{\it CEICO, FZU - Institute of Physics of the Czech Academy of
Sciences}\\
{\it Na Slovance 2, 182 21 Prague 8, Czech Republic}\\
{\tt tchovi@gmail.com}\\

\vspace{2cm}

{\bf Abstract}

\end{center}

We study covariant open bosonic string field theory in lightcone gauge. When lightcone gauge is well-defined, we find two results. First, the vertices of the gauge-fixed action consist of  Mandelstam diagrams with stubs covering specific portions of the moduli spaces of Riemann surfaces. This is true regardless of how the vertices of the original covariant string field theory are constructed (e.g. through minimal area metrics, hyperbolic geometry, and so on). Second, the portions of moduli space covered by gauge-fixed vertices are changed relative to those covered by the original covariant vertices. The extra portions are supplied through the exchange of longitudinal degrees of freedom in scattering processes.

\end{titlepage}

\hypersetup{pageanchor=true}

\tableofcontents

\section{Introduction}

For many years, covariant and lightcone string field theories (SFTs) have stood apart as largely independent approaches to formulating the off-shell dynamics of strings. However, for open bosonic strings, recent work has shown that they can be related with the gauge-fixing condition \cite{lightcone}
\begin{equation}\left(b_0+ip_-\oint_0 \frac{d\xi}{2\pi i} \frac{b(\xi)}{\d X^+(\xi)}\right)\Psi = 0 .\label{eq:intro_lc}\end{equation}
This can be thought of as defining {\it lightcone gauge} in covariant string field theory. When the lightcone gauge condition is satisfied, part of the string field is isomorphic to the space of states of a lightcone-quantized string. The other part is subject to purely algebraic equations of motion (in the lightcone frame), and can be integrated out. In this way we obtain a version of lightcone string field theory by gauge-fixing covariant string field theory. 

However, we do not really understand what form this gauge-fixed theory takes. In fact, the idea that covariant SFT can be fixed to lightcone gauge raises an apparent paradox. Lightcone-quantized strings, as far as is ever considered, only interact through Mandelstam diagrams \cite{Mandelstam}. Interactions in covariant string field theory, however, are different. One recent approach uses Riemann surfaces endowed with metrics of constant negative curvature~\cite{Costello,FiratCubic,FiratQuartic}. But all prescriptions have the property that the parametric length of the string (in the natural metric) is constant, independent of Lorentz frame. In Mandelstam diagrams, string lengths increase with lightcone momentum. It seems impossible for strings in lightcone gauge to interact through the surfaces prescribed by a covariant string field theory.

To understand how this paradox is resolved we must evaluate the vertices of the gauge-fixed action.This is not an easy task, and is the primary goal of this paper. Our analysis reveals three main points:
\begin{itemize}
\item Lightcone-quantized strings do not necessarily have an acceptable off-shell coupling through a covariant string diagram. If one string in the diagram has lightcone momentum which is too small relative to others, the coupling grows exponentially with Virasoro level. The interaction vertex then fails to be normalizable. This is referred to as the {\it soft string problem} of lightcone gauge.
\item If this problem is avoided, lightcone-quantized strings see interaction through a covariant string diagram as {\it identical} to interaction through a Mandelstam diagram. This result goes by the name of the {\it equivalence theorem}, and follows from conformal invariance of DDF operators and the structure of free boson OPEs. The Mandelstam diagrams which emerge through this equivalence always come attached to strips of string called {\it stubs} for each external state. The off-shell coupling is normalizable if and only if the stubs have a length which is not negative. 
\item Each covariant string vertex in lightcone gauge is therefore seen as equivalent to a collection of Mandelstam diagrams with stubs. However, because the geometry of the surfaces is changed, gluing covariant vertices and propagators in lightcone gauge will no longer cover the whole moduli space. We show that the gaps in moduli space are filled by additional vertices representing the exchange of longitudinally polarized string states in the interaction process. These vertices are generated in the process of integrating out the longitudinal part of the covariant string field in lightcone gauge.
\end{itemize}
Fixing lightcone gauge (if it is defined) results in an action for a lightcone string field whose vertices consist of Mandelstam diagrams attached to stubs. The lengths of the stubs and the portions of moduli space contained within vertices are mutually determined so that amplitudes in lightcone gauge cover all of moduli space. 

\subsection{Summary}

Below we summarize of the contents of the paper.

In section \ref{sec:lightcone} we review the formulation of lightcone gauge in open bosonic string field theory. We explain that the covariant string field and the lightcone string field belong to separate chain complexes related by the Aisaka-Kazama transformation \cite{AisakaKazama}. The two chain complexes reflect the distinction between states derived from lightcone quantization and states derived from covariant quantization. The string field is split into transverse and longitudinal parts as characterized by a zero or nonzero eigenvalue of the so-called longitudinal wave operator. The longitudinal wave operator is the BRST variation of an operator at ghost number $-1$ called the longitudinal antighost. Lightcone gauge is defined by demanding that the longitudinal antighost annihilates the string field. When this happens, the longitudinal component of the string field is subject to purely algebraic equations of motion. Eliminating this part of the string field results in a gauge-fixed action involving only a transverse string field of the kind seen in the traditional lightcone string field theory of Kaku and Kikkawa~\cite{KK}. The interactions of the lightcone string field, however, are characterized in a rather obscure way through a sum of Feynman graphs created by contracting covariant vertices through sums over longitudinal intermediate states. In the remainder of the paper we extract the Riemann surface interpretation of these vertices. 

In section \ref{sec:offshell} we discuss how transverse string states interact off-shell in the covariant formulation and in  the lightcone formulation of string theory. We start by reviewing the notion of covariant off-shell amplitude, defined by integrating the covariant measure over a submanifold of the infinite-dimensional fiber bundle $\P_n$ of disks with local coordinates specified around $n$ punctures on the boundary. Next we review the structure of Mandelstam diagrams as composed of strip domains joined through interaction points. We introduce local coordinates on the strip domains, and through the Mandelstam mapping, transform these local coordinates to the upper half plane. With this preparation we introduce the notion of a lightcone off-shell amplitude, which is intended to be the natural notion of off-shell amplitude from the point of view of the lightcone formulation of string theory. The lightcone off-shell amplitude is defined by integrating the lightcone measure over a submanifold of the fiber bundle $\Plc_n$ of Mandelstam diagrams with stubs glued to each of $n$ external strip domains. The fiber bundle $\Plc_n$ is finite dimensional, and can be equivalently characterized as the space of disks with dilatations specified around $n$ punctures on the boundary. The lightcone measure is described in a few different forms. First it is presented in the traditional way as a correlation function on a Mandelstam diagram in the transverse worldsheet theory, carefully accounting for an implicit normalization generated by the conformal anomaly. A second form is as the pullback of the covariant measure onto $\Plc_n$. The needed $b$-ghosts insertions are presented in two ways, first as follows from the Schiffer variation and second as follows from the structure of Feynman graphs of the Kugo-Zwiebach SFT \cite{KZ}. The claim that the lightcone measure can be expressed as the pullback of the covariant measure is justified later in section \ref{sec:freeze}. Next we introduce the {\it replacement formula}, which states that in certain correlation functions the lightlike free boson $X^+$ is proportional to the Mandelstam mapping. Conformal invariance of DDF operators, their characteristic dependence on $X^+$, together with the replacement formula imply the main result of this section, the {\it equivalence theorem}. This asserts that a covariant off-shell amplitude of DDF states is the same as a lightcone off-shell amplitude of the corresponding lightcone-quantized string states. The requisite lightcone amplitude is defined by a collection of Mandelstam diagrams with stubs whose lengths are determined so that the dilatation at each puncture is the same as that defined by the covariant amplitude. In some circumstances this may force stub lengths to be negative, especially when the corresponding string states have sufficiently low lightcone momentum relative to others in the interaction process. When this happens the transverse off-shell amplitude is not normalizable, in the sense that its magnitude increases exponentially with $L_0$ eigenvalue. This is especially a problem when the transverse off-shell amplitude is supposed to define part of an SFT vertex in lightcone gauge. When gluing vertices with propagators, sums over intermediate states will not always converge. This is referred to as the {\it soft string problem} of lightcone gauge.

To derive lightcone gauge vertices we must learn how to evaluate sums over longitudinal intermediate states. A significant aspect of this is dealing with the conformal anomaly, since a sum over longitudinal states in effect creates a propagator strip in the longitudinal worldsheet theory while leaving the surface of the transverse worldsheet theory unchanged. In section \ref{sec:freeze} we show how to deal with this when the sum over longitudinal states connects two Mandelstam diagrams. We find that the longitudinal factor of the worldsheet theory is frozen to the Fock vacuum inside propagator strips of a Mandelstam diagram, a phenomenon we refer to as {\it longitudinal freezing}. Therefore the lengths of propagator strips on a Mandelstam diagram have no effect on the value of the longitudinal correlation function. They can be freely adjusted to match the lengths in the transverse correlation function, where the conformal anomaly cancels. One application is in explaining why the lightcone measure can be expressed as the pullback of the covariant measure. The covariant measure accounts for the exchange of both transverse and longitudinal states, but on a Mandelstam diagram the longitudinal exchange has no effect. We present a proof of longitudinal freezing based on the replacement formula and BRST invariance properties of the string measure. This circumvents the need to regularize and evaluate determinants of Laplacians on Mandelstam diagrams, which is a critical part of the traditional derivation of the lightcone measure as found for example in \cite{GSWII}. 

In section \ref{sec:quartic} we investigate the quartic vertex in lightcone gauge. This has a contribution from the covariant quartic vertex as well as $s$- and $t$-channel diagrams representing the sums of longitudinal states connecting covariant cubic vertices. The computation of the quartic vertex is closely related to the 4-point amplitude in Siegel gauge when external states are transverse. In this circumstance, the equivalence theorem implies that the surfaces of the Siegel gauge amplitude are projected into Mandelstam diagrams with stubs.  We assume that the Siegel gauge amplitude is {\it graphically compatible} with its transverse projection, which means (in part) that the propagator strips of a Siegel gauge diagram are contained inside the propagator strips of the Mandelstam diagram obtained after transverse projection. Under this assumption we compute the sums over longitudinal states in the quartic vertex. We find that the quartic vertex is the same as the transverse Siegel gauge 4-point amplitude except that the length of the propagator strip on the Mandelstam diagram is shortened proportionally to the length of the propagator strip on the Siegel gauge diagram at a given point in moduli space. This ``shortening" can be understood as the result of only accounting for the longitudinal contribution to the sum over intermediate states. Near degeneration the propagator strips on the Mandelstam diagram and the Siegel gauge diagram both become very long, but we show that their difference remains finite and in fact has precisely the correct value to fill the gap in moduli space left by the covariant quartic vertex.  We evaluate the gauge-fixed quartic vertex in closed form when the covariant cubic vertex is defined by $SL(2,\mathbb{R})$ maps. When stub lengths are positive and vertices are normalizable, we find the sums over longitudinal states give a very small correction to the gauge-fixed quartic vertex. But interestingly, the longitudinal sums can sometimes cover and uncover the same part of moduli space more than once.

In section \ref{sec:higher} we generalize to higher vertices in lightcone gauge. The higher order vertices are again given by transverse projection of Siegel gauge amplitudes except that propagator strips are shortened in a similar way as for the quartic vertex. We discuss the quintic vertex in some detail as an example. We also sketch how the structure of the gauge-fixed vertices can change if the Siegel gauge amplitude is not graphically compatible with its transverse projection. Finally we show that the lightcone gauge vertices fill the gaps in moduli space left by diagrams with propagators to all orders. 

After concluding remarks there are a few appendices. In appendix \ref{app:suspension} we discuss the suspension map which relates the Grassmann grading scheme used in this paper and the degree grading scheme used in other recent works. This is helpful for, among other things, determining relative signs of the Feynman graph contributions to the lightcone gauge vertices. In appendix \ref{app:measure} we demonstrate the equivalence of certain forms of the lightcone measure. We start in subappendix \ref{subapp:covtored} by showing that the pullback of the covariant measure (the {\it covariantized measure}) is the same as the traditional form of the lightcone measure as a transverse correlation function on a Mandelstam diagram (the {\it reduced measure}). This is partly a matter of evaluating the longitudinal correlation function, but the main technical hurdle is in calculating the Jacobian determinant which relates the moduli of the Mandelstam diagram to the positions of the punctures on the upper half plane. This is a well-known object in lightcone string theory, but we spell out its computation for completeness. In subappendix \ref{subapp:covtoKZ} we demonstrate by contour deformation that the pullback of the covariantized measure can be described equivalently using $b$-ghost contours around the punctures as determined by the Schiffer variation or by $b$-ghosts in the propagators and on the quartic interaction points as determined by the Feynman graphs of the Kugo-Zwiebach string field theory. 

\subsubsection*{Note added}

While this work was in preparation the paper \cite{Pesando2} appeared, extending other recent work \cite{Pesando1}, which addresses the relation between amplitudes in the lightcone and covariant formulations of string theory. The results are related to the discussion of section \ref{sec:offshell}. 

\subsubsection*{Conventions}
We assume $\alpha'=1$ and set the open string coupling constant to unity.  The 2D Minkowski metric in lightcone coordinates is 
\begin{equation}
ds^2 = 2dx^+dx^-,\nonumber
\end{equation}
and $x^+$ will be identified with lightcone time. Commutators are graded with respect to Grassmann parity and we use the left handed convention in defining open string vertices. 

\section{Lightcone gauge}
\label{sec:lightcone}

We begin by describing the formulation of lightcone gauge in covariant open bosonic string field theory. The discussion is meant to be self-contained, but we refer to \cite{lightcone,AisakaKazama} for proofs of a few results.  

The lightcone gauge condition assumes a worldsheet boundary conformal field theory (BCFT) which can be factorized as
\begin{equation}\BCFT = \BCFT_\perp\otimes \BCFT_\parallel.\label{eq:lcBCFT}\end{equation}
The first factor will be called the {\it transverse} part of the BCFT, and the second factor will be called the {\it longitudinal} part. The transverse  part can be any unitary $c=24$ BCFT. The longitudinal part has central charge $-24$, and consists of one spacelike and one timelike noncompact free boson subject to Neumann boundary conditions,  together with the $bc$ ghost system of central charge $-26$.  The spacelike and timelike free bosons can be combined into a pair of lightlike free bosons, so the worldsheet fields of the longitudinal part are 
\begin{eqnarray} 
\lineup X^+(z,\overline{z}),\ \ X^-(z,\overline{z}),\ \ b(z),\ \ c(z).
\end{eqnarray}
Some conventions:
\begin{itemize}
\item We write 
\begin{equation}X^+(z,\overline{z}) = X^+(z)+X^+(\overline{z}),\label{eq:XXchiral}\end{equation}
where $X^+(z)$ is the chiral free boson.
\item The position zero mode of $X^+(z,\overline{z})$ is denoted $x^+$ and is identified with lightcone time. 
\item The momentum zero mode of $X^-(z,\overline{z})$ is denoted $p_+$ and generates translations in lightcone time. 
\item The momentum zero mode of $X^+(z,\overline{z})$ is denoted $p_-$ and is called the (backwards) lightcone momentum. In Mandelstam diagrams, is proportional to the string length. We always assume~$p_-\neq 0$.  
\end{itemize}
The two component vector of longitudinal momenta will be written $\ppar=(p_+,p_-)$. 

\subsection{Covariant and lightcone vector spaces}

First we need to understand the relation between covariant and lightcone string fields. These occupy respectively the  {\it covariant vector space}, $\Hcov$, and the {\it lightcone vector space}, $\Hlc$.  These vector spaces differ in how they describe physical states. Specifically, they can be characterized as chain complexes, 
\begin{eqnarray}
	\Hcov \lineup = (\H,Q),\\
	\Hlc \lineup = (\H,\Qlc), \label{eq:Hcovlcpar}
\end{eqnarray}
defined over the same graded vector space $\H$ but with different differentials. The graded vector space is taken to be the state space of the total matter+ghost $\BCFT$. The differential in the covariant vector space is the BRST operator $Q$, and physical states are given by its cohomology at ghost number 1. The differential in the lightcone vector space will be called the {\it lightcone BRST operator} $\Qlc$. The cohomology of $\Qlc$ at ghost number 1 is given by states without ghost or lightcone creation operators satisfying the mass shell condition.  

The lightcone BRST operator is defined
\begin{eqnarray}\Qlc = \deltalc+c_0L_0,\label{eq:Qlc}
\end{eqnarray}
where $L_0$ is the zero mode of the total energy-momentum tensor and $\deltalc$ will be called the {\it lightcone differential}, and is given by
\begin{eqnarray}
\deltalc \lineup =p_-\sum_{n\in \mathbb{Z},n\neq 0}c_{-n}\alpha^-_n.\label{eq:deltalc}
\end{eqnarray}
The lightcone differential satisfies
\begin{equation}
\deltalc^2 = 0,\ \ \ \ [\deltalc,c_0]=0,\ \ \ \ [\deltalc,L_0]=0,
\end{equation}
which implies that $\Qlc$ is nilpotent and defines a cohomology. The lightcone BRST operator appears as a contribution to the ordinary BRST operator  
\begin{equation}Q = \Qlc + \mathrm{other\ terms}\end{equation} 
which was the basis for Kato and Ogawa's original proof of the no-ghost theorem \cite{KatoOgawa}. The strongest version of this result was given by Aisaka and Kazama \cite{AisakaKazama}, who showed that covariant and lightcone vector spaces are related by a similarity transformation
\begin{equation}\S: \Hlc \to\Hcov,\ \ \ \ \S^{-1}:\Hcov\to\Hlc.\label{eq:AisakaKazama}\end{equation}
This is a chain map, so it transforms the lightcone BRST operator into the traditional BRST operator,
\begin{equation}\S\Qlc = Q\S.\end{equation}
The transformation also preserves the BPZ inner product,
\begin{equation}\langle \S a,\S b\rangle = \langle a,b\rangle,\ \ \ \ \ \ a,b\in\Hlc.\end{equation}
We do not need to know the explicit form of the Aisaka-Kazama transformation. However, we will need to know how it acts on certain states and operators. For this it will be sufficient to draw on the results of appendix D of~\cite{lightcone}. 

\subsection{Transverse and longitudinal subspaces}

The lightcone vector space is useful because its transverse and longitudinal parts are easy to disentangle. The transverse part consists of states without any ghost or lightcone creation operators. Generally such states take the form 
\begin{equation}V_\perp(0)|-,\kpar\rangle, \ \ \ \ V_\perp(0)|+,\kpar\rangle, \label{eq:Hlc}\end{equation}
where $V_\perp(0)$ is a vertex operator of the transverse BCFT and
\begin{equation}|-,\kpar \rangle = c_1 e^{i\kpar\cdot X(0,0)}|0\rangle, 
\ \ \ \ |+,\kpar\rangle = c_0c_1 e^{i \kpar\cdot X(0,0)}|0\rangle,\label{eq:Fock_vac}
\end{equation}
are the Fock vacua of the total matter+ghost $\BCFT$. These states will be called {\it transverse}, and form the {\it transverse vector space}, denoted $\Hperp$. Transverse states can have ghost numbers 1~or~2. The dynamical field of Kaku and Kikkawa's lightcone string field theory is a transverse state at ghost number 1. The remainder of the lightcone vector space consists of states which contain at least some ghost and lightcone creation operators. These states will be called {\it longitudinal}, and form the  {\it longitudinal vector space}, denoted $\Hpar$. Therefore the lightcone vector space is decomposed into a direct sum
\begin{equation}\Hlc = \Hperp\oplus \Hpar.\label{eq:Hpar}\end{equation} 
Through the Aisaka-Kazama transformation, we infer a similar decomposition of the covariant vector space
\begin{equation}\Hcov = \HDDF\oplus\Hlong.\label{eq:HDDFlong}\end{equation}
The image of the transverse vector space will be called the {\it DDF vector space}, denoted $\HDDF$. This consists of states created from the Fock vacua \eq{Fock_vac} using DDF operators \cite{DDF}. The remainder of the covariant vector space is denoted $\Hlong$. The distinction between transverse and longitudinal states is somewhat obscure in the covariant vector space because DDF states contain lightcone creation operators.  For reference we summarize all of these vector spaces in table \ref{tab:vector}.

\begin{table}[t]
	\begin{center}
		\begin{tabular}{|c|c|c|}
			\hline
			& & \cellcolor{verylight}  \\
			$\begin{matrix}\text{full chain} \\ \text{complex}\end{matrix}$ & $\ \ \Hlc,\  \begin{matrix}\text{lightcone} \\ \text{vector space}\end{matrix}\ \ $ 
			& \cellcolor{verylight} $\ \ \ \ \ \ \ \ \ \Psi\in\Hcov,\  \ \begin{matrix}\text{covariant} \\ \text{vector space}\end{matrix}\ \ \ \ \  \ \ \ \ $\\
			& & \cellcolor{verylight}  \\
			\hline 
			& \cellcolor{verylight}  & \\
			$\ \ \ \begin{matrix}\text{transverse} \\ \text{states}\end{matrix}\ \ \ $  
			&\cellcolor{verylight}  $\ \ \ \Psiperp\in\Hperp,\  \ \ \begin{matrix}\text{transverse} \\ \text{vector space}\end{matrix}\ \ \ $ 
			& $\ \ \ \HDDF,\ \ \ \begin{matrix}\text{DDF} \\ \text{vector space}\end{matrix}\ \ \ $\\
			& \cellcolor{verylight}  & \\
			\hline
			& & \\
			$\ \ \ \begin{matrix}\text{longitudinal} \\ \text{states}\end{matrix}\ \ \ $  
			& $\Hpar,\  \ \ \begin{matrix}\text{longitudinal} \\ \text{vector space}\end{matrix}\ \ \ $ 
			& $\Hlong$\\
			& & \\
			\hline
			\end{tabular}
	\end{center}
	\caption{\label{tab:vector} The columns represent vector spaces related by the Aisaka-Kazama transformation, and the rows represent the decomposition into transverse and longitudinal parts.  The lightcone string field $\Psiperp$ lives in the transverse vector space $\Hperp$, while the covariant string field lives in the covariant vector space $\Hcov$.} 
\end{table}

We will need to describe the transverse/longitudinal decomposition more algebraically. Working in the lightcone vector space, we introduce  operators
\begin{eqnarray}
\bpar \lineup = \frac{1}{\sqrt{2}p_-}\sum_{n\in \mathbb{Z},n\neq 0}\alpha^+_{-n}b_n,\label{eq:bpar}\\
\Lpar  \lineup = \sum_{n=1}^\infty\Big(\alpha^+_{-n}\alpha^-_n + \alpha^-_{-n}\alpha^+_n\Big)+\sum_{n=1}^\infty n\Big(b_{-n}c_n +c_nb_{-n}\Big),\label{eq:Lpar}
\end{eqnarray}
The first operator $\bpar$ will be called the {\it longitudinal antighost}, and  the second $\Lpar$ will be called the {\it longitudinal wave operator}. They satisfy relations
\begin{equation}(\bpar)^2=0,\ \ \ \ [\deltalc,\bpar] = \Lpar, \ \ \ \ [\bpar,c_0] = 0,\ \ \ \ [\bpar,L_0]=0,\end{equation}
which imply
\begin{equation}[\Qlc,\bpar] = \Lpar.\label{eq:bparLpar}\end{equation}
The longitudinal wave operator counts the level created by ghost and lightcone oscillators, with the Fock vacua (at any momentum) counting as level $0$. Therefore, the transverse vector space can be identified with the kernel of $\Lpar$:
\begin{equation}\Hperp = \mathrm{ker}(\Lpar).\end{equation}
 It is readily shown that  transverse states are annihilated by $\deltalc$ and $\bpar$, but cannot be $\deltalc$ or $\bpar$ of something else, since these operators necessarily produce ghost or lightcone creation operators. Meanwhile, any $\deltalc$-closed longitudinal state is also $\deltalc$-exact, since we can write
\begin{equation}a = \deltalc\frac{\bpar}{\Lpar}a,\ \ \ \ \ a\in\Hpar\ \text{and}\ \deltalc a = 0.\end{equation}
Here we use the fact that $\Lpar$ is nonzero when operating on longitudinal states. Likewise, any $\bpar$-closed longitudinal state is also $\bpar$-exact, since we can write
\begin{equation}a = \frac{\bpar}{\Lpar}\deltalc a,\ \ \ \ \ a\in\Hpar\ \text{and}\ \bpar a = 0.\end{equation}
It follows that transverse states can be identified with the cohomology of the lightcone differential $\deltalc$ or the homology of the longitudinal antighost $\bpar$.  

We have an analogous story in the covariant vector space after applying the Aisaka-Kazama transformation. To describe this properly it will be helpful to first introduce transverse counterparts of the operators above so that the following relations hold: 
\begin{eqnarray}
b_0 \lineup =  \bperp+\bpar, \label{eq:blcbpar}\\
L_0 \lineup =  \Lperp+\Lpar.\label{eq:LlcLpar}
\end{eqnarray}
The operator $\bperp$ will be called the {\it transverse antighost} while $\Lperp$ will be called the {\it transverse wave operator}. The operators are related by 
\begin{equation}[\Qlc,\bperp]=\Lperp,\label{eq:bperpLperp}\end{equation}
which follows from \eq{bparLpar} and $[\Qlc,b_0]=L_0$. The transverse wave operator is important because it defines the mass shell condition for the lightcone string field. Because a transverse string field is in the cohomology of $\deltalc$ and is annihilated by $\Lpar$, it is readily seen that these two conditions are equivalent:
\begin{equation}\Qlc\Psiperp = 0\ \ \ \leftrightarrow\ \ \ \Lperp\Psiperp = 0.\end{equation}
The transverse wave operator can be written
\begin{equation}\Lperp = (\ppar)^2 +L_0^\perp -1,\end{equation}
where $L_0^\perp$ is the Virasoro zero mode in the transverse BCFT. Since $b_0$ and $L_0$ are preserved by the Aisaka-Kazama transformation, we have an analogous decomposition in the covariant vector space
\begin{eqnarray}
	b_0 \lineup =  \bDDF+\blong,\label{eq:bDDFblong}\\
	L_0 \lineup =  \LDDF+\Llong, \label{eq:LDDFLlong}
\end{eqnarray}
where
\begin{equation}\bDDF =\S\bperp \S^{-1},\ \ \ \LDDF = \S \Lperp \S^{-1}\end{equation}
will be called the {\it DDF antighost} and the  {\it DDF wave operator}, while 
\begin{equation}\blong =\S\bpar \S^{-1},\ \ \ \Llong = \S \Lpar \S^{-1}\end{equation}
will be called the {\it longitudinal antighost} and {\it longitudinal wave operator} (in the covariant vector space). Transforming \eq{bparLpar} and \eq{bperpLperp} implies
\begin{equation}[Q,\blong] = \Llong,\ \ \ \ [Q,\bDDF] = \LDDF.\end{equation}
From appendix D of \cite{lightcone} we learn that the DDF antighost and DDF wave operator take the form 
\begin{eqnarray}
\bDDF \lineup = -ip_-\oint_0 \frac{d\xi}{2\pi i}\frac{b(\xi)}{\d X^+(\xi)},\\
\LDDF \lineup =  \ppar^2-i p_-\oint_0 \frac{d\xi}{2\pi i}\frac{1}{\d X^+(\xi)}\Big[T^\perp(\xi)-2 \{X^+,\xi\}\Big],\label{eq:LDDF}
\end{eqnarray}
where $T^\perp(\xi)$ is the energy-momentum tensor of the transverse $\BCFT$. The inverse of $\d X^+$ can be defined through a geometric series expansion in powers of the oscillator part of $\d X^+$  when $p_-\neq 0$. The expression
\begin{equation}
\{X^+,\xi\} = \frac{\d^3 X^+(\xi)}{\d X^+(\xi)}-\frac{3}{2}\left(\frac{\d^2 X^+(\xi)}{\d X^+(\xi)}\right)^2
\end{equation}
is the Schwarzian derivative of the chiral lightcone scalar $X^+(\xi)$.  Note that the conformal anomaly of $T^\perp(\xi)$ cancels against the Schwarzian derivative, so the combination 
\begin{equation}T^\perp(\xi)-2 \{X^+,\xi\}\end{equation}
transforms as a primary operator of weight $2$. With the understanding that the inverse of $\d X^+$ counts as a primary of weight $-1$, it follows that the DDF antighost and wave operator are zero modes of weight 1 primaries. Therefore they are conformally invariant, much like the BRST operator. This will be important later. 

\subsection{Lightcone gauge}

Open bosonic string field theory is characterized by a dynamical field $\Psi\in\Hcov$ which is Grassmann odd and ghost number 1. Fixing {\it lightcone gauge} means that the dynamical field is subject to the condition
\begin{equation}\blong\Psi = 0. \label{eq:lc_gauge}\end{equation}
The consequence of lightcone gauge is more transparent when working in the lightcone vector space. We write the string field as 
\begin{equation}\Psi = \S\Psilc,\ \ \ \Psilc \in\Hlc,\label{eq:PsitoPsicovlc}\end{equation}
and further decompose into transverse and longitudinal parts:
\begin{equation}\Psilc = \Psiperp +\Psipar,\ \ \ \ \ \Psiperp\in\Hperp,\ \ \Psipar\in\Hpar.\label{eq:Psilc}\end{equation}
The transverse part is the same kind of string field that appears in Kaku and Kikkawa's lightcone string field theory. 
Let us demonstrate that lightcone gauge is reachable at the linearized level. Because $\Lpar$ is nonzero we can use \eq{bparLpar} to write
\begin{equation}\Psilc = \Psiperp + \left[\Qlc,\frac{\bpar}{\Lpar}\right]\Psipar.\end{equation}
Rearranging the terms gives  
\begin{equation} \Psiperp + \frac{\bpar}{\Lpar} \Qlc\Psipar = \Psilc -\Qlc \left(\frac{\bpar}{\Lpar}\Psipar \right). \end{equation}
On the left hand side the string field satisfies the lightcone gauge condition because $(\bpar)^2=0$. The right hand side shows this is achieved by a linearized gauge transformation of $\Psilc$. Second, let us demonstrate that lightcone gauge completely fixes the gauge at the linearized level. This requires that there is no nonvanishing state  $\Qlc\Lambda$ at ghost number 1 satisfying
\begin{equation}\bpar\Qlc\Lambda = 0. \end{equation} 
Operating with $\Qlc$ implies
\begin{equation}\Lpar\Qlc\Lambda = 0. \end{equation}
Therefore $\Qlc\Lambda$ would have to be a transverse state. But since $\Lpar$ and $\Qlc$ commute, $\Lambda$ itself must be a transverse state up to terms annihilated by $\Qlc$. But there are no nonvanishing states in $\Hperp$ at ghost number 0, so $\Qlc\Lambda$ must vanish. It is worth mentioning that lightcone gauge is a more complete gauge fixing than Siegel gauge. For example the state
\begin{equation}Q\Big( e^{ik\cdot X(0,0)}|0\rangle\Big),\ \ \ k^2=0,\end{equation}
is in Siegel gauge and is BRST exact. 

The kinetic term of the open string field theory action may be expressed as 
\begin{eqnarray}
S_\mathrm{free} \lineup = -\frac{1}{2}\langle\Psi,Q\Psi\rangle\nonumber\\
\lineup = -\frac{1}{2}\langle\Psilc,\Qlc\Psilc\rangle\nonumber\\
\lineup = -\frac{1}{2}\langle\Psiperp,\Qlc \Psiperp\rangle - \frac{1}{2}\langle\Psipar,\Qlc \Psipar\rangle\nonumber\\
\lineup = -\frac{1}{2}\langle\Psiperp,c_0\Lperp\Psiperp\rangle - \frac{1}{2}\langle\Psipar,\Qlc \Psipar\rangle.
\end{eqnarray}
In the first step we substituted \eq{PsitoPsicovlc}, the second \eq{Psilc}, and in the third step we replaced $\Qlc$ with $c_0L_0$ because the lightcone string field is $\deltalc$-closed. We further replace $L_0$ with $\Lperp$ because the lightcone string field is annihilated by $\Lpar$. If we further assume lightcone gauge the longitudinal kinetic term simplifies,
\begin{equation}S_\mathrm{free} = -\frac{1}{2}\langle\Psiperp,c_0\Lperp\Psiperp\rangle - \frac{1}{2}\langle\Psipar,\deltalc\Psipar\rangle.\label{eq:Slcfree}\end{equation}
In lightcone gauge the longitudinal part of the string field is proportional to $\bpar$. This is because the lightcone gauge condition requires that the longitudinal part is $\bpar$-closed, and the absence of homology then implies that it is $\bpar$-exact. This implies that any part of the kinetic operator in the longitudinal sector which commutes with $\bpar$ will drop out, since we can move $\bpar$ from one $\Psipar$ to the other unobstructed to give zero. Since $c_0L_0$ commutes with $\bpar$, we can therefore replace the lightcone BRST operator $\Qlc$ with the lightcone differential $\deltalc$ in the longitudinal kinetic term, as shown in \eq{Slcfree}. The critical point here is that the lightcone differential $\deltalc$ has no lightcone time derivatives. This means that the longitudinal part of the string field is not dynamical in lightcone gauge. It can be eliminated by the equations of motion, but for the moment we will not do this. The fields $\Psiperp$ and $\Psipar$ have different propagators,
\begin{eqnarray}
\text{propagator on }\Hperp\lineup = \frac{\bperp}{\Lperp}, \\
\text{propagator on }\Hpar \lineup =  \frac{\bpar}{\Lpar},
\end{eqnarray}
determined by the condition that they invert the respective kinetic operators on the respective gauge-fixed subspaces. The full propagator in lightcone gauge can be represented by adding these propagators multiplied by a projection onto the respective subspace:
\begin{equation}
\Delta_{\bpar} = \frac{\bperp}{\Lperp}\delta(\Lpar) + \frac{\bpar}{\Lpar},\label{eq:Deltabpar}
\end{equation}
where $\delta(\Lpar)$ is the projector onto the kernel of $\Lpar$. The projection onto longitudinal states in the second term can be seen as implicit since $\bpar$ annihilates transverse states. The first term,
\begin{equation}\Deltaperp =\frac{\bperp}{\Lperp}\delta(\Lpar), \label{eq:Deltaperp}\end{equation}
will be called the {\it transverse propagator} and the second term,
\begin{equation}\Deltapar = \frac{\bpar}{\Lpar},\label{eq:Deltapar}\end{equation}
will be called the {\it longitudinal propagator}. Mapping back to the covariant vector space, the lightcone gauge propagator is 
\begin{equation}
\Delta_{\blong} = \frac{\bDDF}{\LDDF}\delta(\Llong) +\frac{\blong}{\Llong},\label{eq:Deltablong}
\end{equation}
where $\delta(\Llong)$ is the projector onto DDF states. The first term,
\begin{equation}\DeltaDDF = \frac{\bDDF}{\LDDF}\delta(\Llong),\label{eq:DeltaDDF}\end{equation}
will be called the {\it DDF propagator} while the second term,
\begin{equation}\Deltalong = \frac{\blong}{\Llong},\label{eq:Deltalong}\end{equation}
will be called the {\it longitudinal propagator} (in the covariant vector space).

\subsection{Lightcone effective field theory}

The action of covariant open bosonic SFT is 
\begin{eqnarray}
S \lineup = -\frac{1}{2}\big\langle \Psi,Q\Psi\big\rangle- \frac{1}{3}\big\langle\Psi,v_2(\Psi,\Psi)\big\rangle-\frac{1}{4}\big\langle\Psi,v_3(\Psi,\Psi,\Psi)\big\rangle\nonumber\\\lineup\ \ \ \ \ -\frac{1}{5}\big\langle\Psi,v_4(\Psi,\Psi,\Psi,\Psi)\big\rangle+\text{higher orders},\label{eq:Sgen}
\end{eqnarray}
where $\Psi\in\Hcov$ is the dynamical field and 
\begin{equation}
v_n: (\Hcov)^{\otimes n}\to \Hcov\label{eq:vn}
\end{equation}
are a hierarchy of string products defining a cyclic $A_\infty$ algebra. We do not limit ourselves to Witten's string field theory, so the higher order string products can be nonzero. We use the traditional Grassmann grading on the BCFT vector space, and $|X|$ denotes the Grassmann parity of an object $X$. The relation to the degree grading used in \cite{lightcone} and other works is reviewed in appendix \ref{app:suspension}. 

Once we fix lightcone gauge, the longitudinal part of the string field is not dynamical. Therefore we can integrate it out. What is left is a gauge-fixed action for a transverse string field $\Psiperp \in\Hperp$
\begin{eqnarray}
S_\mathrm{lc} \lineup =-\frac{1}{2}\big\langle \Psiperp,c_0\Lperp\Psiperp\big\rangle-\frac{1}{3}\big\langle \Psiperp,\vlc_2(\Psiperp,\Psiperp)\big\rangle-\frac{1}{4}\big\langle\Psiperp,\vlc_3(\Psiperp,\Psiperp,\Psiperp)\big\rangle\nonumber\\
\lineup \ \ \ -\frac{1}{5}\big\langle\Psiperp,\vlc_4(\Psiperp,\Psiperp,\Psiperp,\Psiperp)\big\rangle+\text{higher orders}.\ \ \ \ \ \ \label{eq:lc_eff}
\end{eqnarray}
The vertices of the gauge-fixed action are defined by string products 
\begin{equation}
\vlc_n:(\Hperp)^{\otimes n}\to \Hperp\label{eq:vlcn}
\end{equation} 
which multiply in the transverse vector space. The vertices may be characterized as a sum over Feynman graphs, in a similar manner to the vertices of an effective field theory \cite{Seneff,Okawaeff,Jakubeff}. Each node of the graph represents a vertex of the covariant SFT, each internal line represents a longitudinal propagator $\Deltalong$, and each external line is associated to $\S\Psiperp$. Up to quintic order the gauge-fixed vertices are explicitly 
\begin{eqnarray}
\langle \Psiperp, \vlc_2(\Psiperp,\Psiperp)\rangle \lineup = \Big\langle \S\Psiperp, v_2(\S\Psiperp,\S\Psiperp)\Big\rangle,\phantom{\bigg)}\\
\langle \Psiperp, \vlc_3(\Psiperp,\Psiperp,\Psiperp)\rangle  \lineup =\Big\langle \S\Psiperp,v_3(\S\Psiperp,\S\Psiperp,\S\Psiperp)\Big\rangle\phantom{\bigg)} \nonumber\\
\lineup\ \ \ \ \ \ \ 
 -2\Big\langle v_2(\S\Psiperp,\S\Psiperp), \Deltalong v_2(\S\Psiperp,\S\Psiperp)\Big\rangle,\phantom{\bigg)}\ \ \ \ \ \ \ \ 
\label{eq:cubic_quartic}\\
\langle \Psiperp,\vlc_4(\Psiperp,\Psiperp,\Psiperp,\Psiperp)\rangle \lineup = \Big\langle \S\Psiperp,v_4(\S\Psiperp,\S\Psiperp,\S\Psiperp,\S\Psiperp)\Big\rangle\phantom{\bigg)}\nonumber\\
\lineup\ \ \ \ \ \ 
-5\Big\langle v_3\big(\S\Psiperp,\S\Psiperp,\S\Psiperp\big),\Deltalong v_2(\S\Psiperp,\S\Psiperp)\Big\rangle
\phantom{\bigg)} \nonumber\\
\lineup\ \ \ \ \ \  +5\Big\langle \S\Psiperp,v_2\big(\Deltalong v_2(\S\Psiperp,\S\Psiperp),\Deltalong v_2(\S\Psiperp,\S\Psiperp)\big)\Big\rangle. \phantom{\bigg)}\ \ \ \ \ \ \ \ 
\end{eqnarray}
The quartic vertex has a contribution from diagrams with a single longitudinal propagator. This comes with a factor of $2$ because both $s$- and $t$-channel diagrams must be accounted for. Similarly, the quintic vertex has contributions from diagrams with one or two longitudinal propagators. These come with a factor of five resulting from summing over the five distinct cyclic permutations of these diagrams. Expressions such as these can be derived to any desired order by expanding the homotopy transfer formula given in \cite{lightcone} and translating to the Grassmann grading scheme following appendix \ref{app:suspension}. The structure mimics effective field theory because the gauge-fixed action is derived by integrating out part of the string field. Presently we integrate out the longitudinal part, while in low energy effective field theory we integrate out the high energy states. 

The above in principle completely defines the interactions of covariant SFT in lightcone gauge. But the definition is rather formal and algebraic. It does not tell us how string worldsheets split and join in lightcone gauge. However, this question has been addressed in the special case where the interactions of the covariant string field theory are defined by lightcone-style cubic and quartic vertices \cite{lightcone}. This is the so-called {\it Kugo-Zwiebach string field theory} \cite{KZ}, where fixing lightcone gauge results precisely in the standard lightcone string field theory of Kaku and Kikkawa~\cite{KK}. This means that the geometrical interpretation of the cubic and quartic lightcone vertices is unchanged by the Aisaka-Kazama transformation, and all Feynman graphs with longitudinal propagators evaluate to zero, a surprising phenomenon referred to as {\it transfer invariance}~\cite{lightcone}. In this paper however we want to understand lightcone gauge interactions in general. Now the interactions of the covariant SFT will not be characterized by Mandelstam diagrams, all Feynman graphs will contribute to the gauge-fixed vertices, and transfer invariance will not hold. To set the stage for discussing this, note that each vertex in lightcone gauge has a term which comes directly from the original covariant vertex,
\begin{equation}\big\langle \S\Psiperp,v_{n}(\S\Psiperp,...,\S\Psiperp)\big \rangle.\label{eq:trans_sub}\end{equation}
This can be seen as a direct interaction between transverse degrees of freedom. Therefore we call it the {\it transverse subvertex}. The remaining contributions contain longitudinal propagators. These can be seen as representing an indirect interaction generated through the exchange of longitudinal states. Therefore we call them {\it longitudinal subvertices}. We begin by analyzing the transverse subvertex in the next section. Dealing with the longitudinal subvertices requires more preparation and will be taken up in sections \ref{sec:quartic} and \ref{sec:higher}.

\section{Transverse off-shell amplitudes}
\label{sec:offshell}

Before thinking about string field theory, it will be helpful to understand the connection between covariant and lightcone formalisms at the more primitive level of off-shell amplitudes. There are two natural ways to define off-shell amplitudes between transverse string states. The first is to take a covariant off-shell amplitude, defined in the sense of \cite{Nelson,Sen_off_shell}, and assume that all external states are DDF states. The second is to construct amplitudes from the point of view of lightcone quantization of the string, and extrapolate off-shell using the geometry of Mandelstam diagrams. The result of this section is the {\it equivalence theorem}: These two notions of transverse off-shell amplitude are the same. This result can be viewed as an $n$-point generalization of the computation of the gauge-fixed cubic vertex given in \cite{lightcone}, and essentially the same mechanisms are at play. 

\subsection{Covariant off-shell amplitudes}
\label{subsec:covariant_off}

We are concerned with open bosonic strings at tree level. The $n$-point amplitude is defined by a correlation function on the disk with vertex operators inserted at $n$ points on the boundary (the punctures).  We represent the disk through a global coordinate $u$ on the upper half plane 
\begin{equation}\mathrm{Im}(u)\geq 0.\end{equation}
The punctures are a list of $n$ points on the real axis,
\begin{equation}u_1,u_2,..., u_n, \ \ \ \ \mathrm{Im}(u_i)=0, \label{eq:ui} \end{equation}
which we assume are labeled in cyclic order when tracing from positive to negative values on the real axis (the left handed convention). Two configurations of punctures are equivalent if they differ by M{\"o}bius transformation of the upper half plane. The inequivalent configurations of punctures define the moduli space $M_n$ of disks with $n$ boundary punctures. To define an off-shell amplitude it is necessary to specify local coordinates for the insertion of off-shell vertex operators at the punctures. The local coordinates are $n$ unit half disks,
\begin{equation} \xi_1,\xi_2,...,\xi_n,\ \ \ \ |\xi_i|< 1,\ \mathrm{Im}(\xi_i)\geq 0,\end{equation}
which are related to the global coordinate $u$ on the upper half plane through $n$ real, holomorphic functions, 
\begin{equation}u = f_1(\xi_1),\ \ u= f_2(\xi_2),\ \ ...\ ,\ \ u=f_n(\xi_n), \label{eq:fi}\end{equation}
called local coordinate maps. By convention the puncture always sits at the origin of the local coordinate, which means that the local coordinate maps satisfy
\begin{equation}f_i(0) = u_i.\label{eq:Pn}\end{equation}
The space of local coordinate maps modulo M{\" o}bius transformation defines the {\it covariant fiber bundle} $\P_n$. The base of the fiber bundle is the moduli space $M_n$ of disks with $n$ boundary punctures. The fiber parameterizes the choices of local coordinate maps for a fixed configuration of punctures. We call $\P_n$ ``covariant" to contrast with the ``lightcone" fiber bundle to be introduced shortly. 

The {\it covariant measure} is a linear map which turns $n$ open string states into a differential form living on $\P_n$:
\begin{equation}\langle \Omega_n|:(\Hcov)^{\otimes n}\to \Gamma(\Lambda^\bullet T^*\P_n).\label{eq:Omegan}\end{equation}
Acting on states $\phi_i\in\Hcov$, it can be written as 
\begin{equation}\langle \Omega_n|\phi_1\otimes \phi_2\otimes...\otimes \phi_n = \Omega_n(\phi_1,\phi_2,...,\phi_n).\end{equation}
The defining property of the covariant measure is the {\it BRST identity},
\begin{equation}\langle \Omega_n|{\bf Q} = -d \langle \Omega_n|.\label{eq:BRSTid}\end{equation}
Here $d$ is the exterior derivative on $\P_n$ and ${\bf Q}$ is a sum of BRST operators $Q$ acting on each open string state:
\begin{equation}
{\bf Q} = Q\otimes \mathbb{I}^{\otimes n-1} + \mathbb{I}\otimes Q\otimes \mathbb{I}^{\otimes n-2}+...+\mathbb{I}^{\otimes n-1}\otimes Q.
\end{equation}
The covariant measure (as we are defining it following \cite{ClosedSFT}) is a sum of differential forms of every degree. It also has components at all negative ghost numbers.  The inhomogeneous grading of the measure makes it possible to convert the BRST operator, which is a zero form at ghost number 1, into the exterior derivative, which is a 1-form at ghost number 0. A {\it covariant off-shell amplitude} is a linear map from $n$ open string states into a number,
\begin{eqnarray}
 \langle A_n(\C)|:(\Hcov)^{\otimes n} \lineup \to \mathbb{C}, \label{eq:AnM}
\end{eqnarray}
defined by integrating the covariant measure over an integration cycle $\C$:
\begin{equation}
\langle A_n(\C)| = \int_{\C}\langle \Omega_n|.\label{eq:offshellamp}
\end{equation}
Acting on states $\phi_i\in\Hcov$, it may be written as 
\begin{equation}\langle A_n(\C)|\phi_1\otimes \phi_2\otimes...\otimes \phi_n = A_n(\C,\phi_1,\phi_2,...,\phi_n).\end{equation}
The amplitude depends on a choice of integration cycle (or singular chain) $\C$ in the covariant fiber bundle. The integration cycle $\C$ is defined by a pair $(C,\varphi)$ consisting of an oriented manifold $C$ and an embedding map $\varphi:C\to\P_n$ which places this manifold within the covariant fiber bundle. Integrating the covariant measure over $\C$ is the same as integrating the pullback of the covariant measure over~$C$:
\begin{equation}\int_\C\langle \, \Omega_n|   = \int_{C}\,\varphi^*\langle \Omega_n|. \label{eq:varphi}\end{equation}
We also assume that integration selects the component of the measure with the same form degree as the dimension of $C$. On account of \eq{BRSTid}, the off-shell amplitude satisfies a kind of BRST Ward identity
\begin{equation}\langle A_n(\C)|{\bf Q}=-\langle A_n(\d \C)|,\label{eq:AnQdAn}\end{equation}
where $\d \C$ is the boundary of $\C$. Presently we do not place any restriction on the choice of $\C$, but traditionally an off-shell amplitude is defined by a $\C$ whose bundle projection covers the moduli space. Then the amplitude will be BRST invariant if contributions from the boundary of moduli space can be ignored. 

We now explain how to construct the covariant measure. The zero-form part is a surface state
\begin{equation}\langle \Sigma_n|:(\Hcov)^{\otimes n}\to C^{\infty}(\P_n)\label{eq:Sigman}\end{equation}
which can be thought of as a linear map of $n$ open string fields into a function on the covariant fiber bundle. 
Acting on states $\phi_i\in\Hcov$ it may be written as 
\begin{equation}\langle \Sigma_n |\phi_1\otimes \phi_2\otimes...\otimes \phi_n = \Sigma_n(\phi_1,\phi_2,...,\phi_n).\end{equation}
The surface state is given by an $n$-point correlation function on the upper half plane,
\begin{equation}\Sigma_n(\phi_1,\phi_2, ...,\phi_n) = \big\langle f_1\circ \phi_1(0)\, f_2\circ \phi_2(0)\,...\,f_n\circ \phi_n(0)\big\rangle_\mathrm{UHP}.\end{equation}
This is a function on $\P_n$ through its dependence on the local coordinate maps. The surface state is BRST invariant,
\begin{equation}\langle \Sigma_n|{\bf Q} = 0.\end{equation}
To find the higher form components of the measure we must insert appropriate $b$-ghosts into the correlation function. There are several ways to do this. We describe the approach based on the Schiffer variation \cite{Nelson,Zwiebach}, where $b$-ghosts appear as contour integrals around the punctures. The contour integrals are defined by holomorphic vector fields $v_i$, called Schiffer vector fields, defined on each local coordinate patch which take values as a 1-form on $\P_n$. These represent a variation of the local coordinate maps as a set of diffeomorphisms of the local coordinate patches. In this way the Schiffer vector fields satisfy
\begin{equation}d f_i(\xi) = -v_i(\xi)\d f_i(\xi).\label{eq:vni}\end{equation}
It follows that 
\begin{equation}d v_i(\xi) = \frac{1}{2}[v_i,v_i](\xi),\label{eq:dvni}\end{equation}
where the bracket on the right hand side is the Lie bracket of vector fields. We define a $b$-ghost contour integral around the $i$th puncture
\begin{equation}\mathscr{B}_{i} = \oint_0\frac{d\xi}{2\pi i}v_i(\xi)b(\xi).\label{eq:bni}\end{equation}
To simplify signs we assume that 1-forms on $\P_n$ are uniformly Grassmann odd, which means that they anticommute not only with each other but also with Grassmann odd worldsheet operators~\cite{ClosedSFT}. Therefore the order of the Schiffer vector field and the $b$-ghost in \eq{bni} is meaningful, and the operator $\mathscr{B}_i$ is Grassmann even. We have a similar operator made from the energy-momentum tensor
\begin{equation}\mathscr{T}_{i} = [\mathscr{B}_{i},Q] = \oint_0\frac{d\xi}{2\pi i}v_i(\xi)T(\xi)\end{equation}
which is Grassmann odd. We collect the operators around each puncture into operators which act on all punctures:
\begin{eqnarray}
{\bf B} \lineup = \mathscr{B}_{1}\otimes\mathbb{I}^{\otimes n-1}+ \mathbb{I}\otimes \mathscr{B}_{2}\otimes \mathbb{I}^{\otimes n-2}+\,...\,+\mathbb{I}^{\otimes n-1}\otimes \mathscr{B}_{n},\\
{\bf T} \lineup = \mathscr{T}_{1}\otimes\mathbb{I}^{\otimes n-1}+ \mathbb{I}\otimes \mathscr{T}_{2}\otimes \mathbb{I}^{\otimes n-2}+\,...\,+\mathbb{I}^{\otimes n-1}\otimes \mathscr{T}_{n}.
\end{eqnarray}
These satisfy
\begin{eqnarray}
{\bf T} \lineup = [{\bf B},{\bf Q}],\\
d{\bf B} \lineup = \frac{1}{2}[{\bf B},{\bf T}],\\
d\langle \Sigma_n| \lineup = -\langle\Sigma_n|{\bf T},
\end{eqnarray}
as a consequence of \eq{vni} and \eq{dvni}. One can then show that the covariant measure constructed as 
\begin{equation}\langle \Omega_n| = \langle \Sigma_n|e^{{\bf B}}\end{equation}
will satisfy the BRST identity \eq{BRSTid}. We will often want the measure expressed as a correlation function on the upper half plane: 
\begin{equation}
\Omega_n(\phi_1,...,\phi_n) = \Big\langle \exp\!\big(\mathscr{B}\big)\,f_1\circ \phi_1(0)\,f_2\circ\phi_2(0)\, ... \, f_n\circ\phi_n(0)\Big\rangle_\UHP.
\end{equation}
The relevant $b$-ghost insertion $\mathscr{B}$ is given by transforming each $\mathscr{B}_{i}$ to the upper half plane:
\begin{eqnarray}
\mathscr{B} \lineup = f_1\circ \mathscr{B}_1\, +\,  f_2\circ \mathscr{B}_2 \, +\,  ...\,  +\, f_n\circ \mathscr{B}_n\phantom{\bigg)}\nonumber\\
\lineup  = \oint_{u_1}\frac{du}{2\pi i}V_1(u)b(u) + \oint_{u_2}\frac{du}{2\pi i}V_2(u)b(u)+\ ...\ + \oint_{u_n}\frac{du}{2\pi i}V_n(u)b(u),\ \ \ \ \ \label{eq:Bn}
\end{eqnarray}
where 
\begin{equation}V_i(u) = \frac{d f_i^{-1}(u)}{\d f_i^{-1}(u)}\end{equation}
are the Schiffer vector fields expressed in the upper half plane coordinate.

Let us make a few comments about signs. The integration of the measure is understood to mean the conventional integration of differential forms assuming that all differentials have been commuted to the left through all states, operators, dual states, and outside of correlation functions. Since the surface state $\langle \Sigma_n|$ is Grassmann odd, for consistency we should include a sign when commuting differentials on $\P_n$ to the left outside of a correlation function:
\begin{equation}\langle dt\, ...\, \rangle = -dt\langle \, ...\,\rangle. \end{equation}
Finally, to integrate we have to fix an orientation of the relevant integration cycle. The most important integration cycle is the moduli space $M_n$, and we define its orientation as follows. Using $SL(2,\mathbb{R})$ transformation we fix the locations of the punctures $u_1,u_{n-1},u_n$. The remaining punctures $u_2,...,u_{n-2}$ serve as coordinates on the moduli space. The orientation of the moduli space is defined so that
\begin{equation}
	\int_{M_n}du_2du_3...du_{n-2}\Big(\ ...\ \Big)=\int_{M_n}|du_2du_3...du_{n-2}|\Big(\ ...\ \Big).\label{eq:Mnorient}
\end{equation}
The left hand side represents integration of an $n$-form over $M_n$, and the right hand side represents the traditional Riemann-Lebesgue integral over $M_n$ defined as a set. The absolute value is used to denote the integration density corresponding to a product of basis 1-forms.

\subsection{Mandelstam diagrams}
\label{subsec:Mandelstam}

We now consider {\it lightcone off-shell amplitudes}, a notion which follows naturally from lightcone quantization of the string. In lightcone quantization, the geometry of string interactions is characterized by Mandelstam diagrams. 

\begin{figure}
\begin{center}
\resizebox{3.5in}{1.5in}{\includegraphics{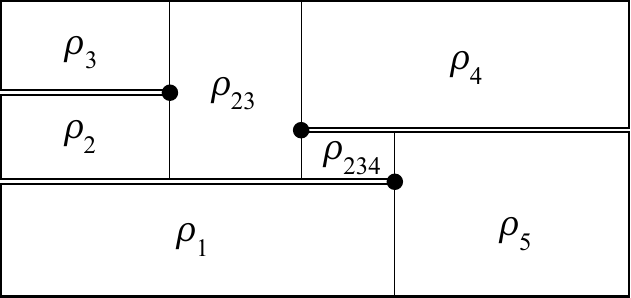}}
\end{center}
\caption{\label{fig:lightcone_gauge9} A Mandelstam diagram is given by gluing together rectangular strip domains $\rho_i$. The strip domains $\rho_1,...,\rho_5$ in this figure represent external states, and should be imagined as extending to plus or minus infinity. The strip domains $\rho_{23}$  and $\rho_{234}$ represent propagators. We have chosen to label the propagators by the list of punctures which are separated from the first puncture at degeneration.}
\end{figure}

If the external states have definite lightcone momentum, a unique Mandelstam diagram can be constructed for every point in the moduli space of Riemann surfaces relevant to a given amplitude~\cite{Giddings}. Presently we are interested in $n$-point open string amplitudes at tree level, where a point in moduli space is characterized as previously by positions of $n$ boundary punctures $u_1,...,u_n$ on the upper half plane (modulo M{\"o}bius transformation). Each puncture has a respective length parameter $\alpha_1,...,\alpha_n$. The length parameter $\alpha_i$ is related to the lightcone momentum of the state at the $i$th puncture as 
\begin{equation}\alpha_i = 2k_-^i.\label{eq:alphai}\end{equation}
With this data, the Mandelstam diagram is obtained by transforming the upper half plane with the Mandelstam mapping, 
\begin{equation}\rho(u) = \sum_{i=1}^n\alpha_i\ln(u-u_i).\label{eq:Mand}\end{equation} 
As shown in figure \ref{fig:lightcone_gauge9}, the Mandelstam diagram is composed of a set of rectangular strip domains~$\rho_i$. If $i\in \{1,...,n\}$, the strip domain $\rho_i$ extends to infinity and contains the image of the puncture $u_i$. The remaining strip domains are internal to the diagram and represent propagators. The number of propagators will be written as $n_p$. For clarity we sometimes use
\begin{eqnarray}
i\lineup \in \puncture = \{1,...,n\},\\
i\lineup \in \propagator \label{eq:puncprop}
\end{eqnarray}
to indicate when the index $i$ labels a puncture or a propagator. Each rectangular strip domain has a vertical height and horizontal width. The vertical height of $\rho_i$ is $\pi\alpha_i$. For propagator strips, the string length $\alpha_i$ is positive (by convention) and otherwise is determined via momentum conservation by the length parameters of external states. The horizontal width of $\rho_i$ will be written $T_i$. For propagator strips $T_i$ is assumed to be positive. If $i$ labels a puncture, it is natural to define $T_i$ as positive and infinite for incoming strips, and negative and infinite for outgoing strips. In this way the signs of the propagator widths and string length parameters agree. 

When gluing the strip domains together there will be $n_c$ points on the Mandelstam diagram where three strip domains touch. These are called {\it cubic interaction points}. There will be $n_q$ points where four strip domains touch. These are called {\it quartic interaction points}. The numbers of punctures, propagators, cubic and quartic interaction points are related as
\begin{eqnarray}
n_p+n_q \lineup = n-3,\\
n_c+2n_q\lineup =n-2.
\end{eqnarray}
The positions of the interaction points on the Mandelstam diagram are given as $\rho(U_I)$, where $U_I$ are roots of the equation 
\begin{equation}\d\rho(U_I)=0.\label{eq:drho}\end{equation}
There are a total of $(n-2)$ roots. We use $I\in \cubic$ to label the roots which are real valued. These are the preimages of the cubic interaction points on the Mandelstam diagram. We use $I\in\quartic$ to label complex roots with positive imaginary part. These are the preimages of the quartic interaction points on the Mandelstam diagram. There are also roots with negative imaginary part, but these are determined through complex conjugation of the roots with positive imaginary part. The Mandelstam diagram describes a scattering process unfolding in (Euclidean) time, where the time coordinate is identified with the real part of the coordinate $\rho$. The strings split and join at specific {\it interaction times} defined by
\begin{equation}\tau_I = \mathrm{Re}[\rho(U_I)],\ \ \ \ I\in \cubic\cup\quartic.\label{eq:tauI}\end{equation}
We need to be able to refer to the interaction points on either side of a propagator strip~$\rho_i$. The interaction point with greater interaction time will be called the {\it successor}, and the interaction point with lesser interaction time will be called the {\it predecessor}. For a strip domain $\rho_i$, the successor will be labeled with $s(i)$ and the predecessor will be labeled with $p(i)$. In this way, the width of the $i$th propagator can be written as 
\begin{equation}T_i = \tau_{s(i)} -\tau_{p(i)},\ \ \ \ i\in\propagator,\label{eq:Ti}\end{equation}
and this is positive. The strip domain of a puncture only touches one interaction point. By convention, we refer to this interaction point as the successor. The strip domain of a puncture does not have a predecessor.

Once we have collected $n+n_p$ strip domains and specified how they are glued together, the geometry of the Mandelstam diagram is determined by $(n-3)$ real parameters. The parameters include the widths $T_i$ of the propagator strips and the displacements of the quartic interaction points from the open string boundary. Up to a shift and factor of two, the latter are equivalent to 
\begin{equation}\theta_I = \mathrm{Im}\big(\rho(U_I) - \rho(U_I^*)\big),\ \ \ \ I\in \quartic. \label{eq:thI}\end{equation} 
The parameters $(T_i,\theta_I)$ are coordinates on the moduli space $M_n$ of disks with boundary punctures. They are related to the positions of the punctures $u_1,...,u_n$ through a complicated coordinate transformation. Note that $(T_i,\theta_I)$ only cover a patch of the moduli space. To cover the whole moduli space we must sum all Feynman graphs which contribute to a given amplitude, and each graph represents a class of Mandelstam diagrams with its own set of parameters $(T_i,\theta_I)$.

\begin{figure}
\begin{center}
\resizebox{6.5in}{2in}{\includegraphics{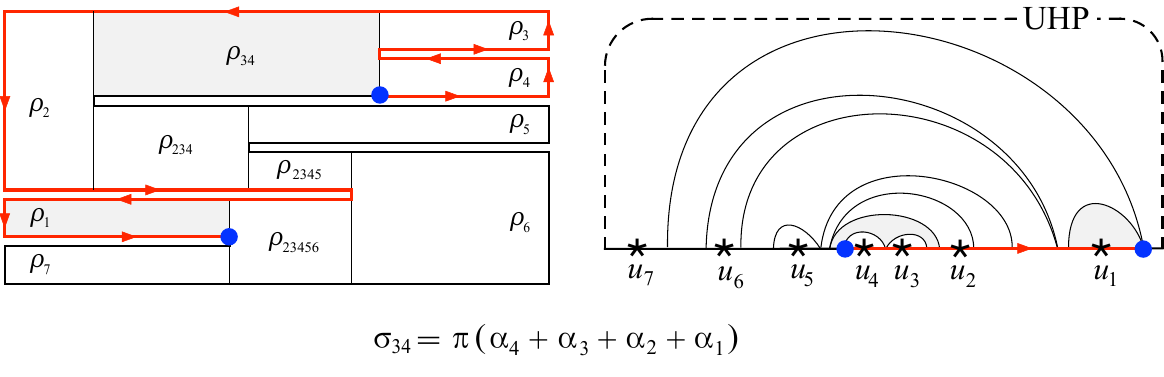}}
\end{center}
\caption{\label{fig:lightcone_gauge8} This figure shows how to construct the vertical displacement $\sigma_{34}$ of the strip domain $\rho_{34}$. First we note that $\rho_{34}$ does not extend to $+\infty$, so we mark a point on the lower right hand corner. Second, we note that $u_1$ is the rightmost puncture in the upper half plane. Since $\rho_1$ does not extend to $+\infty$, we also mark a point on the lower right hand corner. Then we trace a counterclockwise path connecting the marked points. Every time the path crosses a puncture, there is a corresponding contribution to the vertical displacement $\sigma_{34}$. }
\end{figure}

\begin{figure}[t]
\begin{center}
\resizebox{5.5in}{4.3in}{\includegraphics{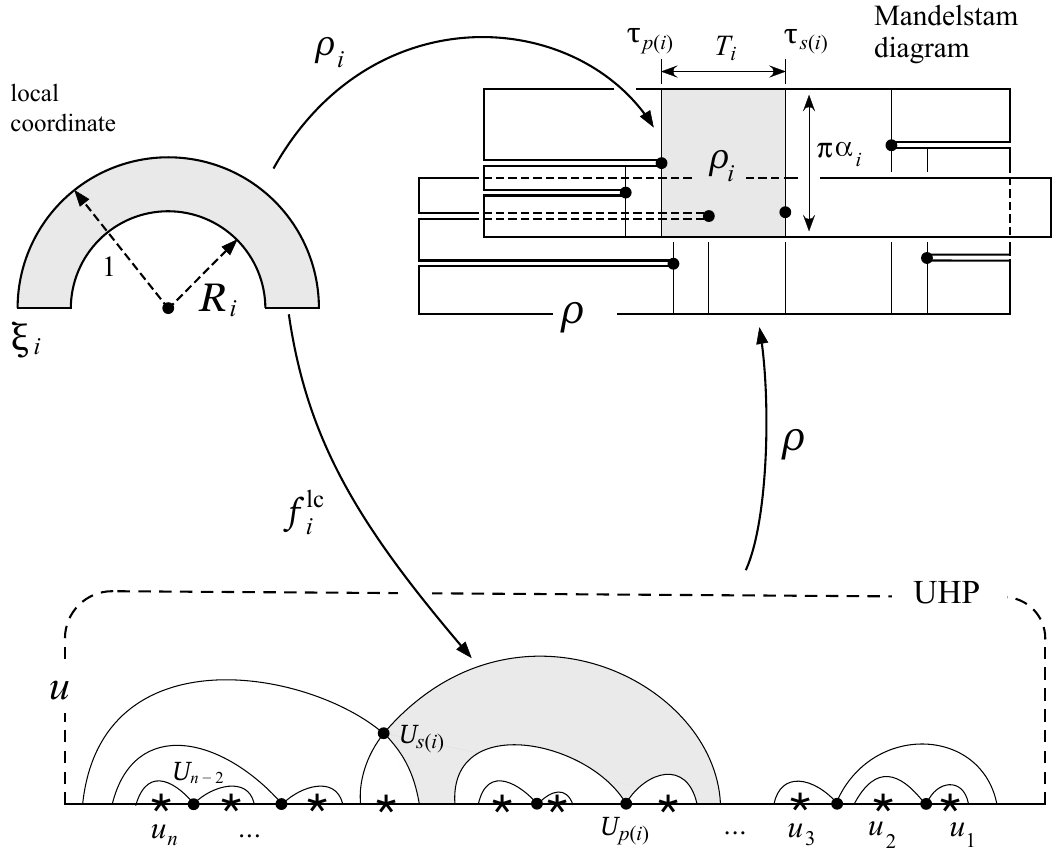}}
\end{center}
\caption{\label{fig:lightcone_gauge1} Conformal transformations between the upper half plane $u$, the Mandelstam diagram $\rho$, and the local coordinates $\xi_i$ on the strip domains $\rho_i$ within the Mandelstam diagram.}
\end{figure}

The geometry of the Mandelstam diagram suggests local coordinates which may be used to extend amplitudes off-shell. Each strip domain $\rho_i$ can be covered by a local coordinate $\xi_i$ which belongs to the unit half disk minus a smaller concentric half-disk:
\begin{equation}R_i\leq |\xi_i|\leq 1,\ \ \ \mathrm{Im}(\xi_i)\geq 0,\end{equation}
where $R_i$ is the radius of the smaller half-disk. The coordinates are mapped to the strip domains through
\begin{equation}\rho_i(\xi_i) = \tau_{s(i)} +i\sigma_i + \alpha_i \ln \xi_i.\label{eq:rhoi}\end{equation}
The real parameters $R_i,\tau_{s(i)},\sigma_i$ are chosen to ensure that the local coordinate covers the entirety of the $i$th strip domain and nothing more. As described before, $\tau_{s(i)}$ is the interaction time of the successor to $\rho_i$. The inner radius is given by
\begin{equation}R_i = e^{-T_i/\alpha_i},\label{eq:Ri}\end{equation}
which is zero for the strip domains of external states. The vertical displacement $\sigma_i$ is determined as follows: 
\begin{description}
\item{(1)} If the strip domain $\rho_i$ extends to $+\infty$, mark a point on the upper left hand corner. Otherwise, mark a point on the lower right hand corner.
\item{(2)} On the upper half plane, find the rightmost puncture $u_{i_\mathrm{max}}$ on the real axis. This will satisfy $u_{i_\mathrm{max}}> u_i$ for all other punctures $u_i$. Consider the corresponding strip domain $\rho_{i_\mathrm{max}}$. If it extends to $+\infty$, mark a point on the upper left hand corner. Otherwise, mark a point on the lower right hand corner. 
\item{(3)} Draw a counterclockwise path on the boundary of the Mandelstam diagram connecting the marked points. Every time the path encounters a puncture $u_j$, add $\pi\alpha_j$. The result is the vertical displacement $\sigma_i$ of the strip domain $\rho_i$.
\end{description}
This is illustrated in figure \ref{fig:lightcone_gauge8}. This procedure expresses the vertical displacements in the form
\begin{equation}\sigma_i = \pi \sum_{j\in\puncture} m_{ij}\alpha_j, \label{eq:sigmai}\end{equation}
where $m_{ij}$ is a rectangular $(n+n_p)\times n$ matrix of 1s and 0s. With this data we can determine local coordinate maps for each strip domain: 
\begin{equation}\flc_i(\xi_i) = \rho^{-1}\circ\rho_i(\xi_i).\label{eq:flc}\end{equation}
We refer to these as the {\it lightcone local coordinate maps}. The inverse maps may be written in closed form
\begin{eqnarray}
(\flc_i)^{-1}(u) \lineup = \exp\!\left(\frac{\rho(u) - (\tau_{s(i)} + i\sigma_i)}{\alpha_i}\right),\label{eq:flcinv}\\
\lineup = \prod_{j=1}^n\left(\frac{(-1)^{m_{ij}}(u-u_j)}{|U_{s(i)}-u_j|}\right)^{\alpha_j/\alpha_i}.
\end{eqnarray}
The relation between the local coordinates $\xi_i$, the strip domains $\rho_i$, the Mandelstam diagram $\rho$, and the upper half plane $u$ is summarized in figure \ref{fig:lightcone_gauge1}. With local coordinates specified we can extend lightcone string amplitudes off-shell. These are the off-shell amplitudes of Kaku and Kikkawa's lightcone string field theory.

\subsection{Lightcone off-shell amplitudes}
\label{subsec:lightcone_off}

However, we are interested in something slightly more general. When going off-shell we allow a scale transformation of each external state generated by the operator
\begin{equation}e^{-\lambda_i L_0}.\label{eq:stubop}\end{equation}
The real constant $\lambda_i$ is called the {\it stub length}. The stub length can be chosen independently for every puncture and at each point in the moduli space. The general lightcone off-shell amplitude is then defined by local coordinate maps of the form
\begin{equation}\flc_i\Big(e^{-\lambda_i}\xi_i\Big) = \flc_i\circ e^{-\lambda_i}(\xi),\ \ \ \ i\in\puncture. \label{eq:flclambda}\end{equation}
The difference from Kaku and Kikkawa's lightcone string field theory is that the local coordinates do not cover the entirety of the respective strip domains. The strip domain $\rho_i$ contains a strip of length 
\begin{equation}\alpha_i\lambda_i\end{equation}
adjacent to the interaction point which is not covered. The uncovered part of the strip domain is called a {\it stub}. The concept of stubs originates from \cite{Zwiebach,Zwiebach1990}, where they were introduced as a device to prevent overcounting of moduli in loops. More recently they have been discussed in connection to effective field theory \cite{Seneff,Stettinger,ErbinFirat,Wilsonian,MaccaferriStubs} and as a mechanism to tame spurious singularities in superstring field theory \cite{SenRev}. The role of stubs in lightcone string theory however is a bit different.
 
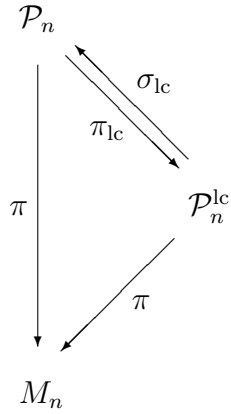
\begin{figure}
\begin{center}
\setlength{\unitlength}{.25cm} 
\begin{picture}(11,21)
\put(.5,20){$\P_n$}
\put(1.5,18){\vector(0,-1){15}}
\put(3,18.5){\vector(1,-1){6}}
\put(9.5,13){\vector(-1,1){6}}
\put(9.5,10){$\Plc_n$}
\put(.5,0){$M_n$}
\put(8.75,8.75){\vector(-1,-1){6}}
\put(0,10){$\pi$}
\put(6.5,5){$\pi$}
\put(4.25,14.25){$\pilc$}
\put(6.75,16.75){$\sigmalc$}
\end{picture}
\end{center}
\caption{\label{fig:lightcone_gauge3} Relation between the covariant fiber bundle, lightcone fiber bundle, and the moduli space.}
\end{figure}

This motivates the definition of the {\it lightcone fiber bundle}, $\Plc_n$. The base is the moduli space $M_n$ of $n$-punctured disks, and the fiber assigns stub lengths to the Mandelstam diagram associated to a given point in $M_n$. Another characterization is as the moduli space of disks with a dilatation specified around each of $n$ boundary punctures. The scaling factor for each dilatation is related to the respective stub length through
\begin{equation}r_i = e^{-\lambda_i} \rlc_i,\ \ \ \ i\in\puncture, \label{eq:ri}\end{equation}
where $\rlc_i$ is the conformal radius of the lightcone local coordinate map,
\begin{eqnarray} \rlc_i = \d\flc_i(0) \lineup = e^{\tau_{s(i)}/\alpha_i}\prod_{j=1,\neq i}^n\frac{1}{ |u_i-u_j|^{\alpha_j/\alpha_i}} \label{eq:dfi}\\
\lineup = |U_{s(i)}-u_i|\prod_{j=1,\neq i}^n\left|\frac{U_{s(i)}-u_j}{u_i-u_j}\right|^{\alpha_j/\alpha_i},\ \ \ \ i\in\puncture. \end{eqnarray}
Let us explain how this relates to the covariant fiber bundle $\P_n$. Consider a power series expansion of the local coordinate maps around the origin:
\begin{equation}f_i(\xi) = u_i+r_i\xi +\frac{1}{2!}\d^2 f_i(0)\xi^2+ ...\ ,\label{eq:fniexp}\end{equation}
where
\begin{equation}u_i = f_i(0),\ \ \ r_i = \d f_i(0).\end{equation}
The constant term defines the location of the $i$th puncture. The linear term specifies a dilatation of the puncture by a scaling factor $r_i$. The scaling factor is also called the {\it conformal radius} of the local coordinate map $f_i(\xi)$. The moduli space $M_n$ appears when we consider only the constant term in the expansion, and ignore everything else. This expresses the existence of a projection 
\begin{equation}\pi:\P_n\to M_n\end{equation}
which maps the covariant fiber bundle down to the moduli space. Similarly, the lightcone fiber bundle appears when we consider the constant {\it and} the linear term in \eq{fniexp}, and ignore everything else. This expresses the existence of a projection
\begin{equation}\pilc: \P_n\to \Plc_n\end{equation}
which maps the covariant fiber bundle down to the lightcone fiber bundle. Of course, the higher order structure of the local coordinate maps is still important in extending lightcone amplitudes off-shell. But this data is determined once the location of the punctures and respective dilatations have been specified. This can be expressed by the existence of a canonical section 
\begin{equation}\sigmalc:\Plc_n\to\P_n\end{equation}
which embeds the lightcone fiber bundle into the covariant fiber bundle by reconstructing the maps \eq{flclambda} from the punctures and dilatations. The section map satisfies
\begin{equation}\pilc\circ\sigmalc = \mathrm{id}.\end{equation}
Finally, we can project the lightcone fiber bundle down to the moduli space by forgetting about the dilatations:
\begin{equation}\pi:\Plc_n\to M_n.\end{equation}
The various bundle maps are summarized in figure \ref{fig:lightcone_gauge3}. One might notice that the lightcone fiber bundle is also the appropriate structure for defining off-shell amplitudes between conformal primary states. This is not a coincidence, as will be explained shortly.


A {\it lightcone off-shell amplitude} is a  multilinear map acting on transverse states,
\begin{equation}\langle \Alc_n(\Clc)|: (\Hperp)^{\otimes n}\to\mathbb{C},\end{equation}
which depends on an integration cycle $\Clc$ in the lightcone fiber bundle. Applied to states $a_i\in \Hperp$, the amplitude is written as
\begin{equation}\langle \Alc_n(\Clc) |a_1\otimes a_2\otimes...\otimes a_n = \Alc_n(\Clc,a_1,a_2,...,a_n).\end{equation}
The amplitude is defined by integrating the {\it lightcone measure},
\begin{equation}\langle\Omegalc_n|:\Hperp^{\otimes n}\to \Gamma(\Lambda^\bullet T^*\Plc_n),\end{equation}
over the integration cycle $\Clc$: 
\begin{equation}\langle \Alc_n(\Clc)| = \int_{\Clc}\langle \Omegalc_n|.\end{equation}
Applied to states $a_i\in \Hperp $, the measure is written as
\begin{equation}\langle \Omegalc_n|a_1\otimes a_2\otimes...\otimes a_n = \Omegalc_n(a_1,a_2,...,a_n).\end{equation}
The construction of the lightcone measure is one of the trickier aspects of the lightcone approach to string perturbation theory. As defined below, the lightcone measure satisfies an analogue of the BRST identity, 
\begin{equation}\langle \Omegalc_n|{\bf Q}_\mathrm{lc}= - d\langle\Omegalc_n|,\end{equation}
but this is not very interesting because lightcone-quantized string states are never $\Qlc$-exact (at ghost number 1). Lorentz invariance is a more meaningful way to understand the consistency of the lightcone measure \cite{Mandelstam,Mandelstam3}, though we will not discuss it here. In section \ref{sec:freeze} we will derive the measure in a different way by showing that longitudinal states are ``frozen" inside the propagator strips of a Mandelstam diagram. For now we describe the lightcone measure without derivation. The lightcone measure can be presented in a number of forms whose equivalence is not obvious. Below we mention three: 
\begin{description}
\item{\bf Reduced measure.} This is the measure as conventionally expressed from the point of view of lightcone quantization of the string. It is ``reduced" in the sense that the longitudinal BCFT is absent from its formulation. We describe the reduced measure only for transverse states at ghost number 1, which may be written as 
\begin{equation}a_i = a_i^\perp(0) |-,\kpar^i\rangle, \end{equation}
where $a_i^\perp(0)$ is a vertex operator in the transverse BCFT.  Then the reduced measure takes the form 
\begin{eqnarray}
\Omegalc_n(a_1,a_2,..., a_n) \lineup = (2\pi)^2\delta^2(\kpar^1+\kpar^2+...+\kpar^n) \label{eq:red_measure}\\
\lineup \ \ \   \times  \left|\prod_{i\in\propagator} dT_i \prod_{I\in\quartic} d\theta_I\,\right| \nonumber\\
\lineup \ \ \  \times \prod_{i\in\puncture} e^{-\lambda_i((\kpar^i)^2-1)}\prod_{i\in \propagator} e^{-k_+^i T_i}\nonumber\\
\lineup\ \ \ \times \Big\langle \big(\rho_1\circ e^{-\lambda_1}\circ a_1^\perp(0)\big)\big(\rho_2\circ e^{-\lambda_2}\circ a_2^\perp(0)\big)...\big(\rho_n\circ e^{-\lambda_n}\circ a_n^\perp(0)\big)\Big\rangle_{(T_i,\theta_I)}^{\BCFT_\perp}.\nonumber 
\end{eqnarray}
On the second line we have a product of differentials formed out of the coordinates \eq{Ti} and \eq{thI} on moduli space. We assume that the reduced measure will be integrated over a local section of $\Plc_n$ with the same orientation as moduli space, which allows us to fix the overall sign and write the second line as an integration density. On the third line we have a product of exponentials originating from the stubs \eq{stubop} and transverse propagators \eq{Deltaperp}. In particular, the transverse propagator may be written
\begin{equation}\frac{1}{\Lperp} = \frac{1}{\alpha_i}\int_0^\infty |dT_i|\, \Big(e^{-k_+^i T_i}\Big)\,e^{- T_i(L_0^\perp-1)/\alpha_i}.\end{equation}
The factor in parentheses appears explicitly as a factor in the reduced measure.  Finally, on the third line we have a correlation function of $n$ vertex operators in the transverse BCFT on the Mandelstam diagram characterized by coordinates $(T_i,\theta_I)$. The correlation function may be mapped to the upper half plane: 
\begin{eqnarray}
\lineup \Big\langle \big(\rho_1\circ e^{-\lambda_1}\circ a_1^\perp(0)\big)\big(\rho_2\circ e^{-\lambda_2}\circ a_2^\perp(0)\big)...\big(\rho_n\circ e^{-\lambda_n}\circ a_n^\perp(0)\big)\Big\rangle_{T_i,\theta_I}^{\BCFT_\perp}\nonumber\\
\lineup \ \ \ \ \ \ \  = Z_n \Big\langle \big(\flc_1\circ e^{-\lambda_1}\circ a_1^\perp(0)\big)\big(\flc_2\circ e^{-\lambda_2}\circ a_2^\perp(0)\big)...\big(\flc_n\circ e^{-\lambda_n}\circ a_n^\perp(0)\big)\Big\rangle_{\UHP}^{\BCFT_\perp}.\ \ \ \ \ \ \ \ 
\label{eq:partition_func}\end{eqnarray}
Since the transverse BCFT has nonzero central charge, the transformation generates a factor $Z_n$ representing the partition function on the Mandelstam diagram. The partition function is related to the determinant of the Laplacian on the Mandelstam diagram. The determinant however is divergent due to curvature singularities on the surface of the Mandelstam diagram, and this divergence must be regularized carefully to give a consistent result. Details can be found in appendix 11.A of Green, Schwarz, and Witten \cite{GSWII}, following the earlier work of Mandelstam~\cite{Mandelstam2}. See \cite{Sonoda,DHoker} for additional discussion which also extends to loops, and \cite{BabaIshibashi} for a derivation based on the anomalous conformal transformation of the transverse energy-momentum tensor. The partition function on the Mandelstam diagram is found to be\footnote{This relates to the expression written in \cite{BabaIshibashi} as
\begin{equation}e^{-\Gamma[\phi]}=\frac{(Z_n)^2}{\alpha_1\alpha_2...\alpha_n}.\end{equation}
The inverse product of $\alpha_i$s is a conventional normalization, and the square appears because we consider open strings, rather than closed strings.}
\begin{equation}Z_n = \frac{\left|\sum_{i=1}^n\alpha_i u_i\right|^2}{\prod_{i=1}^n\Big(\sqrt{|\alpha_i|}\rlc_i\Big)\prod_{I=1}^{n-2}\sqrt{|\d^2\rho(U_I)|}}.\label{eq:Zn}\end{equation} 
\item{\bf Covariantized measure (Schiffer form).} A second representation of the lightcone measure is as the pullback of the covariant measure:
\begin{equation}\langle \Omegalc_n| = (\sigmalc)^* \langle\Omega_n|.\label{eq:covariantized}\end{equation}
Since the covariant and lightcone measures act on different vector spaces, implicit in this equation is a trivial inclusion which relabels one vector space as the other. While this form of the measure looks natural, it is not obvious why it is correct. Various forms of this result are known. It follows from the computation of \cite{KZ} showing that transverse Siegel gauge amplitudes in the Kugo-Zwiebach SFT are the same as the amplitudes of Kaku and Kikkawa's lightcone SFT. It is also related to the procedure of covariantization of lightcone amplitudes discussed by Baba, Ishibashi, and Murakami \cite{BabaIshibashi}. If the covariant measure is expressed through the Schiffer variation, its pullback takes the form
\begin{eqnarray}
\lineup\!\!\! \Omegalc_n(a_1,a_2,...,a_n) \nonumber\\
\lineup = \Big\langle\exp\!\Big[ (\sigmalc)^* \mathscr{B}\Big]\big(\flc_1\circ e^{-\lambda_1}\circ a_1(0)\big)\big(\flc_2\circ e^{-\lambda_2}\circ a_2(0)\big)...\big(\flc_n\circ e^{-\lambda_n}\circ a_n(0)\big)\Big\rangle_\UHP,\ \ \ \ \ \ \ \ \ 
\label{eq:covariantized_Schiffer}
\end{eqnarray}
where the $b$-ghost is
\begin{eqnarray}
(\sigmalc)^* \mathscr{B}  \lineup = d\lambda_1 \big(\flc_1\circ e^{-\lambda_1}\circ b_0\big)+d\lambda_2 \big(\flc_2\circ e^{-\lambda_2}\circ b_0\big)+...+d\lambda_n \big(\flc_n\circ e^{-\lambda_n}\circ b_0\big)\phantom{\bigg)}\nonumber\\
\lineup \ \ \ +\oint_{u_1}\frac{du}{2\pi i}V_1^\mathrm{lc}(u)b(u) + \oint_{u_2}\frac{du}{2\pi i}V_2^\mathrm{lc}(u)b(u)+\ ...\ + \oint_{u_n}\frac{du}{2\pi i}V_n^\mathrm{lc}(u)b(u),\ \ \ \ \ \ 
\label{eq:sigmalcBn}
\end{eqnarray}
where 
\begin{equation}V_i^\mathrm{lc}(u) = \frac{d(\flc_i)^{-1}(u)}{\d (\flc_i)^{-1}(u)},\end{equation}
and $d$ is the exterior derivative on $\Plc_n$. Actually we can simplify this  because lightcone vertex operators always have form form of $c(0)$ or $c\d c(0)$ times a matter operator. This allows us to simplify to
\begin{eqnarray}
(\sigmalc)^* \mathscr{B} \lineup = d\lambda_1(b_0)_{u_1}\,+\,d\lambda_2(b_0)_{u_2}\, +\ ...\ +\  d\lambda_n(b_0)_{u_n} \nonumber\\
\lineup \ \ \  -du_1 (b_{-1})_{u_1}  - du_2 (b_{-1})_{u_2} -\ ...\ -du_n (b_{-1})_{u_n},\label{eq:bsimp}
\end{eqnarray}
where 
\begin{equation}
	(b_k)_{u_i} = \oint_{u_i}\frac{du}{2\pi i}(u-u_i)^{k+1}b(u)\label{eq:bui}
\end{equation}
are $b$-ghost mode operators centered at $u_i$ in the upper half plane. Note that the differentials $d\lambda_i$ only appear for unphysical amplitudes involving states at ghost number~2. Also, the differentials $du_i$ are not linearly independent due to the freedom to perform M{\"o}bius transformation of the upper half plane. 
\item{\bf Covariantized measure (Kugo-Zwiebach form).}
A second representation of the covariantized measure uses a configuration of $b$-ghost insertions which follow from computing Siegel gauge off-shell amplitudes in the Kugo-Zwiebach SFT. In this case an insertion of $b_0$ accompanies each propagator strip, and the quartic lightcone vertex provides additional $b$-ghost insertions at quartic interaction points (as discussed in appendix B.3 of \cite{lightcone}). The measure in this form can be written as a correlation function on the upper half plane:
\begin{eqnarray}
\lineup \!\!\!\!\!\! \Omegalc_n(a_1,a_2,..., a_n) \nonumber\\
\lineup=\Big\langle\exp\!\big(\mathscr{B}^\mathrm{KZ}\big) \big(\flc_1\circ e^{-\lambda_1}\circ a_1(0)\big) \big(\flc_2\circ e^{-\lambda_2}\circ a_2(0)\big)...\big(\flc_n\circ e^{-\lambda_n}\circ a_n(0)\big)\Big\rangle_\UHP,\ \ \ \ \ \ \ \ \ 
\label{eq:covariantized_KZ}
\end{eqnarray}
where the $b$-ghost insertion is
\begin{eqnarray}
\mathscr{B}^\mathrm{KZ} \lineup = \sum_{i\in\propagator}\frac{dT_i }{\alpha_i}\big(\flc_i\circ b_0\big) +\sum_{I\in \quartic}d\theta_I\left(\mathrm{Im}\left[\frac{b(U_I)}{\d^2\rho(U_I)}\right]\right) \nonumber\\
\lineup\ \ \  +\sum_{i\in\puncture} d\lambda_i\big(\flc_i\circ e^{-\lambda_i}\circ b_0\big).
\label{eq:BnKZ}
\end{eqnarray}
The first term represents the expected $b_0$ from the Siegel gauge propagators, and the second term comes from the quartic lightcone vertices. The third comes from varying the stub lengths.  
\end{description} 
We prove the equivalence of these forms of the lightcone measure in appendix~\ref{app:measure}. A fourth representation, called the {\it unreduced measure}, will be discussed in section~\ref{sec:freeze}. 

An integration cycle in the lightcone fiber bundle consists of a pair 
\begin{equation}\Clc = (C,\varphilc),\end{equation}
where $C$ is an oriented manifold which is embedded by $\varphilc: C\to\Plc_n$ into the lightcone fiber bundle.  The embedding map defines an $n$-point Mandelstam diagram with stubs for every point in $C$. Equivalently, it defines the positions of $n$ punctures together with respective dilatations for every point in $C$. 
An important consideration in the choice of integration cycle is that the stubs have positive (or at least not negative) length. This will happen if the dilatations at the punctures do not exceed the lightcone conformal radii
\begin{equation}r_i\leq \rlc_i.\end{equation}
If this condition is always satisfied the integration cycle is termed {\it admissible}. An integration cycle which is not admissible will result in an off-shell amplitude whose magnitude grows exponentially with mass level or conformal weight~\cite{lightcone}. In this sense the amplitude will not be normalizable. 

\subsection{Replacement formula} 

Consider the following:
\begin{description}
\item{\bf Replacement formula.} Consider a correlation function 
\begin{equation}
\Big\langle \mathcal{O}\ e^{i\kpar^1\cdot X(u_1,u_1)}\, ...\, e^{i\kpar^n\cdot X(u_n,u_n)}\Big\rangle_\UHP,\label{eq:rep_corr}
\end{equation}
where $\mathcal{O}$ is any worldsheet operator which is independent of lightcone time $x^+$ and the minus free boson $X^-(u,\overline{u})$. Inside the operator $\mathcal{O}$ it is possible to replace the plus component of the chiral free boson $X^+(u)$ with the Mandelstam mapping as
\begin{equation}X^+(u) = -\frac{i}{2}\rho(u),\end{equation}
without changing the result of the correlation function. The Mandelstam mapping $\rho(u)$ is defined with puncture positions $u_i$ and string lengths $\alpha_i$ as given by the plane wave vertex operators in the correlation function.
\end{description}
The replacement formula follows from standard expressions for free boson correlation functions. A path integral derivation is given in \cite{BabaIshibashi}, where it plays an important role in the covariantization of lightcone amplitudes. Note that 
\begin{equation}\frac{1}{2}\Big(\rho(u)+\rho(\overline{u})\Big)\end{equation}
is the Euclidean time on the Mandelstam diagram corresponding to the point $u$ on the upper half plane. Multiplying by $-i$ converts Euclidean time to Lorentzian time. Therefore, the replacement formula implies that this class of correlation functions equate $X^+(u,\overline{u})$ with Lorentzian time on the string worldsheet. This is exactly what defines {\it lightcone gauge} from the point of view of the symmetries of the worldsheet action. 

\subsection{Equivalence of transverse off-shell amplitudes}
\label{subsec:equivalence}

We are ready to discuss the main result of this section:  
\begin{description}
\item{\bf Equivalence Theorem.} Let $a_1,...,a_n\in \Hperp$ be transverse states in the lightcone vector space and $\S a_1,...,\S a_n\in \HDDF$ be their DDF counterparts. Then the covariant off-shell amplitude evaluated on $\S a_1,...,\S a_n$ is the same as a lightcone off-shell amplitude evaluated on $a_1,...,a_n$,
\begin{equation}A_n(\C,\S a_1,\S a_2,...,\S a_n) = \Alc_n(\Clc,a_1,a_2,...,a_n),\end{equation}
provided that the integration cycles are related through the bundle projection. That is, if $\C$ is defined by embedding $C$ in the covariant fiber bundle with $\varphi$, then $\Clc$ is defined by embedding $C$ in the lightcone fiber bundle with $\pilc\circ\varphi$. We write this more briefly as
\begin{equation}\Clc = \pilc\circ\C.\end{equation}
\end{description}
The equivalence theorem implies that transverse states always see interactions as taking place through Mandelstam diagrams, even in a covariant off-shell amplitude whose local coordinates are defined in a completely unrelated manner.  The only information from the covariant amplitude which is visible from the point of view of transverse degrees of freedom is the set of dilatations at the punctures. These are reinterpreted as stubs of suitable length attached to a Mandelstam diagram. 

The equivalence theorem can be reformulated as a statement about the measure:
\begin{equation}
	\langle\Omega_n|S\otimes...\otimes S = (\sigmalc\circ\pilc)^*\langle\Omega_n|.\ \ \ \ \ \mathrm{on}\ \Hperp\label{eq:measure_equivalence}
\end{equation}
This equality assumes that the lightcone measure is expressed in covariantized form \eq{covariantized}. Contracting with transverse test states, this can be reexpressed as an equality between correlation functions:
\begin{eqnarray}
	\lineup \!\!\!\!\!\!\!\!\! \Big\langle \exp\big[\mathscr{B}\big] \big(f_1\circ \S a_1(0)\big)\big(f_2\circ \S a_2(0)\big)...\big(f_n\circ \S a_n(0)\big) \Big\rangle_\UHP \nonumber\\
	\lineup = \Big\langle \exp\Big[ (\sigmalc\circ\pilc)^*\mathscr{B}\Big] \big(\flc_1\circ e^{-\lambda_1}\circ a_1(0)\big)\big(\flc_2\circ e^{-\lambda_2}\circ a_2(0)\big)...\big(\flc_n\circ e^{-\lambda_n}\circ a_n(0)\big)\Big\rangle_\UHP.\label{eq:corr_equivalence}
\end{eqnarray}
The stub lengths are fixed so that the conformal radii of the local coordinate maps are the same on both sides. The proof of the equivalence theorem is broken into two parts:
\begin{description}
	\item{\bf Vertex operator equivalence}. If $\mathcal{O}$ is any operator which is independent of lightcone time and the minus free boson $X^-(u,\overline{u})$, then
	\begin{eqnarray}
		\lineup\!\!\!\!\!\!\!\!\!\!\!\!\!\!\!\! \Big\langle \mathcal{O}\, \big(f_1\circ \S a_1(0)\big)\big(f_2\circ \S a_2(0)\big)...\big(f_n\circ \S a_n(0)\big) \Big\rangle_\UHP \nonumber\\
		\lineup = \Big\langle\mathcal{O}\, \big(\flc_1\circ e^{-\lambda_1}\circ a_1(0)\big)\big(\flc_2\circ e^{-\lambda_2}\circ a_2(0)\big)...\big(\flc_n\circ e^{-\lambda_n}\circ a_n(0)\big)\Big\rangle_\UHP.\label{eq:vertex_equivalence}
	\end{eqnarray}
	\item{\bf b-ghost equivalence}. If ghost vertex operators $g_i(u)$ take the form of either $c(u)$ or $c\d c(u)$, then 
	\begin{equation}
		\Big\langle \exp\big[\mathscr{B}\big] g_1(u_1) g_2(u_2)...g_n(u_n)\Big\rangle_\UHP^{bc} =  \Big\langle \exp\Big[(\sigmalc\circ\pilc)^*\mathscr{B}\Big] g_1(u_1) g_2(u_2)...g_n(u_n)\Big\rangle_\UHP^{bc}. \label{eq:b_equivalence}
		\end{equation}
	\end{description}
Taken together these imply the equivalence theorem. Both statements however will be useful in other contexts later. 

We start by proving vertex operator equivalence. It will be helpful to compute in a basis. A convenient basis for $\Hperp$ is given by the Verma modules of transverse boundary primaries
\begin{equation}L_{-n_1}^\perp...L_{-n_{N}}^\perp \phi_\perp(0) |\pm,\kpar\rangle \in \Hperp,\label{eq:Vermaperp}\end{equation} 
where $\phi_\perp(0)$ is a primary of $\BCFT_\perp$ and $L_{-m}^\perp$ are Virasoro mode operators made from the transverse energy-momentum tensor. We always have such a basis because the transverse BCFT is unitary. Applying the Aisaka-Kazama transformation gives the basis in the covariant vector space:
\begin{equation}\Big(e^{\frac{in_1}{2k_-}x^+}\mathcal{L}_{-n_1}\Big)...\Big(e^{\frac{in_N}{2k_-}x^+}\mathcal{L}_{-n_{N}}\Big) \phi_\perp(0)|\pm,\kpar\rangle\in \HDDF,\label{eq:VermaDDF}\end{equation}
where we introduce ``DDF Virasoros" $\mathcal{L}_{-n}$ defined according to 
\begin{equation}\S L_{-m}^\perp \S^{-1} = e^{\frac{im}{2p_-}x^+}\mathcal{L}_{-m}.\label{eq:DDFVir}\end{equation}
The zero mode prefactor $e^{\frac{im}{2p_-}x^+}$ cancels the $x^+$ dependence in the DDF Virasoro. Comparing the left and right hand sides of \eq{vertex_equivalence} it is clear that 
\begin{eqnarray}
f_i\circ \Big(\phi_\perp c\, e^{i\kpar \cdot X(0,0)}\Big)\lineup = \flc_i\circ e^{-\lambda_i}\circ\Big(\phi_\perp c\, e^{i\kpar\cdot X(0,0)}\Big),\\
f_i\circ \Big(\phi_\perp c\d c\, e^{i\kpar \cdot X(0,0)}\Big)\lineup = \flc_i\circ e^{-\lambda_i}\circ\Big(\phi_\perp c\d c\, e^{i\kpar\cdot X(0,0)}\Big),
\end{eqnarray}
because the vertex operators are primary and the local coordinate maps on both sides have the same conformal radius. Therefore \eq{vertex_equivalence} holds if we can show that 
\begin{equation}
f_i \circ \Big(e^{\frac{im}{\alpha_i}x^+}\mathcal{L}_{-m}\Big) = \flc_i \circ e^{-\lambda_i}\circ L^\perp_{-m}\label{eq:Virpart}
\end{equation}
inside the correlation function.

To demonstrate this, first we note that the DDF Virasoros take the explicit form (see Appendix~D of \cite{lightcone})
\begin{equation}
\mathcal{L}_{-m} = -ip_-\oint_0\frac{d\xi}{2\pi i}e^{-\frac{im}{p_-}X^+(\xi)}\frac{T^\perp(\xi)-2\{X^+,\xi\}}{\d X^+(\xi)} +\delta_{m=0}. \label{eq:curlyLm}
\end{equation}
As with \eq{LDDF}, the integrand is a primary operator of weight $1$, which means that DDF Virasoros are conformally invariant. It follows that DDF Virasoro descendants of primary states {\it are still primary}. In particular all states in $\HDDF$ are linear combinations of primaries. This explains the origin of the lightcone fiber bundle. The only off-shell data that can enter into a transverse off-shell amplitude is the set of dilatations around each puncture. However this does not explain the relevance of Mandelstam diagrams. This comes about due to the replacement formula. Consider 
\begin{equation}
f_i\circ \mathcal{L}_{-m} = -ik_-^i\oint_{u_i}\frac{du}{2\pi i}e^{-\frac{im}{k_-^i}X^+(u)}\frac{T^\perp(u)-2\{X^+,u\}}{\d X^+(u)}. 
\end{equation}
This differs from \eq{curlyLm} only in that the contour surrounds the $i$th puncture and the lightcone momentum operator evaluates to the lightcone momentum of the $i$th puncture. Next we observe that the correlation function \eq{vertex_equivalence} is of the form where we can apply the replacement formula. Thus we obtain 
\begin{eqnarray}
f_i\circ \mathcal{L}_{-m} \lineup = \alpha_i\oint_{u_i}\frac{du}{2\pi i}e^{-\frac{m}{\alpha_i}\rho(u)}\frac{T^\perp(u)-2\{\rho(u),u\}}{\d\rho(u)}.\label{eq:DDFrep}
\end{eqnarray}
The DDF Virasoro depends on $x^+$, but we can apply the replacement formula anyway since we know this dependence is canceled by the zero mode prefactor (which we deal with in a moment). The right hand side of \eq{DDFrep} can be identified with the image of the transverse Virasoro $L^\perp_{-m}$ after applying a conformal transformation whose inverse is
\begin{equation}\exp\left[\frac{\rho(u)}{\alpha_i}\right].\label{eq:flcnonorm}\end{equation}
Comparing to \eq{flcinv}, this is almost the inverse of the lightcone local coordinate map. There is however a normalization that has to be fixed, which leads to
\begin{equation}
f_i\circ \mathcal{L}_{-m} = e^{-m\left(\frac{\tau_{s(i)}+i\sigma_i}{\alpha_i}-\lambda_i\right)}\flc_i\circ e^{-\lambda_i}\circ L_{-m}^\perp.
\label{eq:fLDDF}\end{equation}
This is how the geometry of the Mandelstam diagram enters. 

We still need to take care of the zero mode prefactor. Applying the local coordinate map,
\begin{equation}f_i\circ e^{\frac{i m}{2k_-^i} x^+} = e^{\frac{i m}{2k_-^i} (f_i\circ\, x^+)}.\end{equation}
The position zero mode $x^+$ is related to the free boson $X^+(\xi,\overline{\xi})$ through 
\begin{equation}x^+ = \frac{1}{\pi}\int_0^\pi d\sigma X^+(e^{i\sigma},e^{-i\sigma}).\end{equation}
After conformal transformation we have
\begin{equation}
f_i\circ x^+ = \frac{1}{\pi}\int_0^\pi d\sigma X^+\big(f_i(e^{i\sigma}),f_i(e^{-i\sigma})\big).
\end{equation}
Now we apply the replacement formula: 
\begin{equation}
f_i\circ x^+ = -\frac{i}{2\pi}\int_0^\pi d\sigma\Big[\rho\big(f_i(e^{i\sigma})\big)+\rho\big(f_i(e^{-i\sigma})\big)\Big].
\end{equation}
To go further we need to substitute the explicit form of the Mandelstam mapping:
\begin{eqnarray}
f_i\circ x^+
\lineup = -\frac{i}{2\pi}\alpha_i\int_0^\pi d\sigma\Big[\ln\big(f_i(e^{i\sigma})-u_i\big)+\ln\big(f_i(e^{-i\sigma})-u_i\big)\Big]\nonumber\\
\lineup \ \ \  -\frac{i}{2\pi}\sum_{j\in \puncture, \neq i} \delta_{j\neq i}\alpha_j\int_0^\pi d\sigma\Big[\ln\big(f_i(e^{i\sigma})-u_j\big)+\ln\big(f_i(e^{-i\sigma})-u_j\big)\Big].\ \ \ \ 
\end{eqnarray}
Here we extracted the term $j=i$ from the sum  since we have to deal with it separately. In the $j=i$ term we modify the argument of the logarithm:
\begin{eqnarray}
f_i\circ x^+
\lineup = -\frac{i}{2\pi}\alpha_i\int_0^\pi d\sigma\left[\ln\left(\frac{f_i(e^{i\sigma})-u_i}{e^{i\sigma}}\right)+\ln\left(\frac{f_i(e^{-i\sigma})-u_i}{e^{-i\sigma}}\right)\right]\nonumber\\
\lineup \ \ \  -\frac{i}{2\pi}\sum_{j\in\puncture,j\neq i} \alpha_j\int_0^\pi d\sigma\Big[\ln\big(f_i(e^{i\sigma})-u_j\big)+\ln\big(f_i(e^{-i\sigma})-u_j\big)\Big].\ \ \ \ 
\end{eqnarray}
The change in the argument of the logarithm cancels between the two terms in the integrand. Note that the two terms in the integrand can be combined by extending the lower limit of integration down to $-\pi$. By further substituting $\xi=e^{i\sigma}$, the result can be expressed as a closed contour integral around the origin:
\begin{equation}
f_i\circ x^+ = -i\alpha_i\oint_0\frac{d\xi}{2\pi i}\frac{1}{\xi}\ln\left(\frac{f_i(\xi)-u_i}{\xi}\right)-i\sum_{j\in\puncture,j\neq i}\alpha_j\oint_0\frac{d\xi}{2\pi i}\frac{\ln\big(f_i(\xi)-u_j\big)}{\xi}.
\end{equation}
The modification of the argument of the logarithm in the $i=j$ term ensures that the integrand has only a simple pole at the origin. We extract the residue to find
\begin{equation}
f_i\circ x^+ = -i \Big(\alpha_i \ln r_i +\sum_{j\in\puncture,j\neq i}\alpha_j \ln(u_i -u_j)\Big).
\end{equation}
Substituting $r_i$ from \eq{ri} and \eq{dfi} gives
\begin{equation}
f_i\circ x^+  = -i\big(\tau_{s(i)}+i\sigma_i-\alpha_i\lambda_i\big).
\end{equation}
From this we can see that the normalization factor in \eq{fLDDF} is canceled, which proves vertex operator equivalence, \eq{vertex_equivalence}. 

Now we prove $b$-ghost equivalence \eq{b_equivalence}. The $b$-ghost may be expanded in mode operators around the punctures \eq{bui},
\begin{equation}
	\mathscr{B} =\left(\sum_{k=-1}^\infty  \omega^1_{k}\, (b_k)_{u_1}\right)+\left(\sum_{k=-1}^\infty  \omega^2_{k}\, (b_k)_{u_2}\right)+...+\left(\sum_{k=-1}^\infty  \omega^n_{k}\, (b_k)_{u_n}\right),\label{eq:phiBexp}
\end{equation}
where $\omega^i_k$ are 1-forms on $\P_n$. Because of the simple ghost dependence of the vertex operators in \eq{b_equivalence}, only the 1-forms that multiply $b_{-1}$ and $b_0$ contribute. These are given by
\begin{equation}
	\omega_{-1}^i = -du_i,\ \ \ \ \ \ \ \omega_{0}^i = -\frac{dr_i}{r_i},\label{eq:omega_equiv}
\end{equation}
where $u_i$ are the locations of the punctures and $r_i$ are the conformal radii of the local coordinate maps $f_i$. Now we can make an expansion analogous to \eq{phiBexp} for $(\sigmalc\circ\pilc)^*\mathscr{B}$. Again only the 1-forms multiplying $b_{-1}$ and $b_0$ will contribute.  But because the local coordinate maps $\flc_i\circ e^{-\lambda_i}$ are associated to the same punctures and dilatations, these 1-forms agree with \eq{omega_equiv}. This establishes $b$-ghost equivalence, and by extension, the equivalence theorem.

\subsection{Transverse subvertex and the soft string problem}
\label{subsec:soft}

A covariant $n$-string vertex is an off-shell amplitude
\begin{equation}\big\langle \Psi ,v_{n-1}(\Psi,...,\Psi)\big\rangle = A_{n}(\V_{n},\Psi,...,\Psi)\label{eq:106}\end{equation}
defined by an integration cycle $\V_{n}$ which forms part of a solution of the geometrical BV equation~\cite{geoBV}. Evaluating this on the transverse string field gives, according to the equivalence theorem,
\begin{eqnarray}
\big\langle \S\Psiperp ,v_{n-1}(\S\Psiperp,...,\S\Psiperp)\big\rangle \lineup = \Alc_{n}(\pilc\circ\V_{n},\Psiperp,...,\Psiperp).\label{eq:107}
\end{eqnarray}
This is the {\it transverse subvertex} of the $n$-string vertex in lightcone gauge. It is characterized by an integration cycle $\pilc\circ \V_{n}$ consisting of Mandelstam diagrams with stubs covering the same portion of moduli space as the original covariant vertex. The total vertex in lightcone gauge also includes contributions from longitudinal subvertices, but we will deal with them later.

Already at the level of the transverse subvertex there is an important issue to address. A covariant integration cycle does not, in general, project to an {\t admissible} integration cycle in the lightcone fiber bundle. The Mandelstam diagrams can have stubs with formally negative length. The difficulty is avoided if the covariant integration cycle $\C$ is chosen to satisfy the inequalities 
\begin{equation}0 <r_i \leq \rlc_i \label{eq:adm}\end{equation}
for each $i\in\puncture$. Unfortunately, this condition is problematic from the point of view of covariant SFT. Typically, vertices in covariant SFT are {\it universal}, meaning that their definition is independent of CFT data. This in particular requires that the conformal radii of a covariant SFT vertex must be independent of the momenta of the external states. Meanwhile, the lightcone conformal radius $\rlc_i$ tends to zero as $\alpha_i$ tends to zero. This is intuitively clear because the strip domain $\rho_i$ becomes infinitely thin in this limit. More explicitly one can compute that
\begin{equation}
\rlc_i = \left|\frac{\alpha_i}{\sum_{j\in\{1,...,n\},j\neq i}\frac{\alpha_j}{u_i-u_j}}\right|+\mathcal{O}(\alpha_i^2),\ \ \ \ \ (\alpha_i<<\alpha_{j}\neq \alpha_i).
\end{equation}
Therefore the inequality \eq{adm} will be violated when at least one string in the vertex has lightcone momentum which is too small relative to the others. When this happens, the vertex is not normalizable. This is not acceptable in SFT since sums over intermediate states must converge when gluing vertices and propagators to construct amplitudes. Therefore lightcone gauge is not always well-defined in covariant SFT. This is called the {\it soft string problem} \cite{lightcone}. In this paper we assume that lightcone gauge is well-defined, which means that the covariant vertices and momenta are chosen favorably so that \eq{adm} holds. Generally speaking, limitations  are minimized if the conformal radii of the covariant vertices are chosen to be small. 

\begin{figure}
\begin{center}
\resizebox{2.2in}{2.2in}{\includegraphics{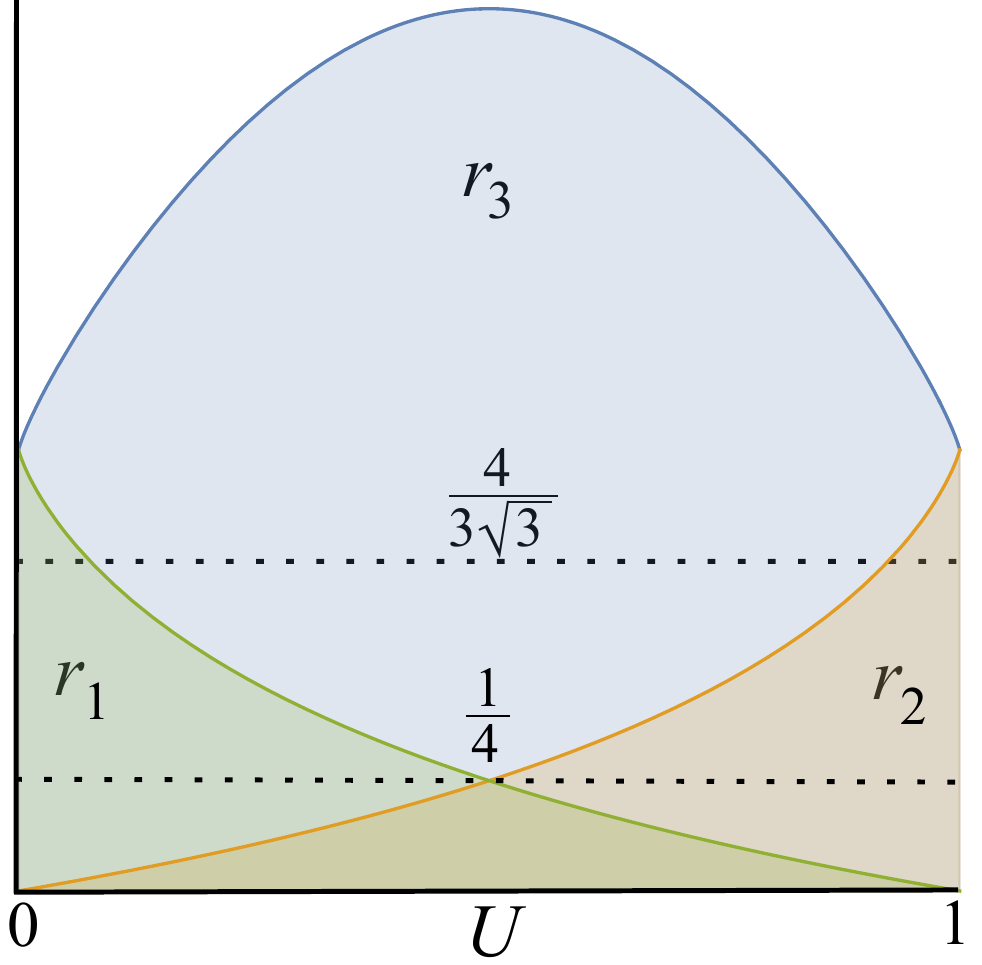}}
\end{center}
\caption{\label{fig:lightcone_gauge4} Assuming by convention that $\alpha_3$ has largest magnitude, the inequalities \eq{s1l}-\eq{s3l} are plotted here as a function of the dimensionless ratio $U = |\alpha_2/\alpha_3|\in [0,1]$. This ratio is equal to the position of the cubic interaction point on the real qxis of the upper half plane. The line $r_1=r_2=r_3= \frac{4}{3\sqrt{3}}$ corresponds to the Witten vertex. In this case there is no interval of $U$ where all three inequalities are obeyed. Below $r_1=r_2=r_3=1/4$, there is an interval containing $U=1/2$ where all stub lengths are positive. The point $U=1/2$ represents the Mandelstam diagram where strings 1 and 2 have equal minus momenta.}
\end{figure}

Let us illustrate the above in the context of the cubic vertex. The vertex is defined by three local coordinate maps $f_1(\xi),f_2(\xi)$ and $f_3(\xi)$ which we assume insert punctures respectively at $1,0$ and $\infty$. We expand
\begin{eqnarray}
f_1(\xi) \lineup = 1 + r_1 \xi + \text{higher orders},\\
f_2(\xi) \lineup = r_2 \xi + \text{higher orders},\\
I\circ f_3(\xi) \lineup = r_3  \xi + \text{higher orders},
\end{eqnarray}
where $I(u)=-1/u$ is the BPZ conformal map. The cubic vertex will be well-defined in lightcone gauge if 
\begin{eqnarray}
0< r_1 \lineup \leq \big(|\alpha_1|^{\alpha_1}|\alpha_2|^{\alpha_2}|\alpha_3|^{\alpha_3}\big)^{\frac{1}{\alpha_1}},\label{eq:s1l}\\
0<r_2 \lineup \leq \big(|\alpha_1|^{\alpha_1}|\alpha_2|^{\alpha_2}|\alpha_3|^{\alpha_3}\big)^{\frac{1}{\alpha_2}},\\
0< r_3\lineup \leq \big(|\alpha_1|^{\alpha_1}|\alpha_2|^{\alpha_2}|\alpha_3|^{\alpha_3}\big)^{\frac{1}{\alpha_3}}.\label{eq:s3l}
\end{eqnarray}
Consider for example Witten's string field theory. The conformal radii are 
\begin{equation} r_1=r_2=r_3= \frac{4}{3\sqrt{3}}\approx .77,\ \ \ \ \text{(Witten vertex)}.\end{equation}
As illustrated in figure \ref{fig:lightcone_gauge4}, there is no configuration of momenta where all three inequalities hold. Therefore lightcone gauge is singular in Witten's SFT for any momenta. However, the Witten vertex presents the worst case scenario. The conformal radii are the largest of any universal cubic vertex \cite{BZ}. The conformal radii can be made smaller in many ways, for example by attaching stubs to the Witten vertex or by deforming into a hyperbolic vertex with smaller systole length. Below a critical value $r_1=r_2=r_3=1/4$  the cubic vertex will be well-defined in lightcone gauge if the minus momenta of two of the three strings in the interaction are close enough to equal. 

\section{Lightcone measure and longitudinal freezing}
\label{sec:freeze} 

The task for the remainder of the paper is to understand how the exchange of longitudinal states contributes to interactions in lightcone gauge. This exchange is represented through the longitudinal propagator, which in the lightcone vector space is expressed as
\begin{equation}\Deltapar = \bpar \int_0^\infty dt\, e^{-t \Lpar}.\end{equation}
The longitudinal wave operator $\Lpar$ is related by a simple shift to the dilatation generator $L_0^\parallel$ in the longitudinal part of the $\BCFT$:
\begin{equation}L_0^\parallel = \big((\ppar)^2 -1\big)+\Lpar. \end{equation}
This means that the longitudinal propagator creates a strip of worldsheet in the longitudinal BCFT while leaving the transverse $\BCFT$ alone. This leads to a situation where the longitudinal and transverse correlation functions are evaluated on different surfaces, and the conformal anomaly does not cancel. Below we explain how to deal with this using the phenomenon of {\it longitudinal freezing}. The claim is that the longitudinal $\BCFT$ is always frozen to the Fock vacuum inside the propagator strips of a Mandelstam diagram. Therefore, changing the length of the propagator strip has no effect on the value of the correlation function in the longitudinal BCFT. One application is to give a derivation of the lightcone measure without explicitly imposing Lorentz invariance or regularizing infinite dimensional determinants. 

To begin let us describe what the lightcone measure {\it ought} to look like based on the form of Kaku and Kikkawa's lightcone SFT. The Kaku-Kikkawa action can be written as
\begin{equation}
S_\mathrm{lc} = \frac{1}{2}\big\langle\Psiperp,c_0 \Lperp\Psiperp\big\rangle + \frac{1}{3}\big\langle \Psiperp,v^
\mathrm{KK}_2(\Psiperp,\Psiperp)\big\rangle+\frac{1}{4}\big\langle\Psiperp,v^
\mathrm{KK}_3(\Psiperp,\Psiperp,\Psiperp)\big\rangle,
\label{eq:KZaction}
\end{equation}
where $\Psiperp\in\Hperp$ is the lightcone string field and 
\begin{equation}v^\mathrm{KK}_2: \Hperp^{\otimes 2} \to\Hperp \ \ \ \ v^\mathrm{KK}_3:\Hperp^{\otimes 3} \to\Hperp \end{equation}
are string products representing cubic and quartic lightcone vertices (without stubs). The vertices are traditionally expressed via the reduced measure in terms of correlation functions on a Mandelstam diagram in the transverse BCFT. The geometrical data of the cubic and quartic lightcone vertices however can also be used to define vertices based on the covariant measure. These are the vertices of the Kugo-Zwiebach SFT, which define string products 
\begin{equation}v^\mathrm{KZ}_2: \Hcov^{\otimes 2} \to\Hcov \ \ \ \ v^\mathrm{KZ}_3:\Hcov^{\otimes 3} \to\Hcov \end{equation}
acting on the covariant vector space. As correlation functions on the upper half plane, the Kugo-Zwiebach vertices are given as 
\begin{eqnarray}
	\big\langle \phi_1,v^\mathrm{KZ}_2(\phi_2,\phi_3)\big\rangle\lineup = \Big\langle\big(\flc_{(3,1)}\circ \phi_1(0)\big)\big(\flc_{(3,2)}\circ \phi_2(0)\big)\big(\flc_{(3,3)}\circ \phi_3(0)\big)\Big\rangle_\UHP,\phantom{\Bigg)}\label{eq:vKZ2}\\
	\big\langle \phi_1,v^\mathrm{KZ}_3(\phi_2,\phi_3,\phi_4)\big\rangle\lineup = -\int_{V_{1234}^\mathrm{lc}}d\theta \left\langle  \mathrm{Im}\left(\frac{b(U)}{\d^2\rho(U)}\right)\big(\flc_{(4,1)}\circ \phi_1(0)\big)\big(\flc_{(4,2)}\circ \phi_2(0)\big)\right.\nonumber\\
	\lineup\ \ \ \ \ \ \ \ \ \ \ \ \ \ \ \ \ \ \ \ \times\big(\flc_{(4,3)}\circ \phi_3(0)\big)\big(\flc_{(4,4)}\circ \phi_4(0)\big)\bigg\rangle_\UHP.\ \ \ \ \ \ \label{eq:vlc3}
\end{eqnarray}
We replace $i\to(n,i)$ to distinguish cubic and quartic local coordinate maps. In the quartic vertex we integrate over the portion of moduli space $V_{1234}^\mathrm{lc}\subset M_4$ characterized by a single quartic interaction point $U$ on the upper half plane. In appendix C of \cite{lightcone} it was shown that the Kaku-Kikkawa vertices, expressed through the reduced measure, are the same as the Kugo-Zwiebach vertices restricted to the transverse sector. Therefore we write Kaku and Kikkawa's action as 
\begin{equation}
	S_\mathrm{lc} = \frac{1}{2}\big\langle\Psiperp,c_0 \Lperp\Psiperp\big\rangle + \frac{1}{3}\big\langle \Psiperp,v^\mathrm{KZ}_2(\Psiperp,\Psiperp)\big\rangle+\frac{1}{4}\big\langle\Psiperp,v^\mathrm{KZ}_3(\Psiperp,\Psiperp,\Psiperp)\big\rangle,
\end{equation}
where the cubic and quartic vertices are identical to those of the Kugo-Zwiebach SFT. From this it is easy to infer the form of the lightcone measure. It is given by gluing Kugo-Zwiebach vertices together through propagators. However, because the lightcone string field is transverse, the relevant propagator is the {\it transverse propagator} 
\begin{equation}\Deltaperp = b_0 \delta(\Lpar) \int_0^\infty dt\, e^{-t \Lperp}.\end{equation}
It follows by inspection that the lightcone measure will be the same as the measure of the Kugo-Zwiebach string field theory except for factors of $\delta(\Lpar)$ which must be inserted on each propagator strip. These factors ensure that only transverse states are exchanged through the propagator. The measure of the Kugo-Zwiebach SFT (generalized to include stubs) is written in \eq{covariantized_KZ}. It follows that the lightcone measure takes the form 
\begin{eqnarray}
\Omegalc_n(a_1,a_2,..., a_n) \lineup = \left\langle\exp\big(\mathscr{B}^\mathrm{KZ}\big) \left(\prod_{i\in\propagator}\flc_i\circ\delta(\Lpar)\right)\right.\nonumber\\
\lineup \ \ \ \ \ \ \ \ \big(\flc_1\circ e^{-\lambda_1}\circ a_1(0)\big) \big(\flc_2\circ e^{-\lambda_2}\circ a_2(0)\big)\ ...\ \big(\flc_n\circ e^{-\lambda_n}\circ a_n(0)\big)\Bigg\rangle_\UHP.\ \ \ \ \ \ \label{eq:unred_measure}
\end{eqnarray}
We call this the {\it unreduced measure}. It is ``unreduced" in the sense that it involves the full matter+ghost worldsheet theory with vanishing central charge. 

Now it is claimed that the unreduced measure is in fact the same as the covariantized measure as originally written in \eq{covariantized_KZ}. The only way this can happen is if the operators $\delta(\Lpar)$ can be ignored for some reason. 
There could be various explanations, but the simplest one turns out to be correct. That is the longitudinal $\BCFT$ is always frozen to the Fock vacuum with $\Lpar = 0$ inside the propagator strip of a Mandelstam diagram. Therefore, insertions $\delta(\Lpar)$ trivially evaluate to $1$. We call this phenomenon {\it longitudinal freezing}. This interpretation has a corollary:
\begin{description}
\item{\bf Freeze Theorem:} Euclidean time evolution generated by the longitudinal wave operator $\Lpar$ is trivial inside propagator strips of a Mandelstam diagram. That is, the expression 
\begin{eqnarray}
\Omegalc_n(a_1,a_2,..., a_n) \lineup = \left\langle\exp\big(\mathscr{B}^\mathrm{KZ}\big) \left(\prod_{i\in\propagator}\flc_i\circ e^{-t_i \Lpar} \right)\right.\nonumber\\
\lineup \ \ \ \ \big(\flc_1\circ e^{-\lambda_1}\circ a_1(0)\big) \big(\flc_2\circ e^{-\lambda_2}\circ a_2(0)\big)\ ...\ \big(\flc_n\circ e^{-\lambda_n}\circ a_n(0)\big)\Bigg\rangle_\UHP\ \ \ \ \label{eq:freeze_measure}
\end{eqnarray}
is independent of the Euclidean time parameters $t_i$.
\end{description}
The covariantized measure appears when each $t_i$ is set to zero, while the unreduced measure appears when each $t_i$ is infinite. The {\it freeze theorem} then implies the equivalence of the unreduced and covariantized measures.

Let us give the proof. The strategy is to evaluate
\begin{equation}
\left.\frac{d}{dt_j}\!\left\langle\!\exp\big(\mathscr{B}^\mathrm{KZ}\big) \! \left(\prod_{i\in\propagator}\flc_i\circ e^{-t_i \Lpar}\right)\right. \! \big(\flc_1\circ e^{-\lambda_1}\circ a_1(0)\big)\ ...\ \big(\flc_n\circ e^{-\lambda_n}\circ a_n(0)\big)\!\Bigg\rangle_{\!\!\!\UHP}\right|_{(T_i,\theta_I)}
\label{eq:419}\end{equation}
and show that it vanishes. The symbol $|_{(T_i,\theta_I)}$ is used to explicitly indicate the dependence of the conformal maps and $b$-ghosts on the moduli. Both here and later, a useful trick for dealing with $\Lpar$ is to make the substitution 
\begin{equation}\Lpar = L_0 - \Lperp,\end{equation}
and let $L_0$ deform the propagator widths in the Mandelstam diagram.  If the Mandelstam diagram is initially described by moduli $(T_i,\theta_I)$ then after $L_0$ deforms the propagator widths the moduli will be 
\begin{equation}(T_i +\alpha_i t_i,\theta_I),\end{equation}
and \eq{419} becomes 
\begin{equation}
\frac{d}{dt_j}\!\!\left.\left\langle\!\exp\big(\mathscr{B}^\mathrm{KZ}\big)\!\! \left(\prod_{i\in\propagator}\!\!\!\flc_i\circ e^{t_i \Lperp}\!\right)\!\! \big(\flc_1\circ e^{-\lambda_1}\circ a_1(0)\big)\,...\,\big(\flc_n\circ e^{-\lambda_n}\circ a_n(0)\big)\!\right\rangle_{\!\!\!\UHP}\right|_{(T_i+\alpha_it_i,\theta_I)}.\label{eq:422}
\end{equation}
In this way we have effectively replaced $\Lpar$ with $-\Lperp$. This is useful because now the free boson $X^-(u,\overline{u})$ and the zero mode $x^+$ only appear in the correlation function through the plane wave vertex operators at the punctures (implicit inside the $a_i$s). This allows us to apply the replacement formula. The first consequence is that we can make use of vertex operator equivalence \eq{vertex_equivalence} to insert the Aisaka-Kazama transformation: 
\begin{equation}\flc_i\circ e^{-\lambda_i}\circ a_i(0) = \flc_i\circ e^{-\lambda_i}\circ \S a_i(0).\end{equation}
The second consequence is that we can replace the lightcone wave operator with the DDF wave operator: 
\begin{equation}\flc_i\circ \Lperp = \flc_i\circ \LDDF. \label{eq:Lperprep}\end{equation}
We will prove this momentarily. After making these substitutions and taking the derivative with respect to $t_j$ we find
\begin{eqnarray}
\lineup -\left\langle\exp\big(\mathscr{B}^\mathrm{KZ}\big)\flc_i\circ\big(L_0 - \LDDF\big)\left(\prod_{i\in\propagator}\flc_i\circ e^{t_i \LDDF}\right)\right.\nonumber\\
\lineup\ \ \ \ \ \ \ \ \ \ \ \ \ \ \ \ \ \ \ \ \ \ \ \ \ \ \ \ \ \ \ \ \  \left.\times \big(\flc_1\circ e^{-\lambda_1}\circ \S a_1(0)\big)\ ...\ \big(\flc_n\circ e^{-\lambda_n}\circ \S a_n(0)\big)\Bigg\rangle_\UHP\right|_{(T_i+\alpha_it_i,\theta_I)}.\ \ \ \ 
\end{eqnarray}
The $L_0$ insertion comes from differentiating the underlying moduli and the $\LDDF$ insertion comes from differentiating the operator $e^{t_j\LDDF}$. Next we write 
\begin{equation}
L_0-\LDDF = Q\cdot(b_0-\bDDF)
\end{equation}
and pull off $Q$ to act on other insertions in the correlator. The BRST operator does nothing to the exponential insertions of $\LDDF$. It replaces $\mathscr{B}^\mathrm{KZ}$ with a corresponding energy-momentum insertion which, through the BRST identity, computes the exterior derivative $d$ on the lightcone fiber bundle. Finally, we have the BRST variation of the vertex operators $\S a_i(0)$. If the states $a_i$ are taken from the Verma module \eq{Vermaperp}, $\S a_i(0)$ is primary and the BRST variation is 
\begin{equation}Q\cdot \S a_i(0) = h_i\d c \S a_i(0),\end{equation}
where $h_i$ is the scaling dimension of $a_i(0)$. In this way we obtain
\begin{eqnarray}
\lineup d\left[\left\langle\exp\big(\mathscr{B}^\mathrm{KZ}\big)\flc_i\circ\big(b_0 - \bDDF\big)\left(\prod_{i\in\propagator}\flc_i\circ e^{t_i \LDDF}\right)\right.\right.\nonumber\\
\lineup\ \ \ \ \ \ \ \ \ \ \ \ \ \ \ \ \ \ \ \ \ \ \ \ \ \ \ \ \ \ \ \ \ \ \ \ \ \left. \left.\times \big(\flc_1\circ e^{-\lambda_1}\circ \S a_1(0)\big)\ ...\ \big(\flc_n\circ e^{-\lambda_n}\circ \S a_n(0)\big)\Bigg\rangle_\UHP\right|_{(T_i+\alpha_it_i,\theta_I)}\right]\nonumber\\
\lineup +\sum_{i=1}^n (-1)^{i+1}h_i \left\langle\exp\big(\mathscr{B}^\mathrm{KZ}\big)\flc_i\circ\big(b_0 - \bDDF\big)\left(\prod_{i\in\propagator}\flc_i\circ e^{t_i \LDDF}\right)\right.\nonumber\\
\lineup \ \ \ \ \left.\times \big(\flc_1\circ e^{-\lambda_1}\circ \S a_1(0)\big)\ ...\ \big(\flc_i\circ e^{-\lambda_i}\circ \d c \S a_i(0)\big)\ ... \ \big(\flc_n\circ e^{-\lambda_n}\circ \S a_n(0)\big)\Bigg\rangle_\UHP\right|_{(T_i+\alpha_it_i,\theta_I)}. \ \ \ \ \
\label{eq:423}
\end{eqnarray}
In all terms we are in a situation where again we can apply the replacement formula. The replacement implies 
\begin{equation}
\flc_i\circ\bDDF = \flc_i\circ b_0,\label{eq:bperprep}
\end{equation}
which makes all terms vanish. This establishes the freeze theorem. 

Let us return to the derivation of \eq{Lperprep}. Conformal transformation of the transverse wave operator gives
\begin{eqnarray}\flc_i\circ \Lperp \lineup =(k_\parallel^i)^2 - 1 +\flc_i\circ L_0^\perp\nonumber\\
\lineup =  (k_\parallel^i)^2 - 1+\oint_{\flc_i\circ C}\frac{du}{2\pi i}\frac{(\flc_i)^{-1}(u)}{\d (\flc_i)^{-1}(u)}\Big[T_\perp(u) -2\{(\flc_i)^{-1},u\}\Big],\label{eq:Lperprep0}
\end{eqnarray}
where $\flc_i\circ C$ is the image of the counterclockwise closed contour surrounding the origin of the unit disk. From \eq{flcinv} we can simplify the integrand using
\begin{eqnarray}
\d (\flc_i)^{-1}(u)\lineup =\frac{\d\rho(u)}{\alpha_i} (\flc_i)^{-1}(u),\\
\{(\flc_i)^{-1},u\}\lineup = \{\rho,u\}-\frac{\big(\d\rho(u)\big)^2}{2\alpha_i^2},
\end{eqnarray}
which gives
\begin{equation}
\flc_i\circ \Lperp =  (k_\parallel^i)^2 - 1+\alpha_i\oint_{\flc_i\circ C}\frac{du}{2\pi i}\frac{1}{\d \rho(u)}\left[T_\perp(u) -2\{\rho,u\}+\frac{\big(\d\rho(u)\big)^2}{\alpha_i^2}\right].
\end{equation}
The last term in the integrand can be simplified and written in the coordinate $\rho$ on the Mandelstam diagram:
\begin{equation}
\flc_i\circ \Lperp =  (k_\parallel^i)^2 - 1+\alpha_i\oint_{\flc_i\circ C}\frac{du}{2\pi i}\frac{1}{\d \rho(u)}\big[T_\perp(u) -2\{\rho,u\}\big]+\frac{1}{\alpha_i} \oint_{\rho_i\circ C}\frac{d\rho}{2\pi i}.
\end{equation}
The integration in the coordinate $\rho$ just gives $2 i$ times the height of the $i$th propagator strip. This cancels the denominator to give $1$. This, in turn, adds with $-1$ to give zero. In total we obtain
\begin{equation}
\flc_i\circ \Lperp =  (k_\parallel^i)^2 + \alpha_i\oint_{\flc_i\circ C}\frac{du}{2\pi i}\frac{1}{\d \rho(u)}\big[T_\perp(u) -2\{\rho,u\}\big].
\end{equation}
Upon using the replacement formula this is the same as the conformal transformation of the DDF wave operator \eq{LDDF}. An analogous and simpler computation proves \eq{bperprep}.

\section{Quartic vertex in lightcone gauge}
\label{sec:quartic}
 
At this point we have a good enough understanding of transverse off-shell amplitudes and the lightcone measure. Now we are ready to think about string field theory in lightcone gauge. We start with the simplest nontrivial vertex, at quartic order. This has two longitudinal subvertices, corresponding to $s$- and $t$-channel Feynman graphs.

\begin{figure}
\begin{center}
\resizebox{5.3in}{1.8in}{\includegraphics{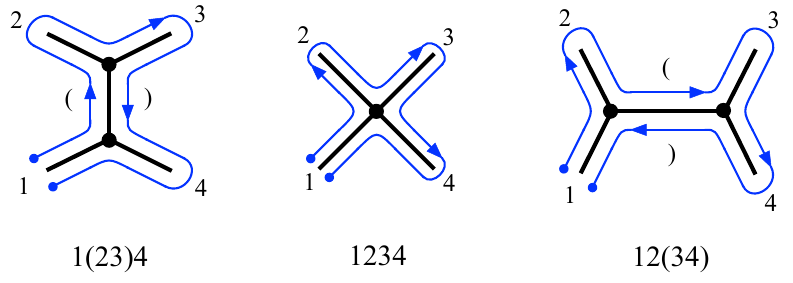}}
\end{center}
\caption{\label{fig:lightcone_gauge12} Starting from the first puncture we form a string of symbols for each Feynman graph by following a clockwise path around the graph. Every time the path wraps an external leg, we append the label of that puncture to the list of symbols. When the path encounters a given propagator in the diagram for the first time, we add an open parentheses to the list of symbols. After the path returns to a given propagator, we add a closed parentheses. We continue adding symbols until we arrive back at the first external leg. In this way we obtain a list of punctures accompanied with a configuration of parentheses that is unique to the diagram.}
\end{figure}

At higher order the number of graphs proliferate. Therefore it is helpful from the outset to devise a systematic way to notate them. For this we take advantage of the well-known correspondence between color ordered, $n$-point tree-level Feynman graphs and the different possible ways of inserting parentheses on a word with $(n-1)$ letters. We take the word to be the punctures listed in order from first to last. The parentheses are inserted following the mnemonic illustrated in figure \ref{fig:lightcone_gauge12}. At quartic order, the $s$-channel, $t$-channel, and quartic vertex graphs are denoted 
\begin{eqnarray}
t\text{-channel} \lineup \ \longleftrightarrow \ 1(23)4,\nonumber\\
\text{quartic vertex} \lineup \ \longleftrightarrow \ 1234,\nonumber\\
s\text{-channel} \lineup\  \longleftrightarrow \ 12(34).
\end{eqnarray}
This notation gives a word with $n$ letters, which is one more than needed. It is easy to see that the first puncture plays no role in distinguishing the graphs, but we write it anyway.

\subsection{Reducing the quartic vertex to a lightcone off-shell amplitude}
\label{subsec:reducing}
 
The longitudinal propagator has some similarity to the Siegel gauge propagator, and for this reason the computation of longitudinal subvertices is closely related to the computation of Siegel gauge amplitudes. A Siegel gauge amplitude is given by integrating over a collection of surfaces formed by gluing covariant vertices with Siegel gauge propagators. We call these surfaces {\it Siegel gauge diagrams}, in analogy to Mandelstam diagrams. The collection of $4$-point Siegel gauge diagrams defines an integration cycle
\begin{equation}\M_4=(M_4,\varphi).\end{equation}
We assume for simplicity that there is one Siegel gauge diagram for every point in moduli space $M_4$, which means that the integration cycle is a section of $\P_4$. The Siegel gauge amplitude is an example of a covariant off-shell amplitude. Therefore it can be written following subsection \ref{subsec:covariant_off} as 
\begin{equation}
	A_4(\M_4,\phi_1,\phi_2,\phi_3,\phi_4) = -\int_{M_4} dm \left\langle\varphi^*\mathscr{B}\left(\frac{\d}{\d m}\right)\big(f_1\circ \phi_1(0)\big)...\big(f_4\circ \phi_4(0)\big)\right\rangle_\mathrm{UHP},\label{eq:A4uni}
\end{equation}
where $\phi_1,...,\phi_4\in\Hcov$ are off-shell states in the covariant vector space at ghost number 1 and $m$ is a global coordinate on moduli space. The  local coordinate maps $f_1,...,f_4$ are determined either by a consistent definition of the quartic vertex or by gluing two cubic vertices with a Siegel gauge propagator, depending on where we are in the moduli space.  The Feynman rules of the string field theory express the Siegel gauge amplitude in a somewhat different form. The three Feynman graphs produce three terms,
\begin{eqnarray}
A_4(\M_4,\phi_1,\phi_2,\phi_3,\phi_4)\lineup = \underbrace{\big\langle \phi_1,v_3(\phi_2,\phi_3,\phi_4)\big\rangle}_{1234}-\underbrace{\big\langle \phi_1,v_2(\Delta_{b_0} v_2(\phi_2,\phi_3),\phi_4))\big\rangle}_{1(23)4}\nonumber\\
\lineup\ \ \   -\underbrace{\big\langle \phi_1,v_2(\phi_2,\Delta_{b_0} v_2(\phi_3,\phi_4))\big\rangle}_{12(34)},\label{eq:SiegelA4}
\end{eqnarray}
corresponding to quartic vertex, $t$-channel, and $s$-channels labeled following figure \ref{fig:lightcone_gauge12}. Each Feynman graph contribution produces a covariant off-shell amplitude, 
\begin{eqnarray}
A_4(\mathcal{M}_4,\phi_1,\phi_2,\phi_3,\phi_4)\lineup =A_4(\V_{1234},\phi_1,\phi_2,\phi_3,\phi_4)+ A_4(\R_{1(23)4},\phi_1,\phi_2,\phi_3,\phi_4) \nonumber\\
\lineup\ \ \ +A_4(\R_{12(34)},\phi_1,\phi_2,\phi_3,\phi_4),
\end{eqnarray}
defined by an integration cycle consisting respectively of quartic vertex, $t$-channel or $s$-channel Siegel gauge diagrams. The integration cycle of the full Siegel gauge amplitude is the union (or formal sum) of these integration cycles:
\begin{equation}\mathcal{M}_4 = \R_{1(23)4}\cup \mathcal{V}_{1234}\cup \R_{12(34)}.\end{equation}
We use $\V$ to denote a contribution from an elementary vertex and $\R$ for a contribution involving propagators. Because $\M_4$ is a section we can write 
\begin{eqnarray}
\R_{1(23)4} = (R_{1(23)4},\varphi),\ \ \ \V_{1234} = (V_{1234},\varphi),\ \ \ \ \R_{12(34)} = (R_{12(34)},\varphi),
\end{eqnarray}
where $R_{1(23)4}$ is the part of moduli space covered by the $t$-channel diagram, $V_{1234}$ is the part of moduli space covered by the quartic vertex, and $R_{12(34)}$ is the part of moduli space covered by the $s$-channel diagram. Together these regions cover the whole moduli space:
\begin{equation}M_4 = R_{1(23)4}\cup V_{1234}\cup R_{12(34)}.\end{equation}
The $t$-channel region $R_{1(23)4}$ has a natural coordinate $t_{23}$ provided by the Schwinger parameter of the Siegel gauge propagator: 
\begin{equation}\Delta_{b_0}= b_0\int_0^\infty |dt_{23}|e^{-t_{23}L_0}.\label{eq:Deltat23}\end{equation}
Likewise, the $s$-channel region $R_{12(34)}$ has a coordinate $t_{34}$ provided by the Schwinger parameter of the Siegel gauge propagator. To keep notation uniform, we also introduce a coordinate $m_{1234}$ on the portion of moduli space covered by the quartic vertex $V_{1234}$. We introduce local coordinates for the propagator strips of the $t$- and $s$-channel diagrams. The local coordinates are written respectively as $\xi_{23}$ or $\xi_{34}$ and are restricted by
\begin{eqnarray}
\lineup e^{-t_{23}}\leq |\xi_{23}|\leq 1,\ \ \ \ \mathrm{Im}(\xi_{23})>0,\label{eq:xi23}\\
\lineup e^{-t_{34}}\leq |\xi_{34}|\leq 1,\ \ \ \ \mathrm{Im}(\xi_{34})>0.
\end{eqnarray}
The local coordinates are transformed to the upper half plane though local coordinate maps
\begin{eqnarray}
\lineup f_{23}(\xi_{23}),\ \ \ m \in R_{1(23)4},\nonumber\\
\lineup f_{34}(\xi_{34}),\ \ \ m\in R_{12(34)}
\end{eqnarray}
defined on the respective regions of moduli space as indicated. With this the Siegel gauge 4-point amplitude can be written as
\begin{eqnarray}
A_4(\M_4,\phi_1,\phi_2,\phi_3,\phi_4) \lineup = -\int_{V_{1234}} dm_{1234} \left\langle\varphi^* \mathscr{B}\left(\frac{\d}{\d m}\right)\big(f_1\circ \phi_1(0)\big)...\big(f_4\circ \phi_4(0)\big)\right\rangle_\mathrm{UHP}\nonumber\\
\lineup\ \ \ -\int_{R_{1(23)4} }dt_{23}\,\Big\langle\big( f_{23}\circ b_0\big)\big( f_1\circ \phi_1(0)\big)...\big(f_4\circ \phi_4(0)\big)\Big\rangle_\mathrm{UHP}\nonumber\\
\lineup\ \ \ -\int_{R_{12(34)}} dt_{34}\,\Big\langle \big(f_{34}\circ b_0\big) \big(f_1\circ \phi_1(0)\big)...\big(f_4\circ \phi_4(0)\big)\Big\rangle_\mathrm{UHP}.\ \ \ \ \ \label{eq:A4broken}
\end{eqnarray}
This is different from the previous expression \eq{A4uni} because in the propagator regions of moduli space we integrate over the Schwinger parameters instead of the global coordinate $m$, and the measure is defined by the $b_0$ of the Siegel gauge propagator instead of the $b$-ghost of the Schiffer variation. To arrange a consistent sign between the terms, the integration over the Schwinger parameters is implemented as integration of a  1-form on moduli space instead of as integration of a density.\footnote{The second term in \eq{SiegelA4} acquires a different sign from the others when commuting the $b$-ghost of the Siegel gauge propagator to the left past the vertex operator in the correlation function.  However, $R_{1(23)4}$ inherits the orientation of moduli space where according to \eq{Mnorient} $du_2$ is equated with a positive integration density $|du_2|$. Since increasing the Schwinger parameter $t_{23}$ decreases $u_2$---as it brings it closer to $u_3$, which is less than $u_2$ in the left handed convention---there is a sign relating $dt_{23}$ and $du_2$. This cancels the sign flip from commuting the $b$-ghost in the second term of \eq{SiegelA4}.}

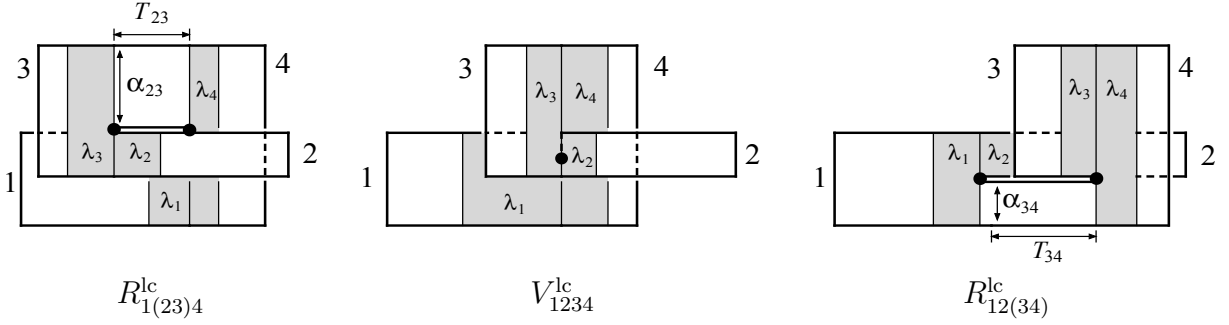
\begin{figure}
\begin{center}
\resizebox{6.5in}{1.5in}{\includegraphics{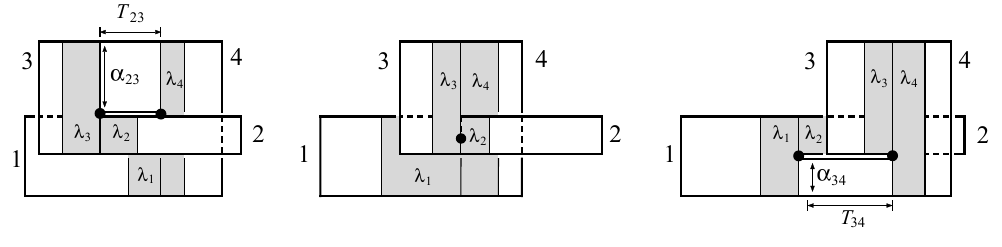}}
\setlength{\unitlength}{.25cm} 
\begin{picture}(60,0)
\put(4,1){$R_{1(23)4}^\mathrm{lc}$}
\put(26,1){$V_{1234}^\mathrm{lc}$}
\put(49,1){$R_{12(34)}^\mathrm{lc}$}
\end{picture}
\vspace{-.4cm}
\end{center}
\caption{\label{fig:lightcone_gauge14} Mandelstam diagrams with stubs characterizing the transverse projection of the Siegel gauge $4$-point amplitude. The $t$-channel, quartic vertex, and $s$-channel Mandelstam diagrams contribute in respective regions of moduli space $R_{1(23)4}^\mathrm{lc}$, $V_{1234}^\mathrm{lc}$, and $R_{12(34)}^\mathrm{lc}$.}
\end{figure}

Now consider the {\it transverse} Siegel gauge amplitude. Specifically, we restrict to external states of the form 
\begin{equation}\phi_i = S a_i,\end{equation}
where $a_i\in\Hperp$ are transverse and carry ghost number 1. The equivalence theorem implies that the Siegel gauge diagrams will project down to Mandelstam diagrams with stubs, resulting in an expression for the amplitude of the form 
\begin{eqnarray}
\lineup A_4(\M_4,Sa_1,Sa_2,Sa_3,Sa_4) \phantom{\Big)}\nonumber\\
\lineup \ \ \ \ \ \ \  = \Alc_4(\pilc\circ \M_4,a_1,a_2,a_3,a_4)\nonumber\\
\lineup \ \ \ \ \phantom{\Bigg)}= \int_{M_4} dm \left\langle(\sigmalc\circ\pilc\circ\varphi)^*\mathscr{B}\left(\frac{\d}{\d m}\right)\big(\flc_1\circ e^{-\lambda_1}\circ a_1(0)\big)...\big(\flc_4\circ e^{-\lambda_4}\circ a_4(0)\big)\right\rangle_{\mathrm{UHP}},
\end{eqnarray}
where the stub lengths  $\lambda_1,...,\lambda_4$ are determined so that conformal radii are preserved. The transverse Siegel gauge amplitude will have its own decomposition into $t$-channel, $s$-channel and quartic vertex Mandelstam diagrams, as illustrated in figure \ref{fig:lightcone_gauge14}. The diagrams appear in respective regions of moduli space:
\begin{equation}M_4= R_{1(23)4}^\mathrm{lc}\cup V_{1234}^\mathrm{lc}\cup R_{12(34)}^\mathrm{lc}.\end{equation}
It should be emphasized that this decomposition of moduli space is unrelated to the Siegel gauge Feynman rules of the original covariant string field theory. A Mandelstam diagram in the $t$-channel might come from projecting a Siegel gauge diagram in the covariant quartic vertex, or {\i vice versa}. The transverse Siegel gauge amplitude breaks into three pieces containing the three types of Mandelstam diagram: 
\begin{eqnarray}  
	\lineup A_4(\M_4,Sa_1,Sa_2,Sa_3,Sa_4) \nonumber\\
	\lineup \ \ \ \ = -\int_{V_{1234}^\mathrm{lc}} d\theta_{1234} \left\langle \left[\mathrm{Im}\left(\frac{b(U_{1234})}{\d^2(U_{1234})}\right)\right]\big(\flc_1\circ e^{-\lambda_1}\circ a_1(0)\big)...\big(\flc_4\circ e^{-\lambda_4}\circ a_4(0)\big) \right\rangle_{\UHP}\nonumber\\
	\lineup\ \ \ \ \ \ \ -\int_{R_{1(23)4}^\mathrm{lc}} \frac{dT_{23}}{\alpha_{23}}\, \Big\langle\big( \flc_{23}\circ b_0\big)\big(\flc_1\circ e^{-\lambda_1}\circ a_1(0)\big)...\big(\flc_4\circ e^{-\lambda_4}\circ a_4(0)\big) \Big\rangle_{\UHP}\nonumber\\
	\lineup\ \ \ \ \ \ \ - \int_{R_{12(34)}^\mathrm{lc}} \frac{dT_{34}}{\alpha_{34}}\, \Big\langle \big(\flc_{34}\circ b_0\big) \big(\flc_1\circ e^{-\lambda_1}\circ a_1(0)\big)...\big(\flc_4\circ e^{-\lambda_4}\circ a_4(0)\big) \Big\rangle_{\UHP}.
\end{eqnarray}
The quartic lightcone vertex contribution comes from Mandelstam diagrams with a quartic vertex interaction point whose preimage in the upper half plane is $U_{1234}$ and whose displacement parameter is $\theta_{1234}$ . The $t$- and $s$-channel contributions come from Mandelstam diagrams with a propagator strip in the respective channel of width $T_{23}$ and $T_{34}$. We use the Kugo-Zwiebach measure and drop the differentials $d\lambda_i$ since they do not contribute at ghost number 1.   

We are now ready to discuss the quartic vertex in lightcone gauge. Like the 4-point amplitude, the quartic vertex has three terms corresponding to the three Feynman graphs in figure \ref{fig:lightcone_gauge12}:
\begin{eqnarray}
\langle a_1,\vlc_3(a_2,a_3,a_4)\rangle\lineup = \underbrace{\big\langle \S a_1,v_3(\S a_2,\S a_3,\S a_4)\big\rangle}_{1234}-\underbrace{\big\langle \S a_1,v_2(\Deltalong v_2(\S a_2,\S a_3),\S a_4))\big\rangle}_{1(23)4}\nonumber\\
\lineup\ \ \   -\underbrace{\big\langle \S a_1,v_2(\S a_2,\Deltalong v_2(\S a_3,\S a_4))\big\rangle}_{12(34)}.
\end{eqnarray}
The vertex is written here acting on four generic transverse states $a_1,a_2,a_3,a_4\in\Hperp$ at ghost number 1. The first term is the transverse subvertex, and the second two are the $t$- and $s$-channel longitudinal subvertices. We will demonstrate that all three terms can be expressed as lightcone off-shell amplitudes:
\begin{eqnarray}
\langle a_1,\vlc_3(a_2,a_3,a_4)\rangle\lineup = \Alc_4(\Slc_{1234},a_1,a_2,a_3,a_4) + \Alc_4(\Slc_{1(23)4},a_1,a_2,a_3,a_4)\nonumber\\
\lineup\ \ \   +\Alc_4(\Slc_{12(34)},a_1,a_2,a_3,a_4),\label{eq:v3perpAlc}
\end{eqnarray}
where $\Slc_{1234},\Slc_{1(23)4}$ and $\Slc_{12(34)}$ are respective integration cycles in the lightcone fiber bundle $\Plc_4$. We write $\mathcal{S}$ to indicate that these define {\it subvertices} of the full vertex in lightcone gauge. The integration cycle of the full vertex is given by taking their union: 
\begin{equation}
\Vlc_{1234} = \Slc_{1(23)4}\cup \Slc_{1234}\cup \Slc_{12(34)}. \label{eq:Vperp1234}
\end{equation}
As discussed earlier, the equivalence theorem determines the transverse subvertex as 
\begin{equation}\Slc_{1234} = \pilc\circ \V_{1234},\end{equation}
where $\V_{1234}$ is the integration cycle defining the covariant quartic vertex. What remains is to compute the longitudinal subvertices. 

\begin{figure}[t]
\begin{center}
\resizebox{4.7in}{4.3in}{\includegraphics{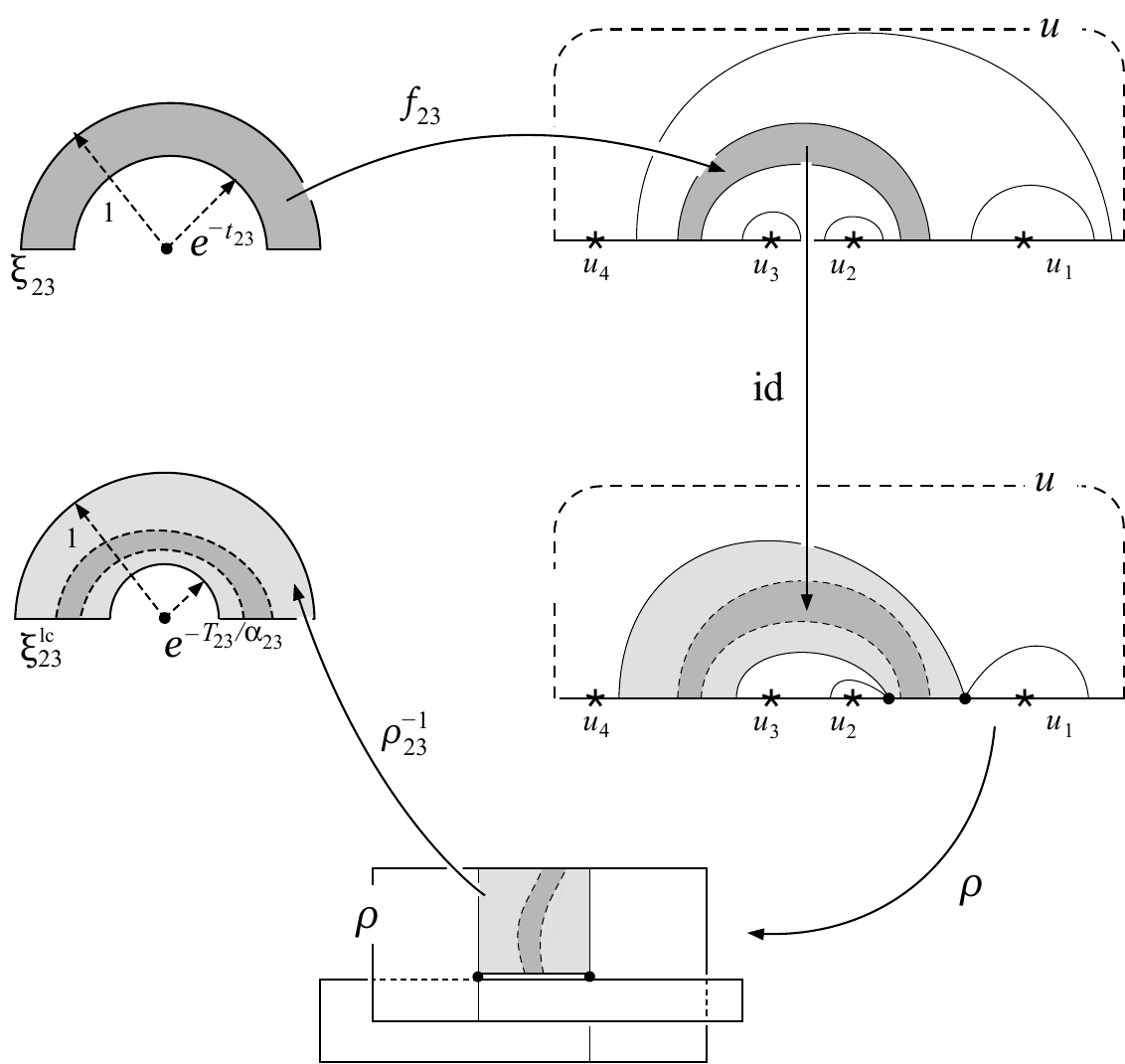}}
\end{center}
\caption{\label{fig:lightcone_gauge16} We assume that the Siegel gauge amplitude is {\it graphically compatible} with its transverse projection. This is reflected above  in three ways. First, when the surface of the 4-point amplitude has a $t$-channel propagator strip, the Mandelstam diagram at the same point in moduli space also has a $t$-channel propagator strip. Second, the propagator strip of the former fits inside the latter. Third, the inner radius of the local coordinate on the Siegel gauge propagator strip is larger than the inner radius of the local coordinate on the Mandelstam diagram's propagator strip.}
\end{figure}

Using cyclicity we restrict attention to the $t$-channel contribution, which may be written as a correlation function in the upper half plane: 
\begin{equation}
-\big\langle \S a_1,v_2(\Deltalong v_2(\S a_2,\S a_3),\S a_4))\big\rangle = \left.\Big\langle \big(f_{23}\circ \Deltalong\big)\big(f_1\circ \S a_1(0)\big)\,...\, \big(f_4\circ \S a_4(0)\big)\Big\rangle_\UHP\right|_{t_{23}=0}.\label{eq:long1}
\end{equation}
Here the local coordinate maps $f_1,...,f_4$ and $f_{23}$ are those of the $t$-channel Siegel gauge diagram when the propagator strip has zero length. $\Deltalong$ is represented as an operator insertion on the collapsed propagator strip. Writing    
\begin{equation}\Deltalong = (b_0-\bDDF)\int_0^\infty |dt_{23}|e^{-t_{23}(L_0-\LDDF)},\label{eq:long15}\end{equation}
we note that $\Deltalong$ contains a factor $e^{-t_{23}L_0}$ whose effect is equivalent to increasing the length of the propagator strip. Then \eq{long1} can be rewritten as
\begin{eqnarray}
\lineup\!\!\!\!\!\!\!\!\!\!\!\!\!\!\! -\big\langle \S a_1,v_2(\Deltalong v_2(\S a_2,\S a_3),\S a_4))\big\rangle\nonumber\\
\lineup 
 = -  \int_{R_{1(23)4}}\!\!\!\! dt_{23}\Big\langle \big(f_{23}\circ(b_0-\bDDF)\big)\big(f_{23}\circ e^{t_{23}\LDDF}\big)\big(f_1\circ \S a_1(0)\big)\,...\, \big(f_4\circ \S a_4(0)\big)\Big\rangle_\UHP. \nonumber\\ 
\ \ \ \ 
\end{eqnarray}
Now the local coordinate maps $f_1,...,f_4$ and $f_{23}$ are those of the $t$-channel Siegel gauge diagram when the propagator strip has Schwinger parameter $t_{23}$. The exponential insertion of the DDF wave operator is supposed to cancel the exchange of transverse states in the Siegel gauge propagator strip. In a slight further alteration we replace the $b_0$ inside the propagator strip with the $b$-ghost of the Schiffer variation:
\begin{eqnarray}
-\big\langle \S a_1,v_2(\Deltalong v_2(\S a_2,\S a_3),\S a_4))\big\rangle\lineup = -  \int_{R_{1(23)4}}\!\!\!\! dt_{23}\left\langle \left(\varphi^* \mathscr{B}\left(\frac{\d}{\d t_{23}}\right)-f_{23}\circ \bDDF\right)\big(f_{23}\circ e^{t_{23}\LDDF}\big)\right.\nonumber\\
\lineup\ \ \ \ \ \ \ \ \ \ \ \ \ \ \ \times\big(f_1\circ \S a_1(0)\big)\,...\, \big(f_4\circ \S a_4(0)\big)\bigg\rangle_\UHP.
\ \ \ \ \label{eq:beforeass}
\end{eqnarray}
In the next step we make an important assumption. We assume that the Siegel gauge amplitude is {\it graphically compatible} with its transverse projection. Roughly speaking, this means that the propagator strips of Siegel gauge diagrams fit inside the propagator strips of Mandelstam diagrams obtained after transverse projection. More precisely, it means three things: 
\begin{description}
\item{\it (1) Channel compatibility.} Transverse projection of a $t$- or $s$-channel Siegel gauge diagram always produces a $t$- or $s$-channel Mandelstam diagram. That is
\begin{equation}R_{1(23)4}\subseteq R^\mathrm{lc}_{1(23)4}\ \ \ \ R_{12(34)}\subseteq R^\mathrm{lc}_{12(34)}.\end{equation}
This means that there will always be propagator widths $T_{23}$ and $T_{34}$ on Mandelstam diagrams which represent the same point in moduli space as Schwinger parameters $t_{23}$ and $t_{34}$ on Siegel gauge diagrams. 
\item{\it (2) Propagator compatibility.} The propagator strip of a Siegel gauge diagram always fits inside the propagator strip of the Mandelstam diagram obtained upon transverse projection. Explicitly, for every $\xi_{23}$ on the $t$-channel propagator strip \eq{xi23}, there is a corresponding $\xi_{23}^\mathrm{lc}$ on the $t$-channel propagator strip of the Mandelstam diagram which satisfies
\begin{equation}f_{23}(\xi_{23}) = \flc_{23}(\xi_{23}^\mathrm{lc}),\label{eq:ws_emb}\end{equation}
where the positions of the punctures are fixed to be equal on both sides of this equation. See figure \ref{fig:lightcone_gauge16}. The analogous condition is also imposed in the $s$-channel. 
\item{\it (3) Length compatibility.} The Schwinger parameters the Siegel gauge diagram cannot be larger than  those of the Mandelstam diagram obtained upon transverse projection:
\begin{equation}T_{23} \geq \alpha_{23}t_{23},\ \ \ \ \ \ \  T_{34} \geq \alpha_{34}t_{34}.\label{eq:length_compatibility}\end{equation}
\end{description}

\noindent We refer to these as {\it graphical compatibility conditions}. We need them to ensure that the longitudinal subvertices take the form described below. We elaborate further on the motivations in subsection~\ref{subsec:embedding}. Because the DDF antighost and DDF wave operator are conformally invariant, we can equate
\begin{equation}
f_{23}\circ \bDDF = \flc_{23}\circ\bDDF,\ \ \ \ \ f_{23}\circ \LDDF = \flc_{23}\circ \LDDF.
\end{equation}
This is possible because propagator compatibility ensures that every point on a closed contour in the $\xi_{23}$ coordinate has a counterpart on a closed contour in the $\xi_{23}^\mathrm{lc}$ coordinate. We further observe that vertex operator equivalence \eq{vertex_equivalence} allows us to equate
\begin{equation}f_i\circ \S a_i(0) = \flc_i\circ e^{-\lambda_i}\circ a_i(0),\label{eq:lVsec}\end{equation}
where the stub lengths $\lambda_i$ are determined by transverse projection of the $t$-channel Siegel gauge diagram with Schwinger parameter $t_{23}$.   Finally we note that $b$-ghost equivalence \eq{b_equivalence} allows us to replace 
\begin{equation}
\varphi^* \mathscr{B}\left(\frac{\d}{\d t_{23}}\right) = (\sigmalc\circ\pilc\circ\varphi)^*\mathscr{B}\left(\frac{\d}{\d t_{23}}\right).
\end{equation}
The result is that the longitudinal subvertex can be expressed as
\begin{eqnarray}
\lineup \!\!\!\!\!\!\!\!\!\!\!\!\!\!\!\!\!\!\!\!\!\!\!\!
-\big\langle \S a_1,v_2(\Deltalong v_2(\S a_2,\S a_3),\S a_4))\big\rangle\nonumber\\
\lineup = -  \int_{R_{1(23)4}}dt_{23}\left\langle \left((\sigmalc\circ\pilc\circ\varphi)^* \mathscr{B}\left(\frac{\d}{\d t_{23}}\right)-\flc_{23}\circ \bDDF\right)\big(\flc_{23}\circ e^{t_{23}\LDDF}\big)\right.\nonumber\\
\lineup \ \ \ \ \ \ \ \ \ \ \ \ \ \ \ \ \ \ \ \ 
\times \big(\flc_1\circ e^{-\lambda_1}\circ a_1(0)\big)\,...\, \big(\flc_4\circ e^{-\lambda_4}\circ a_4(0)\big)\bigg\rangle_\UHP\bigg|_{T_{23}}.
\ \ \ \ \   \label{eq:afterass}
\end{eqnarray}
The integrand is now a correlation function on the Mandelstam diagram obtained by transverse projection of the $t$-channel Siegel gauge diagram with Schwinger parameter $t_{23}$. Channel compatibility ensures that the Mandelstam diagram will also be in the $t$-channel. The width of the propagator strip on the Mandelstam diagram will be written $T_{23}$, which is some nontrivial function of $t_{23}$. The dependence of the correlation function on the width of the propagator strip will be indicated by writing $|_{T_{23}}$ after the closed bracket. The next step is to observe that the replacement formula allows us to equate
\begin{equation}
\flc_{23}\circ \bDDF = \flc_{23}\circ b_0,\ \ \ \ \flc_{23}\circ \LDDF = \flc_{23}\circ L_\perp,\label{eq:afterass2}
\end{equation}
as explained below \eq{Lperprep0}. We also use the result of appendix \ref{subapp:covtoKZ} to replace the Schiffer form of the $b$-ghost with the Kugo-Zwiebach form. This leads to
\begin{eqnarray}
(\sigmalc\circ\pilc\circ\varphi)^* \mathscr{B}\left(\frac{\d}{\d t_{23}}\right)\lineup 
=(\pilc\circ\varphi)^* \mathscr{B}^\mathrm{KZ}\left(\frac{\d}{\d t_{23}}\right)\nonumber\\
\lineup =\frac{1}{\alpha_{23}} \frac{\d T_{23}}{\d t_{23}} \flc_{23}\circ b_0.
\end{eqnarray}
The differentials $d\lambda_i$ drop out of the measure because the transverse vertex operators are proportional to $c$. Then we can write \eq{afterass} as
\begin{eqnarray}
	-\big\langle\S a_1, \lineup\!\! v_2(\Deltalong v_2(\S a_2,\S a_3),\S a_4))\big\rangle \\
	\lineup = -  \int_{R_{1(23)4}}\!\!\!\! dt_{23}\left(\frac{1}{\alpha_{23}} \frac{\d T_{23}}{\d t_{23}}-1\right)\Big\langle \big(\flc_{23}\circ b_0\big)\big(\flc_{23}\circ e^{t_{23}\Lperp}\big)\nonumber\\
	\lineup \ \ \ \ \ \ \ \ \ \ \ \ \ \ \ \ \ \ \ \ \ \ \ \ \ \ \ \ \ \ \ \ \ \ \ \ \ \ \ \ \ \ \times\big(\flc_1\circ e^{-\lambda_1}\circ a_1(0)\big)\,...\, \big(\flc_4\circ e^{-\lambda_4}\circ a_4(0)\big)\Big\rangle_\UHP\Big|_{T_{23}}.\nonumber
\end{eqnarray}
We observe that the transverse wave operator $\Lperp$ is given by subtracting the longitudinal wave operator from $L_0$. Therefore
\begin{equation}e^{t_{23}\Lperp} = e^{t_{23}L_0} e^{-t_{23}\Lpar}.\label{eq:back_op}\end{equation}
Since the first factor is the dilatation generator in the total BCFT, its effect is to decrease the width of the propagator strip on the Mandelstam diagram. Therefore we can drop this factor while at the same time modifying the width of the $t$-channel propagator strip 
\begin{equation}T_{23}\ \to \ T_{23}-\alpha_{23}t_{23}.\end{equation} 
Length compatibility \eq{length_compatibility} ensures that this quantity is positive. The $t$-channel contribution to the quartic vertex is then
\begin{eqnarray}
-\big\langle \lineup\!\! \S a_1,v_2(\Deltalong v_2(\S a_2,\S a_3),\S a_4))\big\rangle   \label{eq:4pen} \\
\lineup = -  \int_{R_{1(23)4}}\!\!\!\! dt_{23}\left(\frac{1}{\alpha_{23}} \frac{\d T_{23}}{\d t_{23}}-1\right)\Big\langle \big(\flc_{23}\circ b_0\big)\big(\flc_{23}\circ e^{-t_{23}\Lpar}\big)\nonumber\\
\lineup \ \ \ \ \ \ \ \ \ \ \ \ \ \ \ \ \ \ \ \ \ \ \ \ \ \ \ \ \ \ \ \ \ \ \ \ \ \ \ \ \ \ \times\big(\flc_1\circ e^{-\lambda_1}\circ a_1(0)\big)\,...\, \big(\flc_4\circ e^{-\lambda_4}\circ a_4(0)\big)\Big\rangle_\UHP\Big|_{T_{23}-\alpha_{23}t_{23}}.\nonumber
\end{eqnarray}
This is almost the expression we are after. To interpret it as a lightcone off-shell amplitude we must identify the appropriate integration cycle.

The integration cycle will be written as
\begin{equation}\Slc_{1(23)4}=(R_{1(23)4},\varphilc).\end{equation} 
where the embedding map $\varphilc $ defines a Mandelstam diagram with stubs for every $t$-channel Siegel gauge diagram. The Mandelstam diagram has the following properties: 
\begin{itemize}
	\item It is in the $t$-channel.
	\item The width of the $t$-channel propagator is 
	\begin{equation}
	T_{23}^\mathrm{vertex} = T_{23} - \alpha_{23}t_{23},  \label{eq:T23}
	\end{equation}
	where $T_{23}$ is the width of the propagator on the Mandelstam diagram obtained by transverse projection of the Siegel gauge diagram and  $t_{23}$ is the Schwinger parameter of the Siegel gauge diagram.
	\item The stub lengths on the Mandelstam diagram are as determined by transverse projection of the Siegel gauge diagram. 
\end{itemize}
To avoid confusion it is worth explaining that the Mandelstam diagrams in the longitudinal subvertex are parameterized by a region of moduli space $R_{1(23)4}$ (the region covered by $t$-channel Siegel gauge diagrams). But this is {\it not} the region of moduli space contained within those Mandelstam diagrams. That is,
\begin{equation}\pi\circ \Slc_{1(23)4} \neq R_{1(23)4}.\end{equation}
The part of moduli space covered by the longitudinal subvertex depends on how the width of the propagator strip on the Mandelstam diagram varies with $R_{1(23)4}$. The backwards shift from the Siegel gauge Schwinger parameter makes this difficult to determine. We address this question in the next subsection.  Other than the backwards shift, the longitudinal subvertex is in all respects identical to the transverse projection of the $t$-channel Siegel gauge amplitude. The backwards shift is simply the result of removing transverse intermediate states from the propagator. It is remarkable that removing transverse states ends up having such an simple geometrical interpretation. 

To finish everything off we need a few more steps. First, we note that it is possible to relate the $b$-ghost insertion in \eq{4pen} with the pullback of the Kugo-Zwiebach $b$-ghost onto our chosen integration cycle:
\begin{eqnarray}
\left(\frac{1}{\alpha_{23}} \frac{\d T_{23}}{\d t_{23}}-1\right)\big(\flc_{23}\circ b_0\big) \lineup = \frac{1}{\alpha_{23}} \frac{\d (T_{23}-\alpha_{23}t_{23})}{\d t_{23}} \big(\flc_{23}\circ b_0\big)\\
\lineup = (\varphilc)^* \mathscr{B}^\mathrm{KZ}\left(\frac{\d}{\d t_{23}}\right).
\end{eqnarray}
Again we can drop $d\lambda_i$ contributions because states are at ghost number 1. Finally, we note that the time evolution generated by the longitudinal wave operator, 
\begin{equation}e^{-t_{23}\Lpar},\end{equation}
can be ignored on account of the freeze theorem \eq{freeze_measure}. We emphasize the importance of the freeze theorem. Without it, lightcone gauge interactions would not be described by the kind of off-shell amplitudes traditionally considered in lightcone string theory. The result~is 
\begin{eqnarray}
-\big\langle \S a_1,\lineup\!\!v_2(\Deltalong v_2(\S a_2,\S a_3),\S a_4))\big\rangle \\
\lineup  = -  \int_{R_{1(23)4}}dt_{23}\left.\left\langle (\varphilc)^*\mathscr{B}^\mathrm{KZ}\left(\frac{\d}{\d t_{23}}\right)\big(\flc_1\circ e^{-\lambda_1}\circ a_1(0)\big)\,...\, \big(\flc_4\circ e^{-\lambda_4}\circ a_4(0)\big)\right\rangle_\UHP\right|_{T_{23}^\mathrm{vertex}},\nonumber
\end{eqnarray}
which is precisely the lightcone off-shell amplitude defined by the integration cycle $\Slc_{1(23)4}$. By cyclicity, the computation goes in the same way in the $s$-channel.

\subsection{Covering moduli space at quartic order} 
\label{subsec:covering4}

Consider the 4-point amplitude in {\it lightcone gauge}:
\begin{eqnarray}
\Alc_4(\Mlc_4,a_1,a_2,a_3,a_4)\lineup  = \underbrace{\big\langle a_1,\vlc_3(a_2,a_3,a_4)\big\rangle}_{1234}-\underbrace{\big\langle a_1,\vlc_2(\Deltaperp\vlc_2(a_2,a_3),a_4))\big\rangle}_{1(23)4}\nonumber\\
\lineup\ \ \   -\underbrace{\big\langle a_1,\vlc_2(a_2,\Deltaperp\vlc_2(a_3,a_4))\big\rangle}_{12(34)}.\label{eq:lcA4}
\end{eqnarray}
This is different from the Siegel gauge amplitude because the external states $a_1,...,a_4\in \Hperp$ are restricted to be transverse, $\vlc_2$ and $\vlc_3$ are the string products of the lightcone effective field theory, and $\Deltaperp$ is the transverse propagator. There are three terms corresponding to the three Feynman graphs in figure \ref{fig:lightcone_gauge12}. Each term can be described as a lightcone off-shell amplitude with the appropriate integration cycle: 
\begin{eqnarray}
\Alc_4(\Mlc_4,a_1,a_2,a_3,a_4)\lineup =\Alc_4(\Vlc_{1234},a_1,a_2,a_3,a_4)+ \Alc_4(\Rlc_{1(23)4},a_1,a_2,a_3,a_4) \nonumber\\
\lineup\ \ \ +\Alc_4(\Rlc_{12(34)},a_1,a_2,a_3,a_4).
\end{eqnarray}
The integration cycle $\Vlc_{1234}$ defines the quartic vertex in lightcone gauge, as discussed in the last subsection. $\Rlc_{1(23)4}$ and $\Rlc_{12(34)}$ represent of $t$- and $s$-channel Mandelstam diagrams derived by connecting cubic vertices in lightcone gauge with a propagator strip. The integration cycle defining the complete amplitude in lightcone gauge can be seen as the union of these, 
\begin{equation}\Mlc_4 = \Rlc_{1(23)4}\cup\Vlc_{1234}\cup\Rlc_{12(34)}.\end{equation}
The question is whether this integration cycle is continuous and covers the whole moduli space. This is equivalent to asking whether the boundary components of $\Mlc_4$ sit on the boundary of moduli space. This is also the question of whether $\Vlc_{1234}$ satisfies the geometrical BV equation.

Expanding $\Vlc_{1234}$ further into subvertices, $\Mlc_4$ consists of five components: 
\begin{equation}
\Mlc_4 = \Rlc_{1(23)4}\cup\Slc_{1(23)4}\cup\Slc_{1234}\cup\Slc_{12(34)}\cup\Rlc_{12(34)}.
\end{equation}
It is manifest that the Mandelstam diagrams in each component vary continuously over the underlying manifold. Therefore, boundaries can only appear if there are gaps between the components. Let us consider first the components that make up the quartic vertex. Note that the geometrical BV equation guarantees that the integration cycle of the Siegel gauge amplitude,
\begin{equation}\M_4 = \R_{1(23)4}\cup\V_{1234}\cup\R_{12(34)},\end{equation}
has no gaps. Considering the transverse Siegel gauge amplitude, we learn that 
\begin{equation}\big(\pilc\circ\R_{1(23)4}\big)\cup \Slc_{1234}\cup\big(\pilc\circ\R_{12(34)}\big)\end{equation}
has no gaps. Next we observe that the Mandelstam diagram at the interface between $\pilc\circ\R_{1(23)4}$ and $\Slc_{1234}$ is the result of transverse projection of a Siegel gauge diagram with a collapsed propagator strip in the $t$-channel. This Mandelstam diagram also appears in $\Slc_{1(23)4}$ because the backwards shift in the propagator width has no effect when the Schwinger parameter vanishes. Therefore $\Slc_{1(23)4}$ connects to $\Slc_{1234}$, and by a similar argument, on to $\Slc_{12(34)}$ without any gaps. In particular, $\Vlc_{1234}$ is connected.

The more difficult question is whether there is a gap between $\Vlc_{1234}$ and the integration cycles defined by the transverse propagators. This requires finding the relation between the propagator width on the Mandelstam diagram and the Schwinger parameter on the Siegel gauge diagram near degeneration.  Let us focus on the $t$-channel. The $t$-channel Feynman graph is formed by connecting two cubic vertices like this:
\begin{equation}
	\resizebox{.8in}{1.2in}{\includegraphics{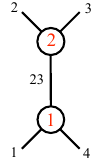}}.
\label{eq:tgraph}
\end{equation}
The punctures on this graph are labeled in order as 1,2,3,4. This defines labels on the punctures of the constituent cubic vertices. The first cubic vertex has punctures labeled in order as $1,23,4$, while the second has punctures labeled $23,2,3$.  The puncture $23$ of the first cubic vertex is connected through the $t$-channel propagator to the puncture $23$ of the second cubic vertex. The Siegel gauge diagram in the $t$-channel defines a surface state 
\begin{equation}
	\big\langle \phi_1,v_2\big(e^{-t_{23}L_0}v_2(\phi_2,\phi_3),\phi_4\big)\big\rangle=\big\langle\big(f_1\circ \phi_1(0)\big)\big(f_2\circ \phi_2(0)\big)\big(f_3\circ \phi_3(0)\big)\big(f_4\circ \phi_4(0)\big) \big\rangle_\UHP.
\label{eq:tchannellong}
\end{equation}
Transverse projection, meanwhile, gives a Mandelstam diagram with stubs defining the surface state
\begin{equation}
	\big\langle\big(\flc_1\circ e^{-\lambda_1}\circ \phi_1(0)\big)\big(\flc_2\circ e^{-\lambda_2}\circ \phi_2(0)\big)\big(\flc_3\circ e^{-\lambda_3}\circ \phi_3(0)\big)\big(\flc_4\circ e^{-\lambda_4}\circ \phi_4(0)\big). \big\rangle_\UHP\label{eq:tchannellonglc}
\end{equation}
What is required is to understand how the local coordinate maps in \eq{tchannellong} depend on the Schwinger parameter $t_{23}$. Then by equating the positions of the punctures between \eq{tchannellong} and \eq{tchannellonglc} we learn how the propagator width on the Mandelstam diagram $T_{23}$ varies as a function of $t_{23}$. The problem is that the puncture locations are usually difficult to determine explicitly as functions of the Schwinger parameter. Thankfully we are only really concerned with the boundary of moduli space, where things simplify. The idea is to write the surface state of the Siegel gauge diagram as 
\begin{equation}
	\big\langle \phi_1,v_2\big(e^{-t_{23}L_0}v_2(\phi_2,\phi_3),\phi_4\big)\big\rangle=\big\langle\big(f_1^1\circ \phi_1(0)\big)\big(f_{23}^1\circ e^{-t_{23}}\circ \phi_{23}(0)\big)\big(f_4^1\circ \phi_4(0)\big)\big\rangle_\UHP+\mathrm{subleading},\label{eq:tcub1}
\end{equation}
where $f_i^1$ are the local coordinate maps of the first cubic vertex, with punctures labeled following the prescription described above. The key here is to find the vertex operator $\phi_{23}(0)$ which represents the fusion of punctures 2 and 3 through a very long $t$-channel propagator strip. It may be defined as the leading order contribution to the product of the two states, 
\begin{equation}
	e^{-t_{23}L_0}v_2(\phi_2,\phi_3) = e^{-t_{23}L_0}\phi_{23} +\mathrm{subleading},
\end{equation}
as seen through a very long propagator strip. As will be clear below, the expansion here exclusively concerns the local coordinate maps, and we do not evaluate the OPE of the vertex operators. In particular, $\phi_{23}(0)$ is not simply the projection of the product onto the kernel of $L_0$. Contracting both sides with a test state $\psi$, this may be written as 
\begin{equation}
	\langle \psi,e^{-t_{23}L_0}\phi_{23}\rangle = \big\langle\big(f_{23}^2\circ e^{-t_{23}}\circ \psi(0)\big)\big(f_2^2\circ \phi_{2}(0)\big)\big(f_3^2\circ \phi_3(0)\big)\big\rangle_\UHP+\mathrm{subleading},
\end{equation}
where $f_i^2$ are the local coordinate maps of the second cubic vertex, with punctures labeled following the prescription described above. The important observation is that the first local coordinate map of the second vertex can be approximated by a linear function near degeneration:
\begin{equation}
	f_{23}^2\circ e^{-t_{23}}(\xi) = L_{23}^2\circ e^{-t_{23}}(\xi)+ \mathrm{subleading},
\end{equation}
where
\begin{equation}L_{23}^2(\xi) =  u_{23}^2+r_{23}^2\xi,  \end{equation}
and $u_{23}^2,r_{23}^2$ are the puncture location and conformal radius of the first puncture in the second cubic vertex. With this one can transform the correlation function with the map 
\begin{equation}e^{-t_{23}}\circ I\circ (L_{23}^2)^{-1}\label{eq:trans_bLv2}\end{equation}	
to arrive at a formula for the (nonlocal) vertex operator 
\begin{equation}
	\phi_{23}(0) =  I\circ (L_{23}^2)^{-1}\circ\Big[\big(f_2^2\circ \phi_{2}(0)\big)\big(f_3^2\circ \phi_3(0)\big)\Big].
\end{equation}
Plugging this back into \eq{tcub1}, we recognize that it is also possible to approximate the second local coordinate map of the first vertex by a linear function:
\begin{equation}
	f_{23}^1\circ e^{-t_{23}}(\xi) = L_{23}^1\circ e^{-t_{23}}(\xi)+ \mathrm{subleading},
\end{equation}
where
\begin{equation}L_{23}^1(\xi) =  u_{23}^1+r_{23}^1\xi,  \end{equation}
and $u_{23}^1,r_{23}^1$ are the puncture location and conformal radius of the second puncture in the first cubic vertex. The result is that we can determine the local coordinate maps of the $t$-channel Siegel gauge diagram near degeneration explicitly in terms of the local coordinate maps of the cubic vertices:
\begin{eqnarray}
	f_1(\xi)\lineup = f_1^1(\xi)+\mathrm{subleading},\label{eq:f41approx}\\
	f_2(\xi)\lineup = \mu\circ f_2^2(\xi)+\mathrm{subleading}, \label{eq:f42approx}\\
	f_3(\xi)\lineup = \mu\circ f_3^2(\xi)+\mathrm{subleading},\label{eq:f43approx}\\
	f_4(\xi)\lineup = f_4^1(\xi)+\mathrm{subleading},\label{eq:f44approx}
\end{eqnarray}
where $\mu(u)$ is a M{\"o}bius transformation
\begin{eqnarray}
	\mu(u) \lineup = L_{23}^1 \circ e^{-t_{23}}\circ I\circ (L_{23}^2)^{-1}(u)\nonumber\\
	\lineup = u_{23}^1-\frac{r_{23}^1r_{23}^2e^{-t_{23}}}{u-u_{23}^2}. \label{eq:mu}
\end{eqnarray}
In this limit the dependence of the maps on the Schwinger parameter can be made explicit.

\begin{figure}[t]
\begin{center}
\resizebox{4.5in}{2in}{\includegraphics{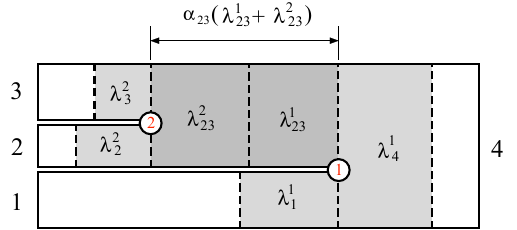}}
\end{center}

\caption{\label{fig:lightcone_gauge17} Mandelstam diagram formed by gluing a cubic vertices in lightcone gauge in the $t$-channel. The stubs on the cubic vertices are labeled according to the corresponding punctures as described in \eq{tgraph}.}
\end{figure}

Due to simplifications in the Mandelstam mapping near degeneration (see subsection \ref{subsec:continuity}), it is not difficult to show that the transverse projection of the Siegel gauge diagram has an analogous simplification near degeneration:
\begin{eqnarray}
	\flc_1\circ e^{-\lambda_1}(\xi)\lineup = f_1^{\mathrm{lc},1}\circ e^{-\lambda_1^1}(\xi)+\mathrm{subleading},\\
	\flc_2\circ e^{-\lambda_2}(\xi)\lineup = \mu\circ f_2^{\mathrm{lc},2}\circ e^{-\lambda_2^2}(\xi)+\mathrm{subleading}, \\
	\flc_3\circ e^{-\lambda_3}(\xi)\lineup = \mu\circ f_3^{\mathrm{lc},2}\circ e^{-\lambda_3^2}(\xi)+\mathrm{subleading},\\
	\flc_4\circ e^{-\lambda_4}(\xi)\lineup = f_4^{\mathrm{lc},1}\circ e^{-\lambda_4^1}(\xi)+\mathrm{subleading},
\end{eqnarray}
where $f_i^{\mathrm{lc},1},f_i^{\mathrm{lc},2}$ are the lightcone local coordinate maps of the first and second cubic vertices, and $\lambda_i^1,\lambda_i^2$ are the stub lengths obtained from transverse projection of the first and second covariant cubic vertices. Here we have the same M{\"o}bius transformation $\mu$ which tells us how the lightcone local coordinate maps depend on the Schwinger parameter of the Siegel gauge diagram. To understand how they depend on the propagator width $T_{23}$, note that the argument of the previous paragraph applied to Mandelstam diagrams will lead to the same formula for the lightcone local coordinate maps near degeneration but with a M{\"o}bius transformation of the form
\begin{eqnarray}
	\mu^\mathrm{lc}(u) = u_{23}^1-\frac{r_{23}^{\mathrm{lc},1}r_{23}^{\mathrm{lc},2}e^{-T_{23}/\alpha_{23}}}{u-u_{23}^2}, \label{eq:mulc}
\end{eqnarray}
where $r_{23}^{\mathrm{lc},1},r_{23}^{\mathrm{lc},2}$ are the conformal radii of the lightcone local coordinate maps $f_{23}^{\mathrm{lc},1},f_{23}^{\mathrm{lc},2}$. It follows that the propagator width and Schwinger parameter must be related as
\begin{equation}
\frac{T_{23}}{\alpha_{23}} + \ln(r_{23}^{\mathrm{lc},1})+\ln(r_{23}^{\mathrm{lc},2}) = t_{23}+ \ln(r_{23}^1)+\ln(r_{23}^2)+\mathrm{subleading},
\end{equation}
which means
\begin{equation}\frac{T_{23}}{\alpha_{23}} -t_{23} = \lambda_{23}^1+\lambda_{23}^2+\mathrm{subleading}.\end{equation}
It follows that the Mandelstam diagram of the quartic vertex in lightcone gauge at the $t$-channel degeneration has moduli and stubs given by
\begin{eqnarray}
\lim_{t_{23}\to\infty}T_{23}^\mathrm{vertex} \lineup = \alpha_{23}\big(\lambda_{23}^1+\lambda_{23}^2\big),\\
\lim_{t_{23}\to\infty}\lambda_1 \lineup = \lambda_1^1,\\
\lim_{t_{23}\to\infty}\lambda_2 \lineup = \lambda_2^2,\\
\lim_{t_{23}\to\infty}\lambda_3 \lineup = \lambda_3^2,\\
\lim_{t_{23}\to\infty}\lambda_4 \lineup = \lambda_4^1.
\end{eqnarray}
Comparing to figure \ref{fig:lightcone_gauge17} it is clear that this Mandelstam is precisely the same as what you get by gluing two cubic vertices in lightcone gauge through the $t$-channel. A similar result applies in the $s$-channel. Since there are no gaps, the $4$-point amplitude in lightcone gauge covers the whole moduli space. 
	
\subsection{An example}
\label{subsec:example}

It is instructive to compute longitudinal subvertices in an example. We work with a covariant open bosonic SFT whose cubic vertex is defined by $SL(2,\mathbb{R})$ local coordinate maps \cite{heterotic_tadpole,ErbinSuvajit}
\begin{eqnarray}
f_{(3,1)}(\xi)\lineup = \frac{3+e^{-\ell}\xi}{3-e^{-\ell}\xi},\label{eq:f31sl}\\
f_{(3,2)}(\xi)\lineup = \frac{2e^{-\ell}\xi}{3+e^{-\ell}\xi},\\
f_{(3,3)}(\xi)\lineup = -\frac{3-e^{-\ell}\xi}{2e^{-\ell}\xi},\label{eq:f33sl}
\end{eqnarray}
which place punctures respectively at $1,0$ and infinity. The punctures here are labeled as $1,2,3$ and we use $i\to(3,i)$ to indicate that these are the maps of the cubic vertex. The parameter $\ell$ represents the length of a stub attached to each entry of the vertex. This choice of vertex is useful because it permits closed form expressions for the longitudinal subvertices. Ultimately this is because $SL(2,\mathbb{R})$ maps preserve the upper half plane. However, the explicit computation below demonstrates that the qualitative form of lightcone gauge interactions likely does not depend very much on the choice of covariant vertex. If the conformal radii of the covariant vertex are small enough to ensure that lightcone gauge is well-defined, the higher order structure of the local coordinate maps will have minor impact on the results. 

We will focus on the $t$-channel longitudinal subvertex. We also assume that the first three string length parameters $\alpha_1,\alpha_2,\alpha_3$ are positive, as shown in figure \ref{fig:lightcone_gauge17}, since this leads to some interesting variations in behavior. The local coordinate maps of the Siegel gauge diagram in the $t$-channel are
\begin{eqnarray}
f_1(\xi)\lineup = N\circ f_{(3,1)}(\xi),\\
f_2(\xi)\lineup = N\circ M \circ f_{(3,2)}(\xi), \\
f_3(\xi)\lineup = N\circ M \circ f_{(3,3)}(\xi),\\
f_4(\xi)\lineup = N\circ f_{(3,3)}(\xi),
\end{eqnarray}
where
\begin{eqnarray}
	M(u) \lineup = f_{(3,2)}\circ e^{-t_{23}}\circ I\circ f_{(3,1)}^{-1}(u)\nonumber\\
	\lineup = \frac{2}{1-9 e^{2\ell+t_{23}}\frac{u-1}{u+1}}
\end{eqnarray}
can be thought of as an exact version of the map $\mu(u)$ in equation \eq{mu}, and $N(u)$ is a scale transformation around $u=1$ given as
\begin{equation}N(u) = 1+\frac{9 e^{2\ell+t_{23}}-1}{9 e^{2\ell+t_{23}}+1}(u-1),\end{equation}
which ensures that the first, third, and fourth punctures are canonically inserted at $1,0$ and infinity respectively. The derivation of these maps follows the construction described (in the context of the closed string) in \cite{ClosedSFT}. Presently we index the maps of the cubic vertex according to \eq{f31sl}-\eq{f33sl}, but to align with the puncture labels of the $t$-channel diagram following \eq{tgraph} we should identify
\begin{eqnarray}
\lineup f_1^1 = f_{(3,1)},\ \ \ f_{23}^1 = f_{(3,2)},\ \ \ f_4^1 = f_{(3,3)},\nonumber\\
\lineup f_{23}^2 = f_{(3,1)},\ \ \ f_{2}^2 = f_{(3,2)},\ \ \ f_3^2 = f_{(3,3)}.
\end{eqnarray}
The location of the second puncture of the Siegel gauge diagram is 
\begin{equation}u_2=f_2(0) = \frac{36 e^{2\ell+t_{23}}}{(9 e^{2\ell+t_{23}}+1)^2}. \end{equation}
Also important is the local coordinate map of the propagator strip on the Siegel gauge diagram:
\begin{equation}
f_{23}(\xi) = N\circ f_{(3,2)}(\xi),\ \ \ \ \ \ 1\geq|\xi|\geq e^{-t_{23}},\ \ \mathrm{Im}(\xi)\geq 0.
\end{equation}
Meanwhile, the local coordinate maps of the $t$-channel Mandelstam diagram can be determined from the formulas given in subsection \ref{subsec:Mandelstam}. Equating the positions of the punctures gives the propagator width $T_{23}$ on the Mandelstam diagram exactly as a function of the Schwinger parameter $t_{23}$ on the Siegel gauge diagram.  The respective conformal radii determine the stub lengths $\lambda_1,...,\lambda_4$ as functions of $t_{23}$. Thus we are able to obtain completely explicit formulas for the integration cycle $\Slc_{1(23)4}$ defining the $t$-channel longitudinal subvertex. We do not write the formulas because they are complicated and not enlightening.

\begin{figure}[t]
	\begin{center}
		\resizebox{6.5in}{1.8in}{\includegraphics{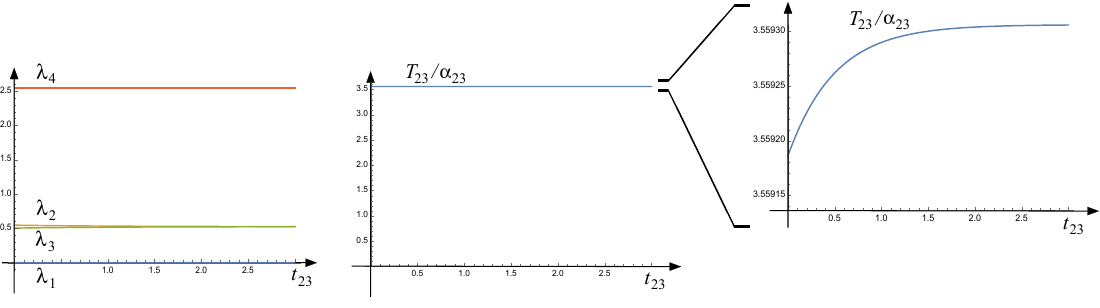}}
	\end{center}
	\caption{\label{fig:lightcone_gauge18} Assuming that the string lengths are in the configuration \eq{ex1} and that the bound \eq{ex1_bound} is saturated at $\ell =1.505$, the plots above show the stub lengths $\lambda_1,...,\lambda_4$ and the propagator width $T_{23}$ as functions of $t_{23}$. The rightmost plot zooms in to show more precisely how the width of the propagator varies in the longitudinal subvertex.}
\end{figure}

We need to impose some conditions to ensure that the longitudinal subvertex is well-defined. First of all, $\Slc_{1(23)4}$ must be an {\it admissible} integration cycle in the lightcone fiber bundle, which means that all stub lengths are zero or positive:
\begin{equation}\lambda_1,...,\lambda_4\geq 0,\ \ \ \text{for all} \ t_{23}\geq 0.\label{eq:ex_adm}\end{equation}
Second, we must ensure that the Siegel gauge amplitude is {\it graphically compatible} with its transverse projection. These conditions are expected to hold if the stub length $\ell$ of the covariant cubic vertex is large enough.

\begin{figure}[t]
	\begin{center}
		\resizebox{3.8in}{2in}{\includegraphics{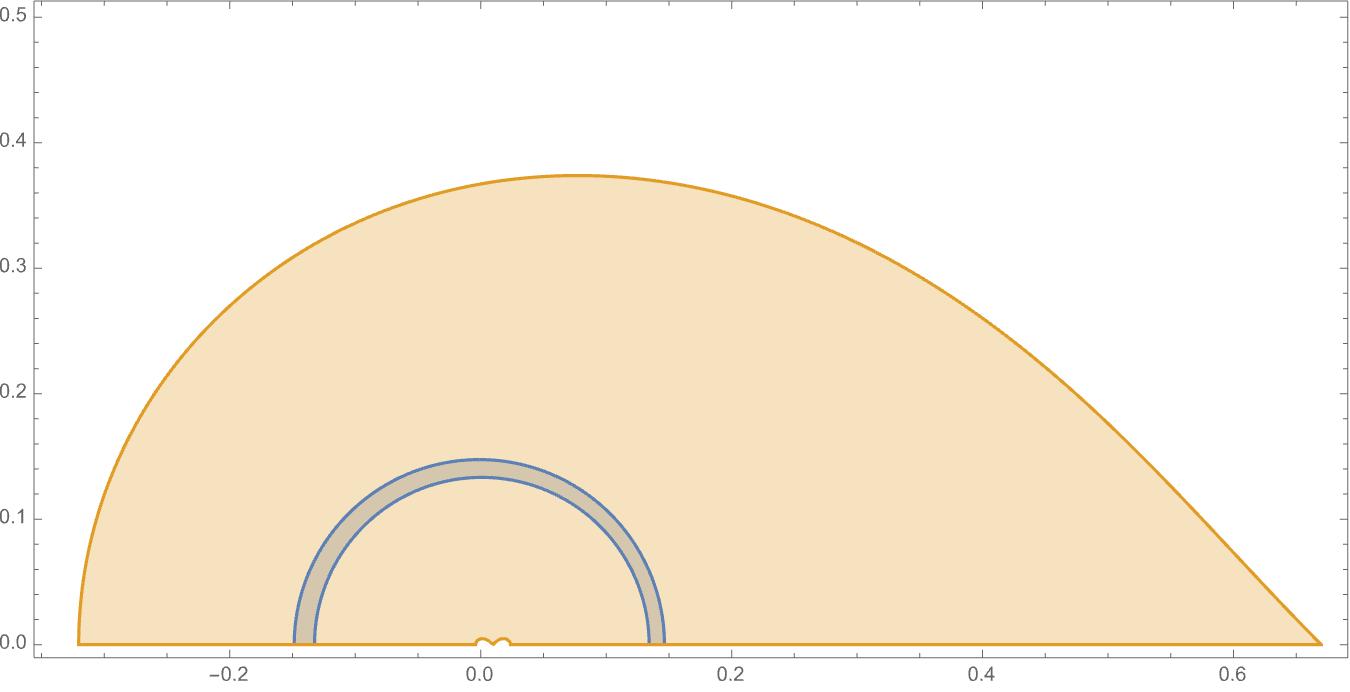}}
	\end{center}
	\caption{\label{fig:lightcone_gauge19} Propagator strips of the Mandelstam diagram and Siegel gauge diagram in the upper half plane coordinate assuming four bits of data: 1)  The string lengths are fixed according to \eq{ex1}. 2) The Schwinger parameter on the $t$-channel Siegel gauge diagram is $t_{23}=0.1$, and the propagator width $T_{23}$ on the Mandelstam diagram is fixed accordingly. 3)~The stub length on the $SL(2,\mathbb{R})$ vertex is chosen with $\ell=1.505$ to saturate the bound \eq{ex1_bound}. 4) The first, third and fourth punctures have been fixed to $1,0$ and infinity respectively. The blue region is the image of the propagator strip on the Siegel gauge diagram, and the orange region is the image of the propagator strip on the Mandelstam diagram. The blue region is generously contained within the orange region, and one can check that this continues to be the case for all $t_{23}\geq 0$. The small bumps on the lower boundary of the orange region near $u=0$ are nearly invisible in this conformal frame, but represent the boundary of the lightcone local coordinate patches of the second and third punctures.}
\end{figure}

As a starting example let us assume that the three incoming strings all have the same length
\begin{equation}\alpha_1=\frac{1}{3}|\alpha_4|,\ \ \ \alpha_2=\frac{1}{3}|\alpha_4|,\ \ \ \alpha_3=\frac{1}{3}|\alpha_4|.\label{eq:ex1}\end{equation}
Working through the conditions of the previous paragraph, it turns out that the limiting one is that the stub of the first puncture has zero or positive length at $t_{23}=0$:
\begin{equation}\lambda_1\geq 0,\ \ \ \text{at} \ t_{23}= 0.\end{equation}
This requires that the stub of the $SL(2,\mathbb{R})$ vertex must be chosen so that
\begin{equation}\ell\gtrsim  1.505.\label{eq:ex1_bound}\end{equation}
Assuming this bound is saturated, we plot the stub lengths and the propagator width as functions of $t_{23}$ in figure \ref{fig:lightcone_gauge18}. What is apparent is that all of these quantities are nearly constant as functions of $t_{23}$. Upon closer inspection, we find some variation of $\lambda_2$ and $\lambda_3$ on the order of one percent of $\lambda_4$ (which defines a natural scale in this context), while $\lambda_1$ and $\lambda_4$ vary on the order of a thousandth of a percent. The propagator width $T_{23}$ also varies on the order of a thousandth of a percent, and is a strictly increasing function of $t_{23}$. What this means is that the $t$-channel longitudinal subvertex is a very small contribution to the 4-point amplitude in lightcone gauge. This is ultimately a consequence of the fact that the $SL(2,\mathbb{R})$ vertex must be chosen with fairly long stubs. Decreasing $\ell$ makes the dependence on $t_{23}$ more apparent, though not all consistency conditions are met. In passing, let us confirm the graphical compatibility conditions in the present scenario. It can be seen directly from figure \ref{fig:lightcone_gauge18} that channel and length compatibility conditions hold. Propagator compatibility is illustrated in figure \ref{fig:lightcone_gauge19}. Even when the bound \eq{ex1_bound} is saturated, the propagator strip on the Siegel gauge diagram easily fits inside the propagator strip on the Mandelstam diagram.

\begin{figure}[t]
	\begin{center}
		\resizebox{3.1in}{2.1in}{\includegraphics{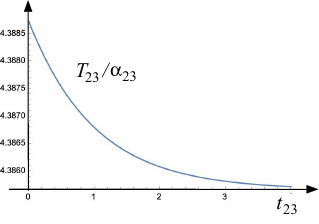}}
	\end{center}
	\caption{\label{fig:lightcone_gauge20} Plot of the propagator width $T_{23}$ as a function of $t_{23}$ assuming the string lengths in the configuration \eq{ex2} and the bound \eq{ex2_bound} saturated at $\ell = 1.844$.}
\end{figure}

Next consider incoming strings with lengths arranged as
\begin{equation}
\alpha_1=\frac{1}{4}|\alpha_4|,\ \ \ \alpha_2=\frac{1}{4}|\alpha_4|,\ \ \ \alpha_3=\frac{1}{2}|\alpha_4|.\label{eq:ex2}
\end{equation}
Again the limiting condition is that the stub of the first puncture has zero or positive length, but in this case when $t_{23}$ is infinite
\begin{equation}\lambda_1\geq 0\ \ \ \text{at}\ t_{23}=\infty.\end{equation}
This limits the stub length of the $SL(2,\mathbb{R})$ vertex as 
\begin{equation}\ell\gtrsim 1.844.\label{eq:ex2_bound}\end{equation}
In this example the stub and propagator lengths are again nearly constant as functions of $t_{23}$, though the variation is slightly more, on the order of a tenth of a percent. What is interesting in this example is that the propagator width on the Mandelstam diagram is {\it decreasing} as a function of $t_{23}$, as shown in figure \ref{fig:lightcone_gauge20}. This means that the longitudinal subvertex uncovers part of the moduli space covered by the transverse subvertex. This must then be covered again by the transverse propagator contribution to the lightcone gauge amplitude. Therefore the amplitude is not characterized by a section of the relevant fiber bundle. To our knowledge this is the natural example of such an amplitude appearing in a string field theory calculation. It is not yet established whether amplitudes defined by hyperbolic string vertices are characterized by sections \cite{Costello}.

\begin{figure}[t]
	\begin{center}
		\resizebox{3.1in}{2.1in}{\includegraphics{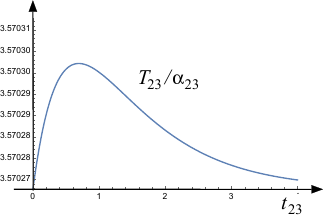}}
	\end{center}
	\caption{\label{fig:lightcone_gauge21} Plot of the propagator width $T_{23}$ as a function of $t_{23}$ assuming the string lengths in the configuration \eq{ex3} and  $\ell = 1.509$.}
\end{figure}

A final example is the configuration
\begin{equation}
\alpha_1 \approx 0.3321|\alpha_4|,\ \ \ \alpha_2\approx 0.3321|\alpha_4|,\ \ \ \alpha_3\approx 0.3357|\alpha_4|,\label{eq:ex3}
\end{equation}
with the stub length on the $SL(2,\mathbb{R})$ vertex fixed to
\begin{equation}\ell\approx 1.509.\end{equation}
In this case the stub length $\lambda_{1}$ is zero when $t_{23}=0$, and all other consistency conditions are satisfied. What is interesting in this example is that the longitudinal subvertex actually covers {\it nothing} on the moduli space. The width of the propagator on the Mandelstam diagram is the same at $t_{23}=0$ as it is at $t_{23}=\infty$. Nevertheless the width $T_{23}$ is varying for intermediate values of $t_{23}$ (on the order of a thousandth of a percent), as shown in figure \ref{fig:lightcone_gauge21}. Since $T_{23}$ is not monotonic, the longitudinal subvertex is already by itself not represented by a section.

\section{Higher vertices in lightcone gauge}
\label{sec:higher}

The generalization to higher vertices is straightforward. The main complication is that there are many more kinds of moduli, integration cycles, Siegel gauge diagrams and Mandelstam diagrams involved, and one has to set some conventions for talking about them. Mainly this reduces to the problem of labeling Feynman graphs and their punctures, vertices and propagators. So to begin we explain our conventions for dealing with this. 

Specifying a color ordered, tree-level, $n$-point Feynman graph $\F$ begins by specifying an ordered list of $n$ symbols representing the punctures. Then, as sketched in figure \ref{fig:lightcone_gauge12}, the configuration of vertices and propagators within the graph is specified by listing the punctures in order and inserting parentheses. We assume that the data of the Feynman graph includes a set of symbols for the propagators and vertices of the diagram. We use
\begin{equation}\puncture(\F)\end{equation}
to denote the set of symbols for the punctures of the Feynman graph,
\begin{equation}\propagator(\F)\end{equation}
to denote the set of symbols for the propagators of the Feynman graph, and
\begin{equation}\cubic(\F),\ \ \ \quartic(\F),\ \ \ \mathrm{quintic}(\F),\ \ \ ...\end{equation}
to denote the set of symbols for the cubic vertices, quartic vertices, quintic vertices, and so on contained within the graph.

\begin{figure}[t]
	\begin{center}
		\resizebox{6in}{2in}{\includegraphics{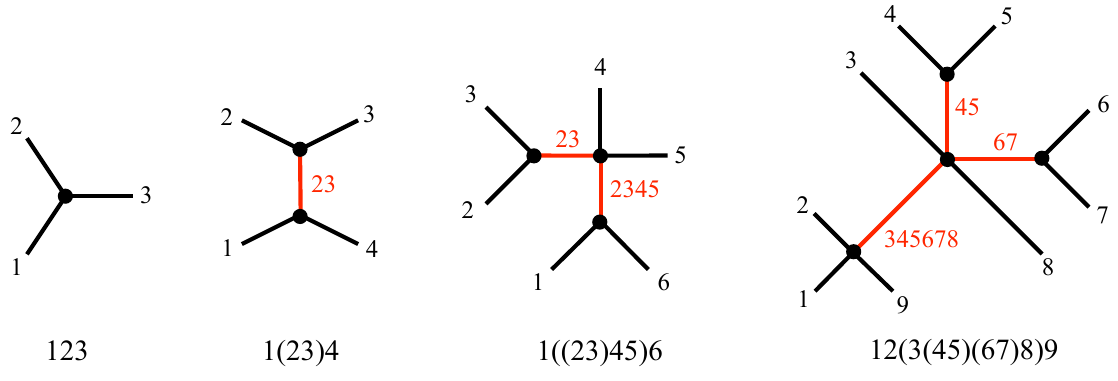}}
	\end{center}
	\caption{\label{fig:lightcone_gauge22} Feynman graphs corresponding to labels given in \eq{prop_ex}, and their associated labeled propagators.}
\end{figure}

\begin{figure}[t]
	\begin{center}
		\resizebox{6in}{2in}{\includegraphics{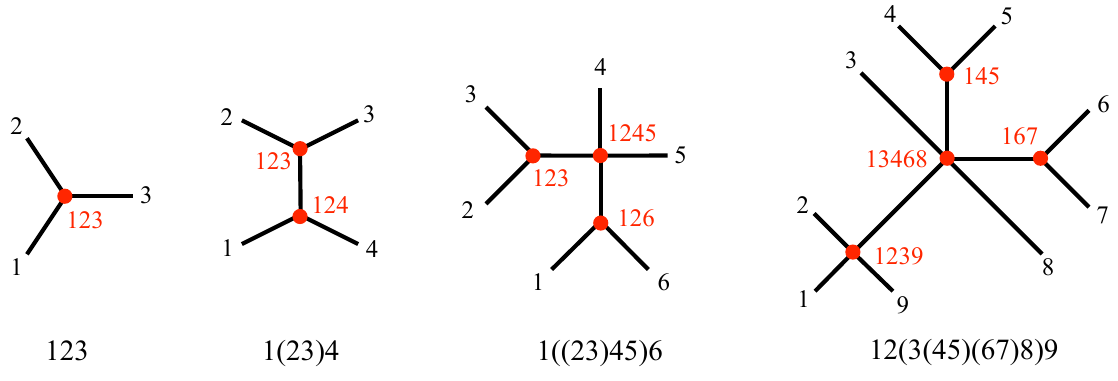}}
	\end{center}
	\caption{\label{fig:lightcone_gauge31} Labels for the vertices of the Feynman graphs in figure \ref{fig:lightcone_gauge22}.}
\end{figure}

When appropriate, we generate the symbols of the Feynman graph by the following default prescription. The punctures listed in order will be $1,2,..,n$. The propagators will be specified by listing the set of punctures in order which are separated from the first puncture at degeneration. The vertices will be specified by counting from 1 and attaching a new number every time the corresponding puncture is connected to a new input of the vertex through the graph. We illustrate the prescription for a few sample Feynman graphs in figures  \ref{fig:lightcone_gauge22} and \ref{fig:lightcone_gauge31}. The propagators are
\begin{eqnarray}
	\propagator\big(123\big)\lineup = \varnothing, \nonumber\\
	\propagator\big(1(23)4\big)\lineup = \{23\},\nonumber\\
	\propagator\big(1((23)45)6\big)\lineup = \{23,2345\};\nonumber\\
	\propagator\big(12(3(45)(67)8)9\big)\lineup = \{45,67,345678\}, \label{eq:prop_ex}
\end{eqnarray}
and the nonempty sets of vertices are
\begin{eqnarray}
	\cubic\big( 123\big) \lineup = \{123\},\nonumber\\
	\cubic\big(1(23)4\big)\lineup = \{123,124\},\nonumber\\
	\cubic\big(1((23)45)6\big)\lineup = \{123,126\},\ \ \ \ \nonumber\\
	\cubic\big(12(3(45)(67)8)9\big)\lineup = \{145,167\},\nonumber\\
	\quartic\big(1((23)45)6\big) \lineup = \{1245\},\nonumber\\
	\quartic\big(12(3(45)(67)8)9\big) \lineup = \{1239\},\nonumber\\
	\mathrm{quintic}\big(12(3(45)(67)8)9\big)\lineup =\{13468\}.
\end{eqnarray}

\subsection{Vertex theorem}
\label{subsec:main}

The computation of the quartic vertex in subsection \ref{subsec:reducing} is straightforward to generalize. The result is an $n$-string vertex given as follows:
\begin{description}
\item{\bf Vertex theorem:} Assume that the $n$-point amplitude in Siegel gauge is graphically compatible with its transverse projection (see subsection \ref{subsec:embedding}). Then the $n$-string vertex in lightcone gauge is a lightcone off-shell amplitude defined by an integration cycle $\Vlc_{12...n}$ which specifies a Mandelstam diagram with stubs for every Siegel gauge diagram which enters the Siegel gauge $n$-point amplitude. The Mandelstam diagram is related to the Siegel gauge diagram as follows:
	\begin{itemize}
		\item The Siegel gauge diagram forms a Feynman graph $\F$. The corresponding Mandelstam diagram in the vertex forms a Feynman graph $\Flc$. The graph $\Flc$ is the same as that of the Mandelstam diagram produced upon transverse projection of the Siegel gauge diagram. The punctures and shared propagator channels of $\F$ and $\Flc$ will be labeled in the same way. 
		\item The moduli on the Mandelstam diagram are related to those of the Siegel gauge diagram according to 
		\begin{eqnarray}
			T_i^\mathrm{vertex} \lineup = T_i - \alpha_i t_i,\ \ \ \ i\in\propagator(\F),\label{eq:Tivertex}\\
			T_i^\mathrm{vertex}\lineup = T_i,\ \ \ \ \ \ \ \ \ \ \ \ i\in \propagator(\Flc)-\propagator(\F),\\
			\theta_I^\mathrm{vertex}\lineup = \theta_I,\ \ \ \ \ \ \ \ \ \ \ \ I\in \quartic(\Flc),
		\end{eqnarray}
		where $t_i$ are the Schwinger parameters on the Siegel gauge diagram and $(T_i,\theta_I)$ are the moduli of the Mandelstam diagram obtained by transverse projection of the Siegel gauge diagram.
		\item The stub lengths on the Mandelstam diagram are
		\begin{equation}\lambda_i^\mathrm{vertex} = \lambda_i,\end{equation}
		where $\lambda_i$ are the stub lengths on the Mandelstam diagram obtained by transverse projection of the Siegel gauge diagram.
	\end{itemize}
\end{description}
The $n$-string vertex in lightcone gauge is the same as the transverse $n$-point amplitude in Siegel gauge except for \eq{Tivertex}, which says that some propagator strips on each Mandelstam diagram are shortened proportionally to the Schwinger parameter of the corresponding Siegel gauge diagram. A propagator strip is shortened if and only if the Siegel gauge diagram has a propagator strip in the same channel, so that the relevant Schwinger parameter can be defined. The set of Mandelstam diagrams in the vertex are parameterized by the set of Siegel gauge diagrams. If the amplitude in Siegel gauge is represented by a section of the covariant fiber bundle $\P_n$, the set of Siegel gauge diagrams is isomorphic to the moduli space $M_n$ of disks with $n$ boundary punctures. In this situation the Mandelstam diagrams in the vertex will be parameterized by $M_n$, even though the diagrams themselves do not cover moduli space. The boundary of moduli space is mapped to the boundary of the vertex region of moduli space.

To prove the vertex theorem we need to compute subvertices with possibly several longitudinal propagators connecting covariant vertices which may themselves carry moduli. But in all important respects the computation at higher order follows that of the quartic vertex, and it will not be necessary to elaborate the general argument in detail.

\subsection{Quintic vertex}
\label{subsec:quintic}

Nevertheless it is helpful to at least sketch the computation of the quintic vertex, paralleling that of the quartic vertex in subsection \ref{subsec:reducing}. We start with the Siegel gauge 5-point amplitude. The amplitude involves five covariant states  $\phi_1,...,\phi_5\in\Hcov$ at ghost number 1. Gluing covariant vertices with Siegel gauge propagators in various combinations gives the amplitude as a sum of eleven terms (not writing them all explicitly),
\begin{eqnarray}
	A_5(\M_5,\phi_1, \lineup \!\! \phi_2,\phi_3,\phi_4,\phi_5)=\underbrace{ \Big\langle \phi_1,v_4(\phi_2,\phi_3,\phi_4,\phi_5)\Big\rangle}_{12345}\label{eq:Siegel5op}\\
	\lineup -\underbrace{\Big\langle\phi_1,v_3\big(\phi_2,\Delta_{b_0}v_2(\phi_3,\phi_4),\phi_5\big)\Big\rangle}_{12(34)5}
	-...-\underbrace{\Big\langle\phi_1,v_2\big(\phi_2,\Delta_{b_0}v_3(\phi_3,\phi_4,\phi_5)\big)\Big\rangle}_{12(345)} \nonumber\\
	\lineup  +\underbrace{\Big\langle \phi_1,v_2\big(\phi_2,\Delta_{b_0}v_2(\Delta_{b_0}v_2(\phi_3,\phi_4),\phi_5\big)\Big\rangle}_{12((34)5)}
	+...+\underbrace{\Big\langle \phi_1,v_2\big(\phi_2,\Delta_{b_0}v_2(\phi_3,\Delta_{b_0}v_2(\phi_4,\phi_5)\big)\Big\rangle}_{12(3(45))},\nonumber
\end{eqnarray}
corresponding to the eleven cyclically inequivalent color-ordered 5-point Feynman graphs.  Each term represents a collection of Siegel gauge diagrams which form that particular Feynman graph. The Siegel gauge amplitude is then a covariant off-shell amplitude whose integration cycle is the union of eleven components:
\begin{equation}\M_5 = \V_{12345}\cup\Big[\R_{12(34)5}\cup ... \cup \R_{12(345)}\Big]\cup\Big[\R_{12((34)5)}\cup...\cup\R_{12(3(45))}\Big].\label{eq:calM5comp}\end{equation}
$\V_{12345}$ represents the contribution from the covariant quintic vertex. The contributions in the first bracket come from diagrams with one propagator connecting a cubic and quartic vertex, and those inside the second bracket come from diagrams with two propagators connecting three cubic vertices. For the sake of discussion we assume $\M_5$ is a section of the covariant fiber bundle. Therefore applying the bundle projection to \eq{calM5comp} implies that the Feynman graphs of the Siegel gauge amplitude decompose the moduli space into  eleven regions:
\begin{equation}
	M_5 = V_{12345}\cup\Big[ R_{12(34)5}\cup ... \cup R_{12(345)}\Big]\cup\Big[R_{12((34)5)}\cup...\cup R_{12(3(45))}\Big],
\end{equation}

\begin{figure}[t]
	\begin{center}
		\resizebox{3in}{2.9in}{\includegraphics{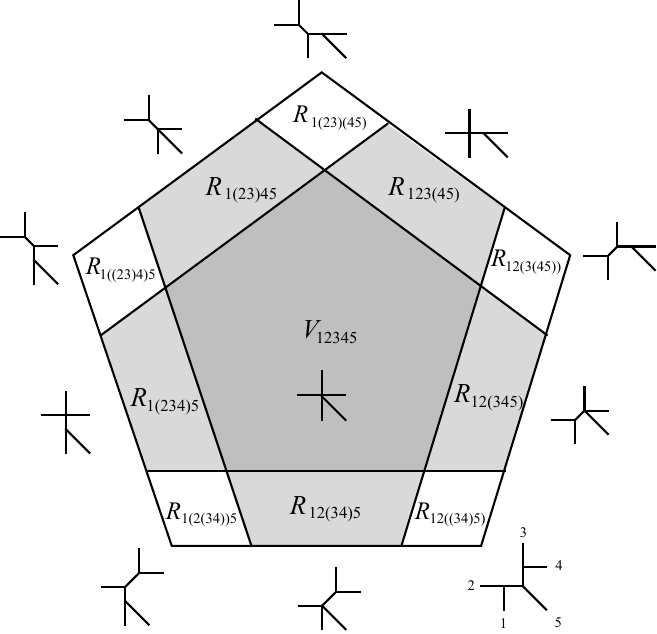}}
	\end{center}
	\caption{\label{fig:lightcone_gauge25} Decomposition of the moduli space $M_5$ into regions covered by respective Feynman graphs of the Siegel gauge 5-point amplitude.}
\end{figure}

\noindent as shown in figure \ref{fig:lightcone_gauge25}. The moduli space may be visualized as a pentagon. Feynman graphs with two propagators connecting three cubic vertices cover the five corners of the pentagon; those with one propagator connecting a cubic and quartic vertex cover the five edges; and, finally, the quintic vertex covers the interior.  Each term in the Feynman graph expansion of the 5-point amplitude \eq{Siegel5op} represents integration of a worldsheet correlation function over the corresponding portion of moduli space. If written out explicitly this reads 
\begin{eqnarray}
	\lineup \!\!\!\!\!\!\! A_5(\M_5,\phi_1,...,\phi_5)\nonumber\\
	\lineup	= \int_{V_{12345}}  dm_{12345}^1 dm_{12345}^2 \left\langle \varphi^*\mathscr{B}\left(\frac{\d}{\d m_{12345}^2}\right)\varphi^*\mathscr{B}\left(\frac{\d}{\d m_{12345}^1}\right)\big(f_1\circ \phi_1(0)\big) .... \big(f_5\circ\phi_5(0)\big)\right\rangle_{\UHP}\nonumber\\
	\lineup\ \ \ +\Bigg[\int_{R_{12(34)5}} dt_{34} dm_{1235} \Big\langle \big(\mathscr{B}_{1235}\big)\big(f_{34}\circ b_0 \big)\big(f_1\circ \phi_1(0)\big) .... \big(f_5\circ\phi_5(0)\Big)\Big\rangle_{\UHP}\nonumber\\ 
	\lineup\ \ \ \ \ \ \ +\,...+\int_{R_{12(345)}} dt_{345} dm_{1345} \Big\langle \big(\mathscr{B}_{1345}\big)\big(f_{345}\circ b_0 \big)\big(f_1\circ \phi_1(0)\big) .... \big(f_5\circ\phi_5(0)\big)\Big\rangle_{\UHP}\Bigg]\nonumber\\
	\lineup\ \ \ +\Bigg[\int_{R_{12((34)5)}} dt_{34}dt_{345}\Big\langle\big(f_{345}\circ b_0\big)\big(f_{34}\circ b_0\big)\big(f_1\circ \phi_1(0)\big) .... \big(f_5\circ\phi_5(0)\big)\Big\rangle_{\UHP}\nonumber\\
	\lineup\ \ \ \ \ \ \ +\, ...+ \int_{R_{12((34)5)}} dt_{45}dt_{345}\Big\langle\big(f_{345}\circ b_0\big)\big(f_{45}\circ b_0\big)\big(f_1\circ \phi_1(0)\big) .... \big(f_5\circ\phi_5(0)\big)\Big\rangle_{\UHP}\Bigg].\label{eq:Siegel5pt}
\end{eqnarray}
$f_1,...,f_5$ are the local coordinate maps appropriate to the Siegel gauge diagram in that part of moduli space, and in the propagator regions, there are also local coordinate maps for the propagator strips. In the first term we integrate over moduli $(m^1_{12345},m^2_{12345})$ of the covariant quintic vertex. In the second group of terms, we integrate over the Schwinger parameter of the propagator together with the modulus of the covariant quartic vertex. Finally, in the third group of terms we integrate over the two Schwinger parameters of the two propagators.  In all terms the measure is expressed in the form directly implied by \eq{Siegel5op}. The insertions $\mathscr{B}_{1235},...,\mathscr{B}_{1345}$ in the second group of terms represent the $b$-ghost of the covariant quartic vertex, as defined for example through the Schiffer variation, transformed to the upper half plane coordinate of the 5-punctured disk. All objects above are indexed according to the punctures, propagators, and vertices of the Feynman graph in that portion of moduli space as defined by the default labeling convention described at the beginning of this section.

Next consider the {\it transverse} Siegel gauge amplitude. We restrict to external states of the form $\phi_i=Sa_i$, where $a_i\in\Hperp$ are transverse and carry ghost number 1. By the equivalence theorem, the amplitude will be expressed in terms of Mandelstam diagrams. These Mandelstam diagrams define their own Feynman graph decomposition of the amplitude, one that is {\it a priori} unrelated to the graphs generated by Siegel gauge Feynman rules.  Mandelstam diagrams do not produce every type of Feynman graph, first of all because there is no quintic lightcone vertex, but also because the quartic lightcone vertex only contributes in special configurations where minus momenta alternate in sign. Let us assume a situation which produces the maximum number of Mandelstam diagrams of different kinds, which results from choosing length parameters with signs
\begin{equation}(\alpha_1,\alpha_2,\alpha_3,\alpha_4,\alpha_5) = (+,-,+,+,-),\end{equation}
and additionally
\begin{equation}\alpha_2+\alpha_3<0,\ \ \ \ \alpha_4+\alpha_5<0.\end{equation} 
In this scenario there are three Mandelstam diagrams with a quartic vertex, corresponding to graphs $12(34)5$, $1(23)45$ and $123(45)$.  Including the five graphs with two propagator strips, Mandelstam diagrams can therefore produce at most eight 5-point Feynman graphs. These decompose the moduli space into eight regions:
\begin{equation}
	M_5 = \Big[R_{12(34)5}^\mathrm{lc}\cup R_{1(23)45}^\mathrm{lc}\cup R_{123(45)}^\mathrm{lc}\Big]\cup  \Big[R_{12((34)5)}^\mathrm{lc}\cup...\cup R_{12(3(45))}^\mathrm{lc}\Big],
\end{equation}
\begin{figure}[t]
	\begin{center}
		\resizebox{3.8in}{3in}{\includegraphics{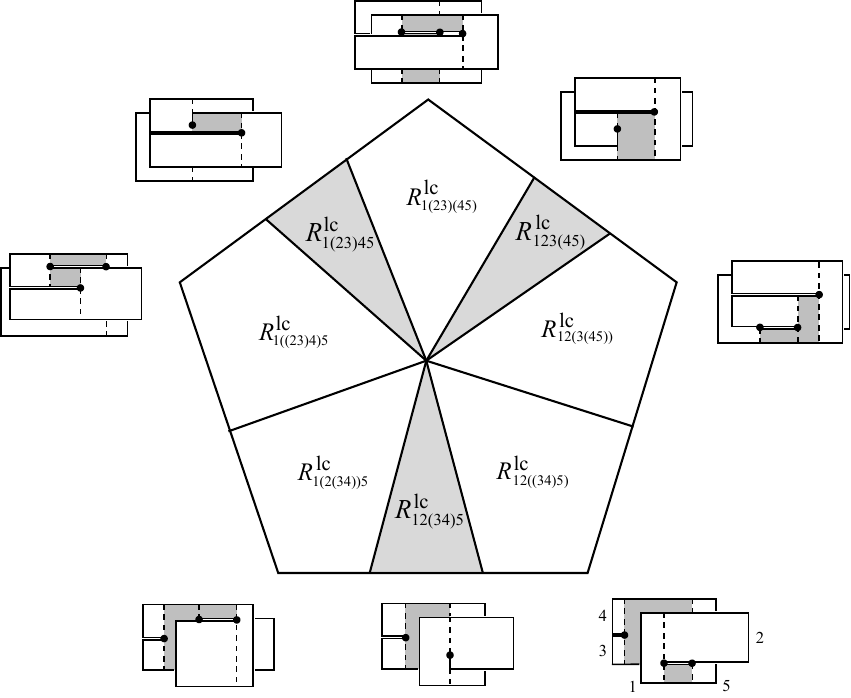}}
	\end{center}
	\caption{\label{fig:lightcone_gauge28} Decomposition of the moduli space $M_5$ into regions characterized by Feynman graphs formed by Mandelstam diagrams.}
\end{figure}

\noindent as shown in figure \ref{fig:lightcone_gauge28}. The transverse Siegel gauge amplitude can be expressed as a sum of contributions from these eight regions:
\begin{eqnarray}
	\lineup A_5(\M_5,Sa_1,Sa_2,Sa_3,Sa_4,Sa_5) \label{eq:transA5}\\
	\lineup	= \Bigg[\int_{R^\mathrm{lc}_{12(34)5}} \! \frac{dT_{34}}{\alpha_{34}} d\theta_{1235} \left\langle\!  \left[ \mathrm{Im}\left(\frac{b(U_{1235})}{\d^2\rho(U_{1235})}\right)\right]\big(\flc_{34}\circ b_0 \big)\big(\flc_1\! \circ\! e^{-\lambda_1}\!\circ\! a_1(0)\big) ... \big(\flc_5\! \circ\! e^{-\lambda_5}\! \circ\! a_5(0)\big)\! \right\rangle_{\UHP}\nonumber\\ 
	\lineup\ \ \  +\int_{R^\mathrm{lc}_{1(23)45}} \! \frac{dT_{23}}{\alpha_{23}} d\theta_{1245}\left\langle \!  \left[\mathrm{Im}\left(\frac{b(U_{1245})}{\d^2\rho(U_{1245})}\right)\right]\big(\flc_{23}\circ b_0 \big)\big(\flc_1\! \circ\! e^{-\lambda_1}\!\circ\! a_1(0)\big) ... \big(\flc_5\! \circ\! e^{-\lambda_5}\! \circ\! a_5(0)\big)\! \right\rangle_{\UHP}\nonumber\\ 
	\lineup\ \ \ +\int_{R^\mathrm{lc}_{123(45)}} \! \frac{dT_{45}}{\alpha_{45}} d\theta_{1234} \left\langle\!  \left[ \mathrm{Im}\left(\frac{b(U_{1234})}{\d^2\rho(U_{1234})}\right)\right]\big(\flc_{45}\circ b_0 \big)\big(\flc_1\! \circ\! e^{-\lambda_1}\!\circ\! a_1(0)\big) ... \big(\flc_5\! \circ\! e^{-\lambda_5}\! \circ\! a_5(0)\big)\! \right\rangle_{\UHP}\Bigg]\nonumber\\
	\lineup\ \ \ +\Bigg[\int_{R_{12((34)5)}^\mathrm{lc}} \frac{dT_{34}dT_{345}}{\alpha_{34}\alpha_{345}}\Big\langle\big(\flc_{345}\circ b_0\big)\big(\flc_{34}\circ b_0\big)\big(\flc_1\! \circ\! e^{-\lambda_1}\!\circ\! a_1(0)\big) ... \big(\flc_5\! \circ\! e^{-\lambda_5}\! \circ\! a_5(0)\big)\Big\rangle_{\UHP}\nonumber\\
	\lineup\ \ \ \ \ \ \ +\, ...+ \int_{R_{12((34)5)}^\mathrm{lc}} \frac{dT_{45}dT_{345}}{\alpha_{45}\alpha_{345}}\Big\langle\big(\flc_{345}\circ b_0\big)\big(\flc_{45}\circ b_0\big)\big(\flc_1\! \circ\! e^{-\lambda_1}\!\circ\! a_1(0)\big) ... \big(\flc_5\! \circ\! e^{-\lambda_5}\! \circ\! a_5(0)\big)\Big\rangle_{\UHP}\Bigg].\nonumber
\end{eqnarray}
The stub lengths $\lambda_1,...,\lambda_5$ are determined by transverse projection of the Siegel gauge diagrams at the corresponding point in moduli space. We integrate over the moduli of the Mandelstam diagram in the given region and use the Kugo-Zwiebach form of the lightcone measure.

\begin{figure}[t]
	\begin{center}
		\resizebox{2.5in}{2.3in}{\includegraphics{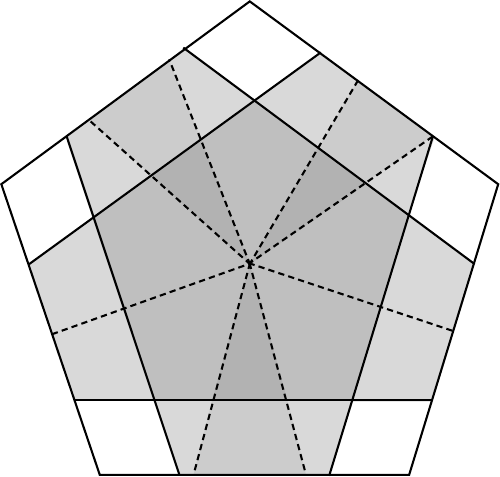}}
	\end{center}
	\caption{\label{fig:lightcone_gauge26} 26 regions distinguished by the Feynman graphs which appear from the Siegel gauge diagram and from the Mandelstam diagram in that part of moduli space.  This amounts to the possible intersections of the regions shown in figures \ref{fig:lightcone_gauge25} and \ref{fig:lightcone_gauge28}. From the point of view of the quintic vertex integration cycle, crossing a dotted line on this figure shifts to a different kind of Mandelstam diagram. Crossing a solid line does not change the kind of Mandelstam diagram, but changes how the moduli of the Mandelstam diagram depend on the moduli of $M_5$.  In particular, a new propagator strip on the Mandelstam diagram will have its width shortened  proportionally to the Siegel gauge Schwinger parameter in the respective channel. }
\end{figure}

Now we discuss to the quintic vertex in lightcone gauge. Like the Siegel gauge amplitude, it is a sum of eleven terms 
\begin{eqnarray}
\lineup \!\!\!  \langle a_1,\vlc_4(a_2,a_3,a_4,a_5)\rangle =\underbrace{ \Big\langle S a_1,v_4(S a_2,Sa_3,Sa_4,Sa_5)\Big\rangle}_{12345}\\
	\lineup \!\!\! -\underbrace{\Big\langle Sa_1,v_3\big(Sa_2,\Deltalong v_2(Sa_3,Sa_4),Sa_5\big)\Big\rangle}_{12(34)5}
	-...-\underbrace{\Big\langle Sa_1,v_2\big(Sa_2,\Deltalong v_3(Sa_3,Sa_4,Sa_5)\big)\Big\rangle}_{12(345)} \nonumber\\
	\lineup \!\!\!  +\underbrace{\Big\langle\! Sa_1,\! v_2\big(Sa_2,\! \Deltalong v_2(\Deltalong v_2(Sa_3,\! Sa_4),\! Sa_5\big)\!\Big\rangle}_{12((34)5)}
	\! +...+\! \underbrace{\Big\langle\! Sa_1,\! v_2\big(Sa_2,\! \Deltalong v_2(Sa_3,\! \Deltalong v_2(Sa_4,\! Sa_5)\big)\!\Big\rangle}_{12(3(45))},\nonumber
\end{eqnarray}
The vertex acts on five transverse states $a_1,a_2,a_3,a_4,a_5\in\Hperp$ at ghost number 1. The first term is the transverse subvertex, and the remaining ten are longitudinal subvertices. The claim is that all terms can be reduced to lightcone off-shell amplitude defined by a respective integration cycle in the lightcone fiber bundle. The union of these integration cycles,
\begin{equation}\Vlc_{12345} = \Slc_{12345}\cup\Big[\Slc_{12(34)5}\cup ... \cup \Slc_{12(345)}\Big]\cup\Big[\Slc_{12((34)5)}\cup...\cup\Slc_{12(3(45))}\Big],\end{equation}
defines the gauge-fixed quintic vertex:
\begin{equation}\big\langle a_1,\vlc_4(a_2,a_3,a_4,a_5)\big\rangle = \Alc_5(\Vlc_{12345},a_1,a_2,a_3,a_4,a_5).\end{equation}
The main issue now is the computation of the longitudinal subvertices. Each longitudinal subvertex may be expressed as a correlation function on a Siegel gauge diagram with an insertion of $\Deltalong$ on a collapsed propagator strip. We then write
\begin{equation}\Deltalong = (b_0-\bDDF)\int_0^\infty |dt_i| e^{-t_i(L_0-\LDDF)}\end{equation}
where $t_i$ can be interpreted as the Schwinger parameter of the Siegel gauge propagator in the appropriate channel. Assuming the appropriate graphical compatibility conditions, from here the computation proceeds pretty much exactly following \eq{long15}. The technical difference now is either: 1) The correlator will carry another insertion $(b_0-\bDDF) e^{t_j\LDDF}$ representing the other longitudinal propagator, together with an integral over the associated Schwinger parameter, or 2)~The correlator will carry a $b$-ghost insertion for the measure of the covariant quartic vertex (e.g. $\mathscr{B}_{1234}$), together with an additional integration over the quartic vertex modulus. Making needed deformations of the $b$-ghosts, which can be justified from uniqueness of the measure, the additional insertions and integration however just follow along for the ride. The result is that each longitudinal subvertex is the same as the transverse projection of the corresponding Feynman graph in the Siegel gauge amplitude except that the propagators on the resulting Mandelstam diagrams are shorted by an amount proportional to the Siegel gauge Schwinger parameter.

Bringing the transverse and longitudinal contributions together gives the full integration cycle:
\begin{equation}\Vlc_{12345} = (M_5,\varphilc),\end{equation}
This specifies a Mandelstam diagram (with stubs) for each Siegel gauge diagram in the 5-point amplitude. Since we assume that the set of Siegel gauge diagrams defines a section of $\P_5$, they can be parameterized by moduli space $M_5$. As shown in figure \ref{fig:lightcone_gauge26}, the integration cycle can be described by breaking the moduli space into 26 regions. The regions are distinguished by which Feynman graphs are formed by the Siegel gauge diagram in that part of moduli space and which Feynman graphs are formed by the Mandelstam diagram.  The two graphs could be the same or different, but graphical compatibility (see next subsection) requires that all channels of the Siegel gauge diagram must be retained after transverse projection to the Mandelstam diagram. This determines the structure of the overlapping regions in figure~\ref{fig:lightcone_gauge26}. From the point of view of the gauge-fixed quintic vertex, the 26 regions are distinguished, first of all, by what kind of Mandelstam diagram appears in that part of moduli space. Secondly, they are distinguished by which propagator strips on the Mandelstam diagram have their widths shortened by the Siegel gauge Schwinger parameter. The width is shortened if and only if the requisite Siegel gauge Schwinger parameter exists in that part of moduli space. Finally, the Mandelstam diagrams of the quintic vertex come with stubs whose lengths $\lambda_1,...,\lambda_5$ are determined by transverse projection of the Siegel gauge amplitude. One might observe that the 26 regions represent a finer decomposition of the gauge-fixed quintic vertex than into transverse and longitudinal subvertices. This is because transverse and longitudinal subvertices can produce more than one kind of Mandelstam diagram.

\subsection{Graphical compatibility conditions}
\label{subsec:embedding}

The vertex theorem assumes the following: 

\begin{description}
\item{\bf Graphical compatibility conditions}. The Siegel gauge $n$-point amplitude is said to be {\it graphically compatible} with its transverse projection if three conditions are satisfied:
\begin{description}
	\item{\it (1) Channel compatibility.} Transverse projection of a Siegel gauge diagram never removes a propagator channel. That is, if the covariant vertices and propagator strips of a Siegel gauge diagram form a Feynman graph $\F$, transverse projection produces a Mandelstam diagram whose strip domains form a Feynman graph $\Flc$ which satisfies
	\begin{equation}\propagator(\F)\subseteq\propagator(\Flc),\end{equation}
	assuming common channels on the two diagrams are labeled in the same way.
	\item{\it (2) Propagator compatibility.} Every propagator strip on a Siegel gauge diagram fits inside the respective propagator strip of the Mandelstam diagram obtained upon transverse projection. Explicitly, for every value of the local coordinate $\xi_i$ on the channel $i$ propagator strip in the Siegel gauge amplitude,
	\begin{equation}1\geq \xi_i\geq e^{-t_i}, \ \ \ \mathrm{Im}(\xi_i)\geq 0,\end{equation}
	there is a corresponding local coordinate $\xi_i^\mathrm{lc}$ on the channel $i$ propagator strip of the Mandelstam diagram,
	\begin{equation}1\geq \xi_i^\mathrm{lc}\geq e^{-T_i/\alpha_i}, \ \ \ \mathrm{Im}(\xi_i^\mathrm{lc})\geq 0,\end{equation}
	which satisfies 
	\begin{equation}f_i(\xi_i) = \flc_i(\xi_i^\mathrm{lc}),\ \ \ \ i\in\propagator(\F),\end{equation}
	where the positions of the punctures are equal on both sides of this equation.
	\item{\it (3) Length compatibility.} The Schwinger parameters of a Siegel gauge diagram cannot be larger than the propagator widths of the Mandelstam diagram obtained upon transverse projection, normalized by the string length:
	\begin{equation}T_i \geq \alpha_it_i,\ \ \ \ i\in \propagator(\F).\end{equation}
\end{description} 
\end{description}
The purpose of these conditions is to limit the range of things that can happen with the operator 
\begin{equation}e^{t_i\LDDF},\ \ \ \ i\in\propagator(\F)\end{equation}
which appears inside the Siegel gauge propagator strip when computing the longitudinal subvertex. 

\begin{figure}[t]
	\begin{center}
		\resizebox{6.75in}{1.7in}{\includegraphics{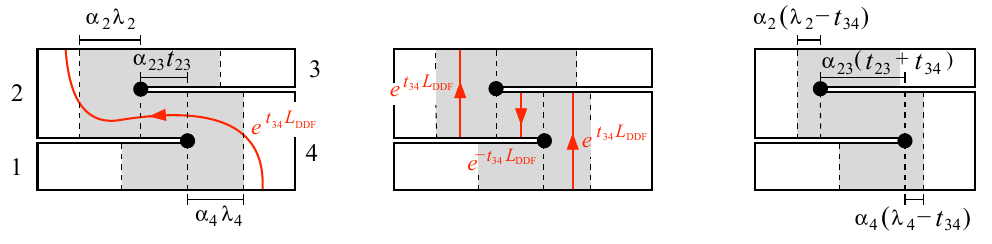}}
	\end{center}
	\caption{\label{fig:lightcone_gauge23} This figure shows the effect of canceling the contribution from transverse states in the $s$-channel when the Mandelstam diagram appears in the $t$-channel.}
\end{figure}

To start, let us suppose that channel compatibility fails. This means that the $i$th channel propagator which exists on the Siegel gauge diagram may not exist on the corresponding Mandelstam diagram. This implies that the contour of the DDF wave operator will cross several strip domains on the Mandelstam diagram.  To see an example, consider the situation in figure \ref{fig:lightcone_gauge23}.  Here the operator $e^{t_{34}\LDDF}$ appears in the $s$-channel of the Siegel gauge diagram while the Mandelstam diagram is in the $t$-channel.  In this scenario the contour can be deformed into a sum of contours through the strip domains of punctures 2 and 4,  subtracted against a contour in the $t$-channel propagator strip. The effect of $e^{t_{34}\LDDF}$ will then be to decrease the stub lengths on punctures $2$ and $4$ by $-t_{34}$ while increasing the Schwinger parameter in the $t$-channel by $+t_{34}$. An analogous contour deformation will always be possible in other examples. So if stub lengths are long enough to allow it, the longitudinal subvertex may be well defined even when channel compatibility fails.  However, the longitudinal subvertex will take a more complicated form. The Schwinger parameters of the Siegel gauge diagram may ``contaminate" the stubs lengths and propagator widths in other channels.

\begin{figure}[t]
	\begin{center}
		\resizebox{6.75in}{1.8in}{\includegraphics{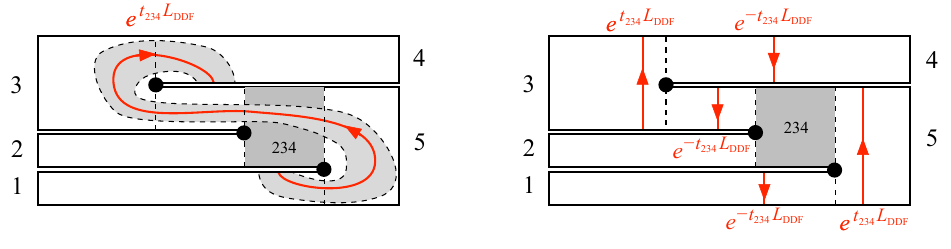}}
	\end{center}
	\caption{\label{fig:lightcone_gauge32} In this 5-point Mandelstam diagram, the dark grey region is the $234$ propagator strip, and the lighter grey region is hypothetically the image of the $234$ propagator strip of the Siegel gauge diagram. Since the former clearly does not contain the latter, propagator compatibility fails. The contour of the DDF wave operator is contained within the $234$ propagator strip of the Siegel gauge diagram, but when deformed to align with the strip domains on the Mandelstam diagram, it leaves the $234$ propagator strip of the Mandelstam diagram unaffected.}
\end{figure}

Next consider propagator compatibility. The purpose of this condition is, first of all, to ensure that the contour of the DDF wave operator represents the same channel in both the Siegel gauge diagram and on the Mandelstam diagram.  A hypothetical example where this fails, but channel compatibility is nevertheless satisfied, is shown in figure \ref{fig:lightcone_gauge32}. Here both the Siegel gauge diagram and the Mandelstam diagram have a propagator strip in the $234$-channel. However, the contour of the DDF wave operator which is in the $234$-channel on the Siegel gauge diagram is not in this channel on the Mandelstam diagram.  In fact, after deforming the contour to align with the strip domains of the Mandelstam diagram, the $234$ propagator strip is completely unaffected by the insertion of $e^{t_{234}\LDDF}$.  Again Schwinger parameters ``contaminate" stub lengths and propagator widths in other channels, making the vertex more complicated to describe. 

\begin{figure}[t]
	\begin{center}
		\resizebox{3.25in}{1.35in}{\includegraphics{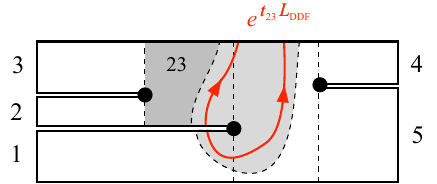}}
	\end{center}
	\caption{\label{fig:lightcone_gauge33} In this 5-point Mandelstam diagram, the dark grey region is the $23$ propagator strip and the lighter grey region is hypothetically the image of the $23$ propagator strip of the Siegel gauge diagram. Since the former is not contained within the later, propagator compatibility fails, and in a particular way where an interaction point of the Mandelstam diagram is contained within the Siegel gauge propagator strip. This means that two contours of  the DDF wave operator which are equivalent on the Siegel gauge diagram implement different deformations of the Mandelstam diagram. The longitudinal subvertex in this case cannot be well-defined.}
\end{figure}

Propagator compatibility however has a second and more important purpose. This is to prevent interaction points of the Mandelstam diagram from entering the Siegel gauge propagator strip. The nature of the problem is illustrated in figure \ref{fig:lightcone_gauge33}.  The cubic interaction point implies that there are two homotopic contours for the DDF wave operator on the Siegel gauge diagram which implement different deformations of the Mandelstam diagram. Therefore there must be an ambiguity in the sum over longitudinal intermediate states, and the longitudinal subvertex is not well-defined. The origin of the problem is the operator
\begin{equation}\frac{1}{\d X^+(z)}\end{equation}
which appears in the integrand of the DDF wave operator. At most points on the worldsheet this operator can be treated as a primary of weight $-1$, which makes the DDF wave operator conformally invariant.  However, there is a problem at points on the worldsheet where $\d X^+(z)=0$. By the replacement formula, this coincides exactly with the interaction points on the Mandelstam diagram.

Finally there is the condition of length compatibility. If it fails, the backwards shift of the Schwinger parameter can put the Mandelstam diagram into a different channel. As when channel compatibility fails, the longitudinal subvertex might still be well-defined but its form will be more complicated.

The graphical compatibility conditions are expected to hold if the conformal radii of  the covariant vertices are small enough. However in the examples of subsection \ref{subsec:example} the limiting factor on the conformal radii was {\it admissibility}---the requirement that stubs on the vertices should have zero or positive length.  This implied graphical compatibility by a wide margin. Therefore in general we do not know whether graphical compatibility really needs to be imposed separately.

\subsection{Covering moduli space}
\label{subsec:continuity}

The  final question is whether lightcone gauge amplitudes cover the moduli spaces of Riemann surfaces. This is implied if the Mandelstam diagrams at the boundary of the lightcone gauge vertex match those obtained by gluing lower order lightcone gauge vertices with a collapsed propagator strip. In other words, the surfaces in the gauge-fixed vertices satisfy the geometrical BV equation.

We consider the $(m+n)$-th order vertex in lightcone gauge with punctures labeled in cyclic order as $1,2,...,m+n$. We assume that Siegel gauge amplitudes are characterized by sections and that graphical compatibility conditions hold. In this case the vertex is defined by an integration cycle which specifies a Mandelstam diagram with stubs for every point in the moduli space $M_{m+n}$. The boundary of the vertex corresponds to the boundary of moduli space. We consider a component of the boundary where $m$ punctures are separated from the others by a very long strip of worldsheet. Using cyclicity, we can assume that punctures $1,2,...,m$ are separated. Our conventions would refer to this as the $(m+1...m+n)$-channel, but we refer to it as the $*$-channel for short. This component of the boundary is represented by Siegel gauge diagrams with a propagator strip in the $*$-channel whose Schwinger parameter $t_*$ is getting very large. It is further broken into several regions corresponding to different graphs for Siegel gauge diagrams and Mandelstam diagrams. We look at a region where the Siegel gauge diagram forms a Feynman graph $\F$ and its transverse projection is a Mandelstam diagram which forms a Feynman graph $\Flc$. Both of these graphs include the propagator strip of the degeneration: 
\begin{equation}
	*\in \propagator(\F)\subseteq\propagator(\Flc).
\end{equation}
The propagator strip separates the Feynman graph $\F$ into two smaller graphs $(\F^1,\F^2)$
\begin{center}
	\resizebox{3.3in}{1in}{\includegraphics{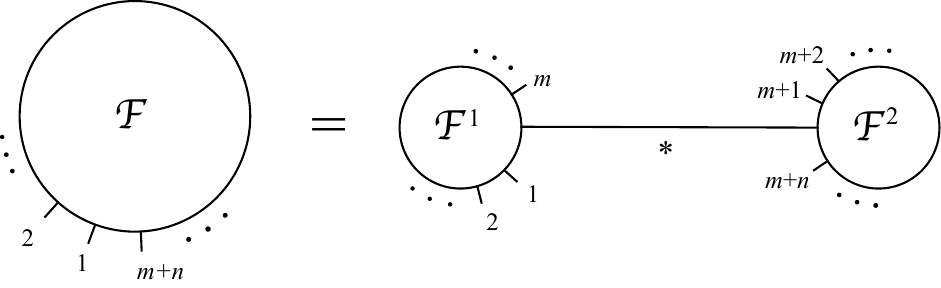}}
\end{center}
and likewise it separates $\Flc$ into $(\Flc^1,\Flc^2)$. The smaller graphs inherit labels on their punctures, vertices, and propagators from the parent graph.  The first graph has $(m+1)$ punctures listed in order as 
\begin{equation}1,\, 2,\, ...,\, m,\,  *.\label{eq:mpunct}\end{equation}
The last symbol refers to the propagator which connects to the second graph. The second graph has $(n+1)$ punctures listed in order as \begin{equation}*,\, m+1,\, m+2,\,  ...,\ m+n.\label{eq:npunct}\end{equation} 
In this case the first symbol refers to propagator which connects back to the first graph. The remaining punctures are labeled consecutively starting from $(m+1)$. Assuming also that vertices and propagators inherit their labels from the parent graph, we note that the labeling prescription for the smaller graphs does not correspond to the default prescription described at the beginning of this section. 

We want to compare the $*$-channel boundary of the vertex with the result of gluing lower order vertices in the $*$-channel. We consider a region of moduli space $M_{n+1}$ of the first lower order vertex where the Siegel gauge diagram forms a Feynman graph $\F^1$ and its transverse projection is a Mandelstam diagram which forms a Feynman graph $\Flc^1$.  We consider a region of moduli space $M_{m+1}$ of the second lower order vertex where the Siegel gauge diagram forms a Feynman graph $\F^2$ and its transverse projection is a Mandelstam diagram which forms a Feynman graph $\Flc^2$.  The vertex theorem implies that the moduli and stubs on the Mandelstam diagrams of the first and second lower order vertices are given by
\begin{eqnarray}
	T_i^{\mathrm{vertex},1} \lineup = T_i^1-\alpha_i t_i^1,\ \ \ \ i\in\propagator(\F^1),\ \ \ \ \ \ \ \ \ \ 
	T_i^{\mathrm{vertex},2} = T_i^2-\alpha_i t_i^2,\ \ \ \ i\in\propagator(\F^2),\nonumber\\
	T_i^{\mathrm{lvertex},1}\lineup = T_i^1,\ \ \ \ \ \ \ \ \ \ \ \ \ i\in\begin{smallmatrix}\propagator(\Flc^1) \\  -\propagator(\F^1)\end{smallmatrix},\ \ \ \ \ \ \ \ \ \ \ \  T_i^{\mathrm{vertex},2}= T_i^2,\ \ \ \ \ \ \ \ \ \ \ \ \ i\in\begin{smallmatrix}\propagator(\Flc^2)\\ -\propagator(\F^2)\end{smallmatrix},\nonumber\\
	\theta_I^{\mathrm{vertex},1}\lineup = \theta_I^1,\ \ \ \ \ \ \ \ \ \ \ \ \ I\in\quartic(\Flc^1),\ \ \ \ \ \ \ \ \ \ \ \ \ \ \ 
	\theta_I^{\mathrm{vertex},2} = \theta_I^2,\ \ \ \ \ \ \ \ \ \ \ \ \ \, I\in\quartic(\Flc^2),\nonumber\\
	\lambda_i^{\mathrm{vertex},1} \lineup = \lambda_i^1,\ \ \ \ \ \ \ \ \ \ \ \ \ i\in \puncture(\Flc^1), \ \ \ \ \ \ \ \ \ \ \ \ \, 
	\lambda_i^{\mathrm{vertex},2}  = \lambda_i^2,\ \ \ \ \ \ \ \ \ \ \ \ \ \,  i\in \puncture(\Flc^2).\nonumber \\
\end{eqnarray}
The index 1 or 2 indicates that these quantities are derived from the moduli space and length parameters of the first or second lower order vertex, respectively. Now we want to glue the Mandelstam diagrams of the first and second lower order vertex through the $*$-channel. The result is a Mandelstam diagram which forms a Feynman graph $\Flc$. This Mandelstam diagram has a propagator strip in the $*$-channel of width 
\begin{equation}\alpha_*(\lambda^1_*+\lambda^2_*), \end{equation}
which is proportional to the sum of the stub lengths of the respective punctures of the first and second lower order vertex. What we need to show is that this Mandelstam diagram is present on the boundary of the higher order lightcone gauge vertex. Explicitly,
\begin{eqnarray}
	\lim_{t_*\to\infty}T_i^\mathrm{vertex} \lineup = T_i^{\mathrm{vertex},1},\ \ \ \ \ \ \ \ \ \ \ \ \ \ \ i\in \propagator(\Flc^1),\label{eq:TvTv1}\\
	\lim_{t_*\to\infty}T_i^\mathrm{vertex} \lineup = T_i^{\mathrm{vertex},2},\ \ \ \ \ \ \ \ \ \ \ \ \ \ \ i\in \propagator(\Flc^2),\label{eq:TvTv2}\\
	\lim_{t_*\to\infty}T_i^\mathrm{vertex} \lineup = \alpha_*(\lambda^1_*+\lambda^2_*),\ \ \ \ \ \ \ \ \ \ i=*, \label{eq:Tvl1l2}\\
	\lim_{t_*\to\infty}\theta_I^\mathrm{vertex} \lineup = \theta_I^{\mathrm{vertex},1},\ \ \ \ \ \ \ \ \ \ \ \ \ \ \ I\in \quartic(\Flc^1),\label{eq:thvthv1}\\
	\lim_{t_*\to\infty}\theta_I^\mathrm{vertex} \lineup = \theta_I^{\mathrm{vertex},2},\ \ \ \ \ \ \ \ \ \ \ \ \ \ \ I\in \quartic(\Flc^2),\\
	\lim_{t_*\to\infty}\lambda_i^\mathrm{vertex} \lineup = \lambda_i^{\mathrm{vertex},1},\ \ \ \ \ \ \ \ \ \ \ \ \ \ \ i\in \puncture(\Flc^1)-\{*\},\\
	\lim_{t_*\to\infty}\lambda_i^\mathrm{vertex} \lineup = \lambda_i^{\mathrm{vertex},2},\ \ \ \ \ \ \ \ \ \ \ \ \ \ \ i\in \puncture(\Flc^2)-\{*\},\label{eq:lvlv2}
\end{eqnarray}
where $(T_i^\mathrm{vertex},\theta_I^\mathrm{vertex})$ and $\lambda_i^\mathrm{vertex}$ are the moduli and stubs in the higher order vertex, and the limit $t_*\to\infty$ takes us to the boundary of moduli space in the $*$-channel. This list of equations is a concrete statement of the geometrical BV equation in the form that is relevant for vertices in lightcone gauge.

We now pursue a generalization of the argument given in subsection \ref{subsec:covering4}. The first step is to understand how the local coordinate maps of the Siegel gauge diagram depend on the Schwinger parameter $t_*$ near degeneration. The surface state of the Siegel gauge diagram will be written 
\begin{equation}
	\langle \Sigma_{m+n}| \phi_1\otimes\phi_2\otimes...\otimes\phi_{m+n} = \big\langle \big(f_{1}\circ\phi_1(0)\big)\big(f_{2}\circ\phi_2(0)\big) ... \big(f_{m+n}\circ\phi_{m+n}(0)\big)\big\rangle_\UHP.\label{eq:Sigmamn}
\end{equation}
Near the $*$-channel degeneration the surface state must factorize as
\begin{equation}
	\langle\Sigma_{m+n}| = \langle \Sigma_{m+1}^1|\otimes \langle \Sigma_{n+1}^2|\bigg[\mathbb{I}^{\otimes \ell}\otimes \Big( - e^{-t_* L_0} \otimes \mathbb{I}|\mathrm{bpz}^{-1}\rangle \Big)\otimes \mathbb{I}^{\otimes n-\ell}\bigg],\label{eq:sigmafact}
\end{equation}
where
\begin{eqnarray}
		\langle \Sigma_{m+1}^1| \phi_1\otimes\phi_2\otimes...\otimes\phi_{m+1} \lineup = \big\langle \big(f_{1}^1\circ\phi_1(0)\big)... \big(f_{m}^1\circ\phi_{m}(0)\big)\big(f_*^1\circ\phi_{m+1}(0)\big)\big\rangle_\UHP,\label{eq:Sigma1}\\
		\langle \Sigma_{n+1}^2| \phi_1\otimes\phi_2\otimes...\otimes\phi_{n+1} \lineup = \big\langle \big(f_*^2\circ\phi_1(0)\big)\big(f_{m+1}^2\circ\phi_2(0)\big)...\big(f_{m+n}^2\circ\phi_{n+1}(0)\big)\big\rangle_\UHP\label{eq:Sigma2}
\end{eqnarray}
are the surface states of lower order Siegel gauge diagrams. The index 1 or 2 indicates that these objects are derived from the moduli space (and possibly length parameters) of the first or second lower order Siegel gauge diagram, respectively. The local coordinate maps are labeled by the punctures of the first and second graphs as defined in \eq{mpunct} and \eq{npunct}. The  surface states in \eq{sigmafact} are tied together with the double ket 
\begin{equation}
	-e^{-t_* L_0} \otimes \mathbb{I}|\mathrm{bpz}^{-1}\rangle
\end{equation}
which connects the two punctures labeled $*$ with a propagator strip whose  Schwinger parameter is $t_*$.  Contracting \eq{sigmafact} with states $\phi_1,...,\phi_{m+n}$, the objective is to arrive at an expression 
\begin{equation}
	\langle \Sigma_{m+n}| \phi_1\otimes...\otimes\phi_{m+n} = \langle \Sigma_{m+1}^1| \phi_1\otimes...\otimes\phi_m\otimes \big(e^{-t_*L_0}\phi_*\big)+\mathrm{subleading},
\label{eq:ntol}
\end{equation}
where $\phi_*$ is a vertex operator expressing the fusion of punctures $m+1,...,m+n$ close to degeneration. This vertex operator is defined by the leading term in the expansion of the other factor,
\begin{equation}
 \mathbb{I}\otimes \langle \Sigma_{n+1}^2|\Big( -e^{-t_* L_0} \otimes \mathbb{I}|\mathrm{bpz}^{-1}\rangle \Big) \otimes \phi_{m+1}\otimes... \otimes\phi_{m+n}
 = e^{-t_* L_0} \phi_* + \mathrm{subleading},
\end{equation}
in the limit where the Schwinger parameter is large. To compute the limit we contract with a test state $\psi$ and use \eq{Sigma2} to obtain
\begin{eqnarray}
	\lineup\langle\psi,e^{-t_*L_0} \phi_*\rangle+\mathrm{subleading}\label{eq:jj1}\\
	\lineup \ \ \ \ \ =\Big\langle \big(f_*^2\circ e^{-t_*}\circ\psi(0)\big)\big(f_{m+1}^2\circ \phi_{m+1}(0)\big)...\big(f_{m+n}^2\circ \phi_{m+n}(0)\big)\Big\rangle_\UHP.\nonumber
\end{eqnarray}
For large Schwinger parameter the first local coordinate map may be approximated by a linear function
\begin{equation}
	f_*^2\circ e^{-t_*}(\xi) = L_*^2\circ e^{-t_*}(\xi)+\mathrm{subleading},
\end{equation}
where 
\begin{equation} 
	L_*^2(\xi) = u_*^2+r_*^2\xi.
\end{equation}
The constant term $u_*^2$ is the location of the $*$th puncture of the second vertex in the upper half plane, and $r_*^2$ is the conformal radius of the local coordinate map around that puncture. Transforming the correlator \eq{jj1} with
\begin{equation}e^{-t_*}\circ I\circ (L_*^2)^{-1}, \end{equation}
we obtain
\begin{equation} \langle\psi,e^{-t_* L_0} \phi_*\rangle = \big\langle\big( I\circ\psi(0)\big)\big(e^{-t_*}\circ \phi_*(0)\big)\big\rangle_\UHP,\end{equation}
where the desired (nonlocal) vertex operator is given as
\begin{equation}
	\phi_{\ell+1...n}(0) = I\circ (L_*^2)^{-1}\circ \Big[\big(f_{m+1}^2\circ \phi_{m+1}(0)\big)...\big(f_{m+n}^2\circ \phi_{m+n}(0)\big)\Big].\label{eq:phi2deg1}
\end{equation}
Now return to \eq{ntol} and write
\begin{equation}
	 \langle \Sigma_{m+1}^1| \phi_1\otimes...\otimes\phi_m\otimes \big(e^{-t_*L_0}\phi_*\big) = \big\langle \big(f_1^1\circ \phi_1(0)\big)...\big(f_m^1 \circ \phi_m(0)\big)\big(f_*^1 \circ e^{-t_*}\circ \phi_*(0)\big)\big\rangle_\UHP. \label{eq:ntol2}
\end{equation}
Near degeneration the last local coordinate map can again be approximated by a linear function,
\begin{equation}
	f_*\circ e^{-t_*}(\xi) = L_*^1\circ e^{-t_*}(\xi)+\mathrm{subleading},
\end{equation}
where 
\begin{equation} 
	L_*^1(\xi) = u_*^1+r_*^1\xi.
\end{equation}
The constant term $u_*^1$ is the location of the $*$th puncture of the first vertex on the upper half plane and $r_*^1$ is the conformal radius of the local coordinate map around that puncture. Substituting the vertex operator $\phi_*$ into \eq{ntol2} we find formulas for the local coordinate maps in terms of those of the lower point diagrams: 
\begin{eqnarray}
	f_1(\xi)\lineup = f_1^1 (\xi) +\mathrm{subleading}, \nonumber\\
	\lineup\,\vdots\nonumber\\
	f_m(\xi) \lineup =  f_m^1(\xi)+\mathrm{subleading},\nonumber\\
	f_{m+1}(\xi) \lineup = \mu \circ f_{m+1}^2 (\xi)+\mathrm{subleading}, \nonumber\\
	\lineup \,\vdots\nonumber\\
	f_{m+n}(\xi)\lineup = \mu \circ f_{m+n}^2(\xi)+\mathrm{subleading},\label{eq:fsdeg}
\end{eqnarray}
where $\mu$ is the M{\"o}bius transformation
\begin{eqnarray}
\mu(u)\lineup =L_*^1\circ e^{-t_*}\circ I\circ  (L_*^2)^{-1}(u)\nonumber\\
\lineup = u_*^1 +\frac{r_*^1 r_*^2 e^{-t_*}}{u-u_*^2}.\label{eq:mugen}
\end{eqnarray}
This shows the dependence on the Schwinger parameter $t_*$ near degeneration.

Transverse projection of the Siegel gauge diagram gives a Mandelstam diagram defined by the Mandelstam mapping 
\begin{equation}\rho(u) = \sum_{i=1}^m \alpha_i \ln(u-u_i) + \sum_{i=m+1}^{m+n}\alpha_i \ln(u-u_i).\end{equation}
The first sum accounts for the punctures of the graph $\Flc^1$ and the second sum the punctures of~$\Flc^2$. According to \eq{fsdeg} the puncture positions near degeneration are given by 
\begin{eqnarray}
u_i \lineup = u_i^1+\mathrm{subleading},\ \ \ \ \ \ \ \ i\in \puncture(\F_1)-\{*\},\label{eq:uF1}\\
u_i \lineup = \mu(u_i^2)+\mathrm{subleading},\ \ \ \ i\in \puncture(\F_2)-\{*\}.\label{eq:uF2}
\end{eqnarray}
Now we distinguish two cases. The first is where $u$ is not close to $u^1_*$ in the degeneration limit. Then the punctures of the second graph can be approximated by  $u_i\approx u_*^1$, and the Mandelstam mapping simplifies to 
\begin{eqnarray}
	\rho(u)\lineup  = \sum_{i=1}^m \alpha_i \ln(u-u_i^1) + \left(\sum_{i=m+1}^{m+n}\alpha_i\right) \ln(u-u_*^1)+\mathrm{subleading}\nonumber\\
	\lineup =\rho^1(u). \label{eq:rhotorho1}
\end{eqnarray}
Now $\rho^1(u)$ defines the Mandelstam diagram produced by transverse projection of the first lower order Siegel gauge diagram. The second case is where $u$ is close to $u_*^1$. Then the sum over the punctures of the first graph is approximately constant and the Mandelstam mapping simplifies to
\begin{eqnarray}
	\rho(u) \lineup = \sum_{i=1}^m \alpha_i \ln(u_*^1-u_i^1) +\sum_{i=m+1}^{m+n}\alpha_i \ln(u-\mu(u_i^2))+\mathrm{subleading}\nonumber\\
	\lineup = \mathrm{constant} +\rho^2\big(\mu(u)\big).\label{eq:rhotorho2}
\end{eqnarray}
Here $\rho^2(u)$ defines the Mandelstam diagram produced by transverse projection of the second lower order Siegel gauge diagram. From this we learn something about how the preimages of the interaction points $U_I$ of the Mandelstam diagram behave near degeneration. Those $U_I$ which belong to the first graph $\Flc^1$ will not be close to $u_*^1$ in the degeneration limit. It follows from \eq{rhotorho1} that these $U_I$ will therefore approach the preimages of the interaction points $U_I^1$ of the first lower order Mandelstam diagram:
\begin{equation}\lim_{t_*\to\infty}U_I = U_I^1,\ \ \ \ \ \ I\in \cubic(\Flc^1)\cup\quartic(\Flc^1).\end{equation}
Meanwhile, the preimages of the interaction points $U_I$ which belong to the second graph $\Flc^2$ are close to $u_*^1$ in the degeneration limit. Therefore it follows from \eq{rhotorho1} that these $U_I$, after the appropriate M{\"o}bius transformation, will approach the preimages of the interaction points $U_I^2$ of the second lower order vertex:
\begin{equation}
	\lim_{t_*\to\infty}\mu^{-1}(U_I) = U_I^2,\ \ \ \ \ \ I\in\cubic(\Flc^2)\cup\quartic(\Flc^2).
\end{equation}
We also learn something about the lightcone local coordinate maps near degeneration. For punctures which belong to the first graph $\Flc^1$ the maps approach those of the first lower order vertex:
\begin{equation}\lim_{t_*\to\infty}\flc_i = f_i^{\mathrm{lc},1},\ \ \ \ \ \ i\in\puncture(\Flc^1)-\{*\}.\end{equation}
For punctures which belong to the second graph $\Flc^2$ the maps, after M{\"o}bius transformation, approach those of the second lower order vertex:
\begin{equation}\lim_{t_*\to\infty}\mu^{-1}\circ\flc_i = f_i^{\mathrm{lc},2},\ \ \ \ \ \ i\in\puncture(\Flc^2)-\{*\}.\label{eq:muflc2}\end{equation}
The behavior of the $U_I$s and local coordinate maps near degeneration determines the behavior of the stubs and moduli near degeneration. In particular we learn that
\begin{eqnarray}
	\lim_{t_*\to\infty}T_i \lineup = T_i^1,\ \ \ \ \ \ \ \ \ \ \ \ \ \ \ i\in \propagator(\Flc^1),\\
	\lim_{t_*\to\infty}T_i \lineup = T_i^2,\ \ \ \ \ \ \ \ \ \ \ \ \ \ \ i\in \propagator(\Flc^2),\\
	\lim_{t_*\to\infty}\theta_I \lineup = \theta_I^1,\ \ \ \ \ \ \ \ \ \ \ \ \ \ \ I\in \quartic(\Flc^1),\\
	\lim_{t_*\to\infty}\theta_I \lineup = \theta_I^2,\ \ \ \ \ \ \ \ \ \ \ \ \ \ \ I\in \quartic(\Flc^2),\\
	\lim_{t_*\to\infty}\lambda_i\lineup = \lambda_i^1,\ \ \ \ \ \ \ \ \ \ \ \ \ \ \ i\in \puncture(\Flc^1)-\{*\},\\
	\lim_{t_*\to\infty}\lambda_i \lineup = \lambda_i^2,\ \ \ \ \ \ \ \ \ \ \ \ \ \ \ i\in \puncture(\Flc^2)-\{*\}.
\end{eqnarray}
These are the stubs and moduli of the Mandelstam diagram obtained by transverse projection of the Siegel gauge diagram on the boundary of moduli space. To obtain the stubs and moduli of the Mandelstam diagram on the boundary of the {\it vertex}, we must shift the propagator widths backwards proportionally to the Schwinger parameters of the Siegel gauge diagram. If $t_i$ are the Schwinger parameters of the higher order Siegel gauge diagram and $t_i^1$ and $t_i^2$ are the Schwinger parameters of the first and second lower order Siegel gauge diagrams, we are free to make the identification
\begin{eqnarray}
	\lim_{t_*\to\infty}t_i \lineup = t_i^1,\ \ \ \ i\in\propagator(\F^1),\nonumber\\
	\lim_{t_*\to\infty}t_i \lineup = t_i^2,\ \ \ \ i\in\propagator(\F^2).
\end{eqnarray}
This implies that the Mandelstam diagrams at the boundary of the lightcone gauge vertex satisfy equations \eq{TvTv1}-\eq{TvTv2} and \eq{thvthv1}-\eq{lvlv2}.

What is left is to show that the Mandelstam diagrams at the boundary of the lightcone gauge vertex satisfy \eq{Tvl1l2}. For this we note that the argument leading to \eq{fsdeg}, applied in the context of Mandelstam diagrams, shows that the lightcone local coordinate maps near degeneration satisfy
\begin{eqnarray}
	\flc_1(\xi)\lineup = f_1^{\mathrm{lc},1} (\xi) +\mathrm{subleading}, \nonumber\\
	\lineup\,\vdots\nonumber\\
	\flc_m(\xi) \lineup =  f_m^{\mathrm{lc},1}(\xi)+\mathrm{subleading},\nonumber\\
	\flc_{m+1}(\xi) \lineup = \mu^\mathrm{lc} \circ f_{m+1}^{\mathrm{lc},2} (\xi)+\mathrm{subleading}, \nonumber\\
	\lineup \,\vdots\nonumber\\
	\flc_{m+n}(\xi)\lineup = \mu^\mathrm{lc} \circ f_{m+n}^{\mathrm{lc},2}(\xi)+\mathrm{subleading},
\end{eqnarray}
where $\mu^\mathrm{lc}$ is the analogue of \eq{mugen} derived from the lightcone surface state
\begin{equation}
\mu^\mathrm{lc}(u) =  u_*^1 +\frac{r_*^{\mathrm{lc},1} r_*^{\mathrm{lc},2} e^{-T_*/\alpha_*}}{u-u_*^2}.
\end{equation}
In particular, $u_*^1,u_*^2$ and $r_*^{\mathrm{lc},1},r_*^{\mathrm{lc},2}$ are the punctures and conformal radii of the lower order lightcone local coordinate maps and $T_*/\alpha_*$ is the Schwinger parameter of the propagator strip in the $*$-channel. Comparing to \eq{muflc2} we must have
\begin{equation}\mu(u) = \mu^\mathrm{lc}(u),\end{equation}
which requires that the propagator width is related to the Siegel gauge Schwinger parameter as
\begin{equation}
	\frac{T_*}{\alpha_*} + \ln(r_*^{\mathrm{lc},1})+ \ln(r_*^{\mathrm{lc},2}) = t_* + \ln(r_*^1)+\ln(r_*^2),
\end{equation}
which implies \eq{Tvl1l2}. This completes the proof.

\section{Concluding remarks}

The goal of this work has been to understand the nature of interactions in lightcone gauge. We have found that strings in lightcone gauge always interact through Mandelstam diagrams, regardless of how the interactions of the original covariant string field theory are defined. The only information from the covariant interactions which survives is the set of dilatations at the punctures, which determine lengths of strips of worldsheet---stubs---attached to each external state in the Mandelstam diagram. An important part of what makes this geometrical transformation possible is that strings in lightcone gauge can interact through the exchange of unphysical ``longitudinally polarized" states. This exchange converts the decomposition of moduli space defined by the original covariant SFT into one which is consistent with the geometry of Mandelstam diagrams. 

The story generalizes in the expected way to classical closed bosonic string field theory. However the extension to the quantum theory meets new complications. Integration over loop momenta will force Mandelstam diagrams into unfavorable kinematic configurations where stub lengths cannot remain positive. To avoid this we can narrow focus to the Kugo-Zwiebach SFT or another theory with non-covariant vertices of the right kind. In a different direction, we can consider the extension to superstring field theory. The lightcone description of superstring interactions is well-known to be problematic due to singular collisions of operators inserted at interaction points on Mandelstam diagrams~\cite{Greensite1,Greensite2,GreenSeiberg}. The divergences are in principle canceled by divergent counterterms in the lightcone Hamiltonian, but the specifics are cumbersome to work out, especially beyond quartic order. However we have found that lightcone gauge in covariant SFT naturally produces a Lagrangian description of lightcone interactions where Mandelstam diagrams come with stubs. Stubs will prevent direct collision of operators on the Mandelstam diagram, raising the prospect that lightcone superstring interactions can be described by an nonpolynomial action with completely finite vertices. Recent work indicating this is possible appears in \cite{Kunitomo}. Moreover, the structure of the vertices should follow from gauge-fixing covariant superstring field theory, whose vertices may be constructed recursively following~\cite{OkawaSS,WittenSS}. So further development in this direction could give a finite and systematic description of lightcone superstring interactions. This could be useful for testing dualities in matrix string theory \cite{Motl,Banks,Dijkgraaf,DijkgraafMotl} and the plane wave limit of AdS/CFT~\cite{BMN,Spradlin}.

An important question we have not addressed is the {\it soft string problem}---the breakdown of lightcone gauge for strings with low lightcone momentum. The problem appears to be nontrivial. One can try to circumvent it by executing a field redefinition to the Kugo-Zwiebach SFT, where lightcone gauge is well-defined and produces the standard lightcone string field theory of Kaku and Kikkawa. A field redefinition of this kind from polyhedral or hyperbolic SFTs has recently been constructed following~\cite{HataZwiebach,Portugal}, generalizing the procedure of~\cite{lightcone,Kaku}. The difficulty is that the field redefinition is not defined acting on transverse string states with low lightcone momentum. This is a different manifestation of the same ``soft string" problem. As it stands there is no map between covariant string field theory and Kaku and Kikkawa's lightcone string field theory that works for all momenta. It would be desirable to understand why this is the case and come to a tractable resolution. At least part of the problem is the conformal invariance of the DDF construction. Perhaps a different characterization of the string spectrum, such as considered in~\cite{Skvortsov}, could point the way to a more robust gauge-fixed description of covariant string field theory.

\subsubsection*{Acknowledgments}

\vspace{.25cm}

\noindent I would like to thank T. Kugo for conversation concerning Siegel gauge and lightcone gauge amplitudes in the Kugo-Zwiebach string field theory. This helped clarify the origin of the lightcone measure and the phenomenon of longitudinal freezing. I would like to thank I. Pesando for conversations on recent work, and A. F{\i}rat and V. Bernardes for comments on the draft. This work was co-funded by the European Union and supported by the Czech Ministry of Education, Youth and Sports (Project No. FORTE–CZ.02.01.01/00/22\_008/0004632) and (Project CoGraDS-CZ.02.1.01/0.0/0.0/15\_003/0000437). The work was also partially supported by grant NSF PHY-2309135 to the Kavli Institute for Theoretical Physics (KITP).

\begin{appendix}

\section{Signs of the suspension}
\label{app:suspension}

Open string field theory can be formulated using one of two $\mathbb{Z}_2$ gradings. The {\it degree} grading (in the terminology of \cite{WittenSS}) leads to simpler expressions of homotopy algebraic relations. However, the {\it Grassmann} grading is more natural in conformal field theory. In this appendix we explain how to translate between these conventions. The procedure is well-known in mathematics. Some accounts can be found in~\cite{Lada,MoellerSachs}.

It is helpful to consider the degree and Grassmann gradings as defining separate (but isomorphic) vector spaces of string fields, $\Hdeg$ and $\Hgrass$. This allows us to consider a single $\mathbb{Z}_2$ grading defined on the direct sum of these vector spaces, 
\begin{equation}\Hdeg\oplus \Hgrass,\end{equation}
which we simply call {\it parity}. For states in $\Hgrass$, the parity is Grassmann parity, while for states in $\Hdeg$, the parity is degree. We use $|X|$ to denote the parity of an object $X$, which in general can be any linear map between tensor products of the direct sum $\Hdeg\oplus \Hgrass$. The vector spaces $\Hdeg$ and $\Hgrass$ are related by an isomorphism with odd parity:
\begin{equation}s:\Hdeg\to\Hgrass,\ \ \ \ s^{-1}:\Hgrass\to \Hdeg.\end{equation}
The map $s$ is called the {\it suspension}. It satisfies
\begin{equation}
ss^{-1} = \mathbb{I}_{\Hdeg},\ \ \ \ s^{-1}s= \mathbb{I}_{\Hgrass},\ \ \ \ |s|=|s^{-1}|=1\ \text{(mod }\mathbb{Z}_2),
\end{equation}
where $\mathbb{I}$ is the identity operator (we drop the subscript when the relevant vector space is clear). Since the suspension has odd parity, we have
\begin{eqnarray}
|s A| \lineup = |A| +1 \ \text{(mod }\mathbb{Z}_2), \ \ \ A\in\Hdeg,\\
|s^{-1} a| \lineup = |a|+1\ \text{(mod }\mathbb{Z}_2), \ \ \ a\in\Hgrass,
\end{eqnarray}
Thus isomorphic vectors in the two spaces are assigned opposite parity. It is worth mentioning that the approach we take here is opposite from \cite{WittenSS} and some other works. That work considers only {\it one} vector space of string fields, but defines {\it two} $\mathbb{Z}_2$ gradings on that vector space (Grassmann parity and degree). Presently, we consider {\it two} vector spaces string fields, but there is only {\it one} $\mathbb{Z}_2$ grading defined on those vector spaces (parity). 

Having introduced the suspension map we can ask about the nature of its construction. However, this question is misguided. It assumes that we have independent definitions of $\Hgrass$ and $\Hdeg$ and wish to understand their connection. In string field theory, however, we only really have one vector space---the vector space of the BCFT---and what is being debated is how to fix the overall parity of this vector space. Let us make a choice, and assume that the vector space of the $\BCFT$ is the Grassmann vector space $\Hgrass$. This choice ensures that states have the same parity as their vertex operators. How then should we understand the origin of the degree vector space $\Hdeg$? A consistent point of view is that the degree vector space is {\it defined} by applying the inverse suspension to $\Hgrass$. So, for example, the tachyon state in $\Hgrass$,
\begin{equation}c_1e^{ik\cdot X(0,0)}|0\rangle\in\Hgrass,\end{equation} 
would be expressed in $\Hdeg$ as
\begin{equation}s^{-1}c_1e^{ik\cdot X(0,0)}|0\rangle \in \Hdeg. \end{equation}
Here we do not wish to ``construct" the suspension map. It is a primitive ingredient defined axiomatically by the fact that it anticommutes and creates a new vector space. No other properties are needed in any meaningful computation.

We are concerned with linear maps between tensor products of the degree vector space,
\begin{equation}M_{mn}: \Hdeg^{\otimes n}\to\Hdeg^{\otimes m},\end{equation}
and linear maps between tensor products of the Grassmann vector space,
\begin{equation}V_{mn} : \Hgrass^{\otimes n}\to\Hgrass^{\otimes m}.\end{equation}
The first index indicates the number of output vectors and the second the number of input vectors. The problem of translating between degree and Grassmann gradings amounts to determining how the maps $M_{mn}$ and $V_{mn}$ should be related if they are regarded as equivalent. We will take equivalence to mean 
\begin{equation}
\underbrace{s\otimes ... \otimes s}_{m\text{ times}} M_{mn} = V_{mn}\underbrace{s\otimes ... \otimes s}_{n\text{ times}}.\label{eq:mv}
\end{equation}
It follows that the parity of the two maps is related according to
\begin{equation}|M_{mn}|+m = |V_{mn}|+n \ \ (\text{mod } \mathbb{Z}_2).\label{eq:MVparity}\end{equation}
If we know $M_{mn}$, we can determine the equivalent $V_{mn}$ by ``pulling" the suspension maps from left to right.  In doing this we must consistently treat the suspension map as an odd object, so that it passes through odd objects in parallel vector spaces with a sign. Therefore, for example
\begin{eqnarray}
s\otimes s \lineup = (s\otimes\mathbb{I})(\mathbb{I}\otimes s) \phantom{\Big)}\nonumber\\
\lineup = - (\mathbb{I}\otimes s)(s\otimes \mathbb{I}).\phantom{\Big)}
\end{eqnarray}
If $X_{mn}$ is a linear map with $n$ input vectors and $m$ output vectors,
\begin{eqnarray}
s\otimes X_{mn}\lineup = (s\otimes \mathbb{I}^{\otimes m})(\mathbb{I}\otimes X_{mn}) \phantom{\Big)}\nonumber\\
\lineup = (-1)^{|X_{mn}|} (\mathbb{I}\otimes X_{mn})(s\otimes \mathbb{I}^{\otimes n}),\phantom{\Big)}\\
X_{mn}\otimes s \lineup = (X_{mn}\otimes \mathbb{I})(\mathbb{I}^{\otimes n}\otimes s) \phantom{\Big)}\nonumber\\
\lineup  = (-1)^{|X_{mn}|}(\mathbb{I}^{\otimes m}\otimes s)(X_{mn}\otimes \mathbb{I}).\phantom{\Big)}
\end{eqnarray}
A possible source of confusion is that \eq{mv} does not have a sign from ``anticommutation" of $s$ through either $M_{mn}$ or $V_{mn}$. Consider for example the BRST operator. We denote it as  $Q$ when operating on either $\Hgrass$ or $\Hdeg$. The definition \eq{mv} implies that the BRST operators on the two vector spaces are related~by
\begin{equation}sQ = Qs.\end{equation}
There is no sign from anticommutation, even though $Q$ and $s$ are odd objects.

Let us explain how the translation works with a few examples. Start with $A_\infty$ relations. Let $Q,m_2,m_3,m_4$ be 1-, 2-, 3- and 4-products of an $A_\infty$ algebra on $\Hdeg$ and let $Q,v_2,v_3,v_4$ be the corresponding products on $\Hgrass$. By \eq{mv} the products are related as 
\begin{eqnarray}
sQ \lineup = Qs,\\
s m_2 \lineup = v_2 (s\otimes s),\\
sm_3\lineup = v_3 (s\otimes s\otimes s),\\
sm_4\lineup = v_4 (s\otimes s\otimes s\otimes s).
\end{eqnarray}
In the degree grading scheme the first four $A_\infty$ relations read
\begin{eqnarray}
0\lineup = Q^2,\\
0\lineup = Qm_2+m_2(Q\otimes \mathbb{I}+\mathbb{I}\otimes Q),\\
0\lineup = Q m_3 + m_3(Q\otimes \mathbb{I}\otimes\mathbb{I}+\mathbb{I}\otimes Q\otimes \mathbb{I}+\mathbb{I}\otimes\mathbb{I}\otimes Q),\nonumber\\
\lineup\ \ \ \ +m_2(m_2\otimes\mathbb{I}+\mathbb{I}\otimes m_2),\\
0\lineup=  Q m_4 + m_4(Q\otimes \mathbb{I}\otimes\mathbb{I}\otimes \mathbb{I}+\mathbb{I}\otimes Q\otimes \mathbb{I}\otimes \mathbb{I}+\mathbb{I}\otimes\mathbb{I}\otimes Q\otimes \mathbb{I}+\mathbb{I}\otimes \mathbb{I}\otimes\mathbb{I}\otimes Q)\nonumber\\
\lineup\ \ \ \ + m_2(m_3\otimes \mathbb{I}+\mathbb{I}\otimes m_3)+m_3(m_2\otimes\mathbb{I}\otimes\mathbb{I}+\mathbb{I}\otimes m_2\otimes\mathbb{I}+\mathbb{I}\otimes\mathbb{I}\otimes m_2).
\end{eqnarray}
To obtain the corresponding relations on $\Hdeg$ we apply the suspension map to these equations and pull through to the right following \eq{mv}. For the first $A_\infty$ relation this is a bit trivial, 
\begin{eqnarray}
0\lineup = sQ^2\nonumber\\
\lineup  = Q^2 s,
\end{eqnarray}
and says that $Q$ is nilpotent in either grading scheme. Applying \eq{mv} to the second $A_\infty$ relation,
\begin{eqnarray}
0\lineup =s\Big(Q m_2 + m_2(Q\otimes \mathbb{I}+\mathbb{I}\otimes Q)\Big)\phantom{\Big)}\nonumber\\
\lineup = Q sm_2 + v_2(s\otimes s)(Q\otimes \mathbb{I}+\mathbb{I}\otimes Q)\phantom{\Big)}\nonumber\\
\lineup = Qv_2 (s\otimes s) +v_2(-sQ\otimes s + s\otimes sQ)\phantom{\Big)}\nonumber\\
\lineup = Qv_2 (s\otimes s) +v_2(-Qs\otimes s + s\otimes Qs)\phantom{\Big)}\nonumber\\
\lineup = Qv_2 (s\otimes s) +v_2(-Q\otimes \mathbb{I} - \mathbb{I}\otimes Q)(s\otimes s)\phantom{\Big)}\nonumber\\
\lineup = \Big(Qv_2 -v_2(Q\otimes \mathbb{I} + \mathbb{I}\otimes Q)\Big)(s\otimes s).\phantom{\Big)}
\end{eqnarray}
Note that in the third and fifth lines we obtain minus signs from commuting $s$ through an odd object (in this case $Q$) in a parallel vector space in the tensor product. Therefore the second $A_\infty$ relation reads
\begin{equation}Qv_2  = v_2(Q\otimes \mathbb{I} + \mathbb{I}\otimes Q),\end{equation}
which is just the Leibniz rule for a (Grassmann even) 2-product. Applying \eq{mv} to the third $A_\infty$ relation,
\begin{eqnarray}
0\lineup = s\Big(Qm_3 + m_3(Q\otimes \mathbb{I}\otimes \mathbb{I}+\mathbb{I}\otimes Q\otimes \mathbb{I} + \mathbb{I}\otimes\mathbb{I}\otimes Q)+m_2(m_2\otimes \mathbb{I}+\mathbb{I}\otimes m_2)\Big)\phantom{\Big)}\nonumber\\
\lineup = Q s m_3 + v_3(s\otimes s\otimes s)(Q\otimes \mathbb{I}\otimes \mathbb{I} + \mathbb{I}\otimes Q\otimes \mathbb{I} +\mathbb{I}\otimes \mathbb{I}\otimes Q)+v_2(s\otimes s)(m_2\otimes \mathbb{I} +\mathbb{I}\otimes m_2)\phantom{\Big)}\nonumber\\
\lineup = Qv_3 (s\otimes s\otimes s)+ v_3(s Q\otimes s\otimes s - s\otimes sQ\otimes s +s\otimes s\otimes sQ)\phantom{\Big)}\nonumber\\
\lineup\ \ \ \ \ \ +v_2(- s m_2\otimes s + s\otimes s m_2 )\phantom{\Big)}\nonumber\\
\lineup = Qv_3 (s\otimes s\otimes s)+ v_3( Qs\otimes s\otimes s - s\otimes Qs\otimes s +s\otimes s\otimes Qs)\phantom{\Big)}\nonumber\\
\lineup\ \ \ \ \ \ +v_2(-  (v_2 (s\otimes s) )\otimes s + s\otimes ( v_2(s\otimes s) )\phantom{\Big)}\nonumber\\
\lineup = \Big(Qv_3 + v_3(Q\otimes \mathbb{I}\otimes \mathbb{I} + \mathbb{I}\otimes Q\otimes \mathbb{I} +\mathbb{I}\otimes \mathbb{I}\otimes Q)+v_2(-v_2\otimes \mathbb{I} +\mathbb{I}\otimes v_2)\Big)(s\otimes s\otimes s).\phantom{\Big)}
\end{eqnarray}
Therefore the third $A_\infty$ relation reads
\begin{equation}
v_2(v_2\otimes \mathbb{I} -\mathbb{I}\otimes v_2) = Qv_3 + v_3(Q\otimes \mathbb{I}\otimes \mathbb{I} + \mathbb{I}\otimes Q\otimes \mathbb{I} +\mathbb{I}\otimes \mathbb{I}\otimes Q).\label{eq:3rdAgrass}
\end{equation}
This is the familiar statement that the failure of $v_2$ to be associative is equal to the failure of the Leibniz rule for $Q$ acting on $v_3$. One can continue in this way to find the fourth $A_\infty$ relation, 
\begin{eqnarray}
\lineup v_2\big(v_3\otimes \mathbb{I} +\mathbb{I}\otimes v_3\big)-v_3\big(v_2\otimes \mathbb{I}\otimes \mathbb{I}-\mathbb{I}\otimes v_2\otimes \mathbb{I}+\mathbb{I}\otimes \mathbb{I}\otimes v_2\big)\nonumber\\
\lineup \ \ \ \ = Q v_4 - v_4\big(Q\otimes \mathbb{I}\otimes\mathbb{I}\otimes \mathbb{I}+\mathbb{I}\otimes Q\otimes \mathbb{I}\otimes \mathbb{I}+\mathbb{I}\otimes\mathbb{I}\otimes Q\otimes \mathbb{I}+\mathbb{I}\otimes \mathbb{I}\otimes\mathbb{I}\otimes Q\big).
\end{eqnarray}
Consider now the inner product. The BPZ inner product is a symmetric bilinear form defined on the Grassmann vector space,
\begin{equation}\langle \bpz|: \Hgrass^{\otimes 2} \to \Hgrass^{\otimes 0}, \end{equation}
while the symplectic form is an antisymmetric bilinear form defined on the degree vector space,
\begin{equation}\langle \omega|: \Hdeg^{\otimes 2} \to \Hdeg^{\otimes 0}.\end{equation}
They are related through \eq{mv} 
\begin{equation}\langle \omega| = \langle \bpz|s\otimes s.\label{eq:om_bpz}\end{equation}
To see that this identification is correct, let us prove that the symplectic form is graded antisymmetric. Operating on a pair of states $A,B\in\Hdeg$ gives
\begin{eqnarray}
\langle \omega| A\otimes B \lineup = \langle \bpz|(s\otimes s) A\otimes B\phantom{\Big)}\nonumber\\
\lineup = (-1)^{|A|}\langle \bpz|sA\otimes sB\phantom{\Big)}\nonumber\\
\lineup = (-1)^{|A|}\langle \bpz|a\otimes b\phantom{\Big)},
\end{eqnarray}
where we have written 
\begin{equation}a=sA,\ \ \ \ b=sB.\end{equation}
Next we use graded symmetry of the BPZ inner product to switch the order of $a$ and $b$: 
\begin{eqnarray}
\langle \omega| A\otimes B \lineup = (-1)^{|A|}(-1)^{|a||b|}\langle \bpz|b\otimes a\phantom{\Big)}\nonumber\\
\lineup = (-1)^{(|A|+1)(|B|+1)+|A|}\langle \bpz|s B\otimes sA\phantom{\Big)}\nonumber\\
\lineup = (-1)^{(|A|+1)(|B|+1)+|A|+|B|}\langle \bpz|(s\otimes s) B\otimes A\phantom{\Big)}\nonumber\\
\lineup = -(-1)^{|A||B|}\langle\omega|B\otimes A,
 \end{eqnarray}
establishing graded antisymmetry. Note that \eq{om_bpz} can be written as
\begin{equation}
\omega(A,B) = (-1)^{|A|}\langle a,b\rangle.
\end{equation}
An equivalent definition appears in \cite{WittenSS}, except in that context we have $A=a$ and $B=b$ because the degree and Grassmann vector spaces are equated. Also $|A|$ must be explicitly identified with the degree of the state $A$. Another thing which is worth explaining is cyclicity. An $n$-product $M_n$ defined on $\Hdeg$ is cyclic if 
\begin{equation}
0=\langle \omega|\big(M_n\otimes\mathbb{I}+\mathbb{I}\otimes M_n\big).\label{eq:Mcyc}
\end{equation}
We have a corresponding product $V_n$ on $\Hgrass$
\begin{equation}sM_n = V_n \underbrace{s\otimes ...\otimes s}_{n\text{ times}}.\end{equation}
The definition of cyclicity for $V_n$ is implied by \eq{Mcyc} and \eq{om_bpz}:
\begin{eqnarray}
0\lineup =\langle \bpz|(s\otimes s)\big(M_n\otimes\mathbb{I}+\mathbb{I}\otimes M_n\big)\phantom{\Big)}\nonumber\\
\lineup =\langle \bpz|\Big((-1)^{|M_n|}s M_n\otimes s+ s\otimes s M_n\Big)\phantom{\Big)}\nonumber\\
\lineup = \langle \bpz|\Big((-1)^{|M_n|} \big(V_n \underbrace{s\otimes ...\otimes s}_{n\text{ times}}\big)\otimes s+ s\otimes \big( V_n \underbrace{s\otimes ...\otimes s}_{n\text{ times}}\big)\Big)\phantom{\Big)}\nonumber\\
\lineup = \langle \bpz|\big((-1)^{|M_n|} V_n \otimes \mathbb{I}+ (-1)^{|V_n|} \mathbb{I}\otimes V_n \big)\underbrace{s\otimes ...\otimes s}_{n+1\text{ times}}\phantom{\Big)}\nonumber\\
\lineup = (-1)^{|M_n|}\langle \bpz|( V_n \otimes \mathbb{I}+ (-1)^{n+1} \mathbb{I}\otimes V_n)\underbrace{s\otimes ...\otimes s}_{n+1\text{ times}}.\phantom{\Big)}
\end{eqnarray}
In the last step we used \eq{MVparity} to relate the parity of $M_n$ and $V_n$. Therefore 
\begin{equation}0 = \langle \bpz|( V_n \otimes \mathbb{I}+ (-1)^{n+1} \mathbb{I}\otimes V_n)\end{equation}
expresses cyclicity in the Grassmann vector space.

The definition \eq{mv} is not the only rule we could adopt for mapping between degree and Grassmann grading schemes. An alternative is
\begin{equation}M_{mn}\underbrace{s^{-1}\otimes ...\otimes s^{-1}}_{n\text{ times}} = \underbrace{s^{-1}\otimes...\otimes s^{-1}}_{m\text{ times}}\widetilde{V}_{mn}.\end{equation}
The linear map $\widetilde{V}_{mn}$ is not the same as $V_{mn}$ in \eq{mv}, because tensor products of $s^{-1}$ do not invert tensor products of $s$. Instead
\begin{equation}(\underbrace{s^{-1}\otimes ...\otimes s^{-1}}_{m\text{ times}})(\underbrace{s\otimes...\otimes s}_{m\text{ times}})=(-1)^{\frac{m(m-1)}{2}}\mathbb{I}^{\otimes m}.\
\end{equation}
This implies that 
\begin{equation}
\widetilde{V}_{mn} = (-1)^{\frac{m(m-1)}{2}+\frac{n(n-1)}{2}}V_{mn}.
\end{equation}
The sign affects the form of the $A_\infty$ relations in the Grassmann grading scheme. For example, the third $A_\infty$ relation now reads 
\begin{equation}
Q\widetilde{v}_3 + \widetilde{v}_3(Q\otimes \mathbb{I}\otimes \mathbb{I} + \mathbb{I}\otimes Q\otimes \mathbb{I} +\mathbb{I}\otimes \mathbb{I}\otimes Q)+\widetilde{v}_2(\widetilde{v}_2\otimes \mathbb{I} -\mathbb{I}\otimes \widetilde{v}_2)=0.
\end{equation}
Compared to \eq{3rdAgrass}, there is an additional sign in front of the associator of 2-string products. The degree grading scheme in principle also suffers from a similar convention ambiguity, but we can fix the convention by requiring that $A_\infty$ relations are expressed without any signs. This resolution is not available in the Grassmann grading scheme. We adopt \eq{mv} since it adheres to the conventions of \cite{WittenSS}.

\section{Lightcone measures}
\label{app:measure}

In section \ref{sec:freeze} we derived the unreduced measure directly from Kaku and Kikkawa's lightcone SFT, and then by proving the freeze theorem inferred the validity of the Kugo-Zwiebach form of the covariantized measure. Here we show that the Kugo-Zwiebach form implies the other expressions for the lightcone measure given in subsection \ref{subsec:lightcone_off}. The chain of reasoning follows the diagram:

\setlength{\unitlength}{.25cm}
\noindent 
\begin{picture}(60,9)
\put(0,4){$\displaystyle{\begin{matrix} \text{unreduced} \\ \text{measure}\end{matrix}}$}
\put(8,4.5){\vector(1,0){11}}
\put(9.5,2){\footnotesize{${\begin{matrix} \text{longitudinal} \\ \text{freezing} \end{matrix}}$}}
\put(20,3.5){$\displaystyle{\begin{matrix} \text{Kugo-Zwiebach} \\ \text{covariantized} \\ \text{measure} \end{matrix}}$}
\put(32,4.5){\vector(1,0){10}}
\put(33,2.5){\footnotesize{appendix \ref{subapp:covtoKZ}}}
\put(43,3.5){$\displaystyle{\begin{matrix} \text{Schiffer} \\ \text{covariantized} \\ \text{measure}\end{matrix}}$}
\put(53.5,4.5){\vector(1,0){10}}
\put(54.5,2.5){\footnotesize{appendix \ref{subapp:covtored}}}
\put(64.5,4){$\displaystyle{\begin{matrix} \text{reduced} \\ \text{measure}\end{matrix}}$}
\end{picture}

\noindent In this way we have a complete derivation of the lightcone measure in all forms used in this paper.

\subsection{Covariantized measure to reduced measure}
\label{subapp:covtored}

We start by relating the covariantized and reduced measures. We consider the Schiffer form of the covariantized measure \eq{covariantized_Schiffer} acting on states of the form
\begin{equation}a_i = a_i^\perp(0) |-,\kpar^i\rangle \in \Hperp,\end{equation}
where $a_i^\perp(0)$ is a vertex operator of the transverse BCFT. We perform a M{\" o}bius transformation so that the punctures $u_1,u_{n-1},u_n$ are at fixed positions and the differentials $du_1,du_{n-1}$ and 
$du_n$ vanish. Then the $b$-ghost \eq{bsimp} removes the $c$ insertion accompanying the $2$nd to the $(n-2)$nd vertex operator.  The covariantized measure then takes the form
\begin{eqnarray}
\lineup \!\!\!\!\!\!\!\!(\sigmalc)^*\Omega_n(a_1,...,a_n) = \frac{du_{2}}{r_{2}}...\frac{du_{n-2}}{r_{n-2}}\Big\langle \Big(\flc_1\circ e^{-\lambda_1}\circ \big(c\, a_1^\perp e^{i\kpar^1\cdot X(0,0)}\big)\Big)\nonumber\\
\lineup\ \ \ \ \ \ \ \ \ \ \ \ \ \ \ \ \ \ \ \ \ \ \ \ \ \   
\times\Big(\flc_2\circ e^{-\lambda_2}\circ \big(a_2^\perp e^{i\kpar^2\cdot X(0,0)}\big)\Big)...\Big(\flc_{n-2}\circ e^{-\lambda_{n-2}}\circ \big(a_{n-2}^\perp e^{i\kpar^{n-2}\cdot X(0,0)}\big)\Big)\nonumber\\
\lineup \ \ \ \ \ \ \ \ \ \ \ \ \ \ \ \ \ \ \ \ \ \ \ \ \ \   
\times\Big(\flc_{n-1}\circ e^{-\lambda_{n-1}}\circ \big(c\, a_{n-1}^\perp e^{i\kpar^{n-1}\cdot X(0,0)}\big)\Big)\Big(\flc_n\circ e^{-\lambda_n}\circ \big(c\, a_n^\perp e^{i\kpar^n\cdot X(0,0)}\big)\Big) \Big\rangle_\mathrm{UHP}.\nonumber\\
\end{eqnarray}
We assume that the reduced measure is integrated over a local section of $\Plc_n$ with the orientation induced from $M_n$. Because the product of differentials $du_2,...,du_{n-2}$ appears in the order prescribed in \eq{Mnorient}, under this assumption we  can replace them with the corresponding integration density. Evaluating the correlator in the longitudinal BCFT then gives
\begin{eqnarray}
(\sigmalc)^*\Omega_n(a_1,...,a_n) \lineup = (2\pi)^2\delta^2(\kpar^1+...+\kpar^n)\nonumber\\
\lineup\ \ \ \times |du_2...du_{n-2}||u_1-u_{n-1}||u_1-u_n||u_{n-1}-u_n|\nonumber\\
\lineup\ \ \ \times \prod_{i\in\puncture} (r_i)^{(\kpar^i)^2-1}\prod_{i,j\in\puncture, i>j}|u_i-u_j|^{2\kpar^i\cdot\kpar^j}\nonumber\\
\lineup \ \ \ \times \Big\langle \big(\flc_1\circ e^{-\lambda_1}\circ a_1^\perp(0)\big)...\big(\flc_n\circ e^{-\lambda_n}\circ a_n^\perp(0)\big)\Big\rangle_\mathrm{UHP}^{\BCFT_\perp}.
\end{eqnarray}
After substituting 
\begin{equation}
r_i= e^{-\lambda_i}\d\flc_i(0) = e^{-\lambda_i+ \tau_{s(i)}/\alpha_i}\prod_{j\in\puncture,j\neq i}\frac{1}{ |u_i-u_j|^{\alpha_j/\alpha_i}},
\end{equation}
the factor on the third line changes to
\begin{eqnarray}
(\sigmalc)^*\Omega_n(a_1,...,a_n) \lineup = (2\pi)^2\delta^2(\kpar^1+...+\kpar^n)\nonumber\\
\lineup\ \ \ \times |du_2...du_{n-2}||u_1-u_{n-1}||u_1-u_n||u_{n-1}-u_n|\nonumber\\
\lineup\ \ \ \times \prod_{i\in\puncture} e^{-\lambda_i ((\kpar^i)^2-1)}\prod_{i\in\puncture} e^{k^i_+ \tau_{s(i)}}\nonumber\\
\lineup \ \ \ 
\times \frac{1}{\prod_{i=1}^n \rlc_i}\Big\langle \big(\flc_1\circ e^{-\lambda_1}\circ a_1^\perp(0)\big)...\big(\flc_n\circ e^{-\lambda_n}\circ a_n^\perp(0)\big)\Big\rangle_\mathrm{UHP}^{\BCFT_\perp}.\ \ \ \ \ 
\end{eqnarray}
The final product of exponentials on the third line originates from the transverse propagators, and is supposed to be reexpressed as
\begin{equation}
\prod_{i\in\puncture} e^{k_+^i\tau_{s(i)}}=\prod_{i\in\propagator}e^{-k_+^i T_i}.\label{eq:B13}
\end{equation}
To see why this holds, we substitute \eq{Ti}, 
\begin{equation}
-\sum_{i\in \propagator} k_+^i T_i = -\sum_{i\in\propagator} k_+^i \big(\tau_{s(i)} - \tau_{p(i)}\big),
\end{equation}
and reindex the sum so that it is carried out over interaction points labeled with $I$ rather than propagators labeled with $i$. With our established conventions, the momentum $k_+^i$ of the $i$th propagator strip flows {\it into} the successor interaction point $s(i)$, but flows {\it out of} the predecessor interaction point $p(i)$. Therefore we have 
\begin{equation}
-\sum_{i\in\puncture} k_+^i T_i= -\sum_{I \in \cubic\cup\quartic}\left(\sum_{i\in \genfrac\{\}{0pt}{2}{\text{propagator momenta}}{\text{flowing into }I}}k_+^i -\sum_{i\in \genfrac\{\}{0pt}{2}{\text{propagator momenta}}{\text{flowing out of }I}}k_+^i \right)\tau_I.
\end{equation}
By momentum conservation, we can replace the sum over propagator momenta with minus the sum over the momenta of external states flowing into $I$:
\begin{equation}
-\sum_{i\in\propagator} k_+^i T_i= \sum_{I \in \cubic\cup\quartic}\left(\sum_{i\in \genfrac\{\}{0pt}{2}{\text{external momenta}}{\text{flowing into }I}}k_+^i \right)\tau_I.
\end{equation}
Apparently, the only interaction points which contribute to this sum are those which touch the external strip domains. This allows us to reindex this as a sum over external states: 
\begin{equation}
-\sum_{i\in \propagator} k_+^i T_i= \sum_{i\in\puncture} k_+^i \tau_{s(i)}.
\end{equation}
Therefore \eq{B13} holds and
\begin{eqnarray}
(\sigmalc)^*\Omega_n(a_1,...,a_n) \lineup = (2\pi)^2\delta^2(\kpar^1+...+\kpar^n)\nonumber\\
\lineup\ \ \ \times |du_2...du_{n-2}||u_1-u_{n-1}||u_1-u_n||u_{n-1}-u_n|\nonumber\\
\lineup\ \ \ \times \prod_{i\in\puncture} e^{-\lambda_i ((\kpar^i)^2-1)}\prod_{i\in\propagator}e^{-k_+^i T_i}\nonumber\\
\lineup \ \ \ 
\times \frac{1}{\prod_{i=1}^n \rlc_i}\Big\langle \big(\flc_1\circ e^{-\lambda_1}\circ a_1^\perp(0)\big)...\big(\flc_n\circ e^{-\lambda_n}\circ a_n^\perp(0)\big)\Big\rangle_\mathrm{UHP}^{\BCFT_\perp}.\ \ \ \ \ \label{eq:B15}
\end{eqnarray}
Comparing to \eq{red_measure} we see much of the expected structure emerging. What remains is to derive the Jacobian relating the integration density on the Mandelstam diagram, 
\begin{equation}\left|\prod_{i\in\propagator}dT_i \prod_{I\in\quartic} d\theta_I\,\right|, \end{equation}
and the integration density on the upper half plane,
\begin{equation}|du_{2}du_3...du_{n-2}|.\end{equation}
The Jacobian factor is known to be 
\begin{eqnarray}
\left|\prod_{i\in \propagator} dT_i \prod_{I\in\quartic} d\theta_I\,\right| \lineup = |du_{2}du_3...du_{n-2}| |u_1-u_{n-1}||u_1-u_n||u_{n-1}-u_n|\nonumber\\
\lineup\ \ \ \times \frac{\prod_{i=1}^n\sqrt{|\alpha_i|}\prod_{I=1}^{n-2}\sqrt{|\d^2 \rho(U_I)|}}{\left|\sum_{i=1}^n \alpha_i u_i\right|^{2}}.\ \ \ \ \ \label{eq:Jacobian}
\end{eqnarray}
A derivation can be found for example in appendix 11.B of \cite{GSWII}. We can easily check that this substitution converts \eq{B15} into the reduced measure \eq{red_measure}, including the correct factor for the partition function on the Mandelstam diagram. Therefore the covariantized and reduced measures are equivalent. 

For completeness we give another derivation of the Jacobian factor~\eq{Jacobian}. The derivation is more straightforward but also lengthier than the argument of \cite{GSWII}. At the first stage, we need to express the integration density on the Mandelstam diagram in terms of the positions of the interaction points and their conjugates. The result can be written~as 
\begin{equation}
\left|\prod_{i\in \propagator} dT_i \prod_{I\in\quartic} d\theta_I\,\right| = \left|\prod_{I=1,\neq *}^{n-2} d\big(\rho(U_I)-\rho(U_*)\big)\right|,\label{eq:B20}
\end{equation}
where we choose an interaction point labeled with $*$ as a reference for measuring the positions of the others. To demonstrate this equality, let us assume that $\rho(U_*)$ is a cubic interaction point. The special case where all interaction points are quartic can be checked independently. The differential of a cubic interaction point is equal to the differential of its real part, 
\begin{equation} d\rho(U_I) = d\tau_I,\ \ \ \ I\in\cubic, \label{eq:drho1}\end{equation}
because the imaginary part (at least locally) does not vary with the moduli. For quartic interaction points, on the other hand, the imaginary part varies as 
\begin{equation}
d\rho(U_I) = d\tau_I + \frac{i}{2}d\theta_I,\ \ \ \ I\in\quartic,
\end{equation}
while for their conjugates
\begin{equation}
d\rho(U_I^*) = d\tau_I - \frac{i}{2}d\theta_I,\ \ \ \ I\in\quartic.\label{eq:drho3}
\end{equation}
We split the integration density into a product over cubic interaction points, quartic interaction points, and their conjugates:
\begin{eqnarray}
\left|\prod_{I=1,\neq *}^{n-2} d\big(\rho(U_I)-\rho(U_*)\big)\right|  \lineup = \left|\prod_{I\in\cubic,\neq *} d\big(\rho(U_I)-\rho(U_*)\big)\right|\nonumber\\
\lineup \ \ \ \times \left|\prod_{I\in\quartic} d\big(\rho(U_I)-\rho(U_*)\big)d\big(\rho(U_I^*)-\rho(U_*)\big)\right|.
\end{eqnarray}
Then substituting \eq{drho1}-\eq{drho3},
\begin{eqnarray}
\left|\prod_{I=1,\neq *}^{n-2} d\big(\rho(U_I)-\rho(U_*)\big)\right|  \lineup = \left|\prod_{I\in\cubic,\neq *} d\big(\tau_I-\tau_*\big)\right|\nonumber\\
\lineup\ \ \ \times \left|\prod_{I\in \quartic}\bigg(d(\tau_I-\tau_*) + \frac{i}{2}d\theta_I\bigg)\bigg(d(\tau_I-\tau_*) - \frac{i}{2}d\theta_I\bigg)\right|.\nonumber\\
\end{eqnarray}
Multiplying out the differentials and dropping the imaginary unit gives 
\begin{equation}
\left|\prod_{I=1,\neq *}^{n-2} d\big(\rho(U_I)-\rho(U_*)\big)\right| =
\left|\prod_{I\in\cubic\,\cup\, \quartic}d\big(\tau_I-\tau_*\big)\right|\left|\prod_{I\in \quartic}d\theta_I\right|. \label{eq:B26}
\end{equation}

\begin{figure}[t]
\begin{center}
\resizebox{5.3in}{4.9in}{\includegraphics{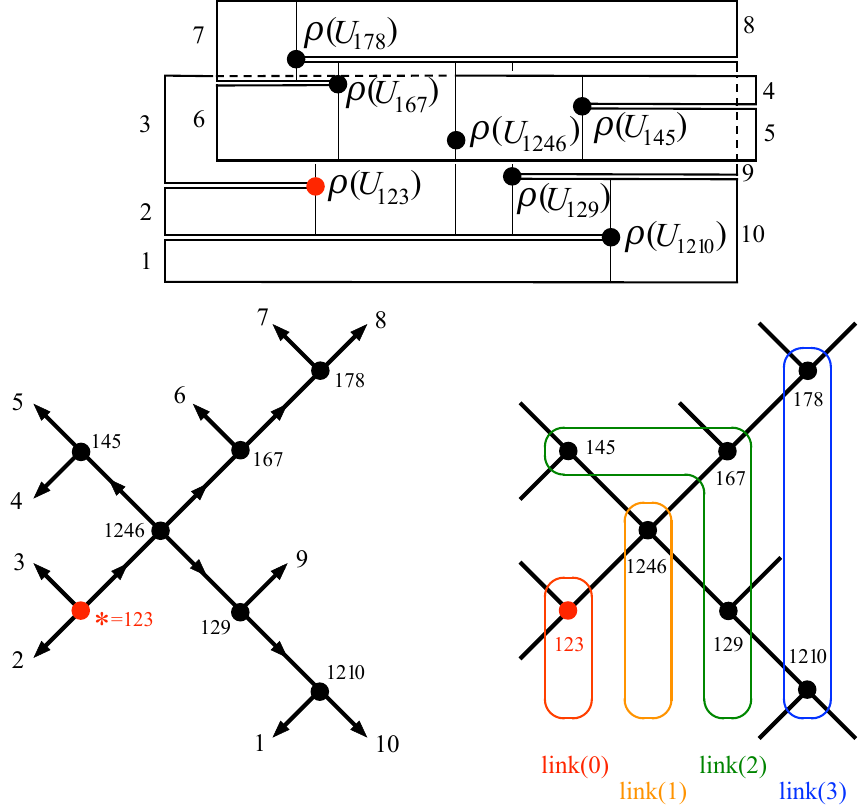}}
\end{center}
\caption{\label{fig:lightcone_gauge5} A Mandelstam diagram and its corresponding Feynman graph. On the bottom left, the red dot indicates the reference interaction point and the arrows indicate the associated partial ordering of links and vertices on the graph. Below right shows the vertices partitioned into subsets according to link number. The vertices are labeled according to the prescription described in figure~\ref{fig:lightcone_gauge31}.}
\end{figure} 

\noindent The first factor on the right must be expressed in terms of propagator widths. It is not difficult to see how this works by inspection. However, we will take some effort to explain it precisely because the setup will be useful later. First we need a bookkeeping device for the purposes of setting up a recursion. Consider the Feynman graph of the Mandelstam diagram under consideration. The vertices of the graph can be identified with the interaction points on the diagram, and the links can be identified with strip domains. A choice of reference interaction point $*$ defines a partial ordering on the set of vertices and links of this graph. We say that $a$ occurs {\it before} $b$ if the path from $*$ to $b$ inside the Feynman graph includes $a$. Equivalently, we can say that $b$ occurs {\it after} $a$. The partial ordering implies a notion of successor and predecessor which is different from the one discussed earlier.\footnote{The earlier definition of successor and predecessor is associated with a different partial ordering on the Feynman graph, but this partial ordering generally will have more than one maximal element. This makes it less convenient for setting up a recursion.} The interaction point which occurs immediately {\it after} a strip domain (as defined by this partial ordering) will be called the {\it successor}, while the one which occurs immediately {\it before} will be called the {\it predecessor}. The successor and predecessor to $\rho_i$ defined in this sense will be labeled respectively as $s_*(i),p_*(i)$, to distinguish from $s(i),p(i)$ used earlier. Since these definitions differ by at most an interchange of successor and predecessor, the propagator widths may be written
\begin{equation}T_i = \tau_{s(i)}-\tau_{p(i)} = (-1)^{\eps_*(i)}(\tau_{s_*(i)}-\tau_{p_*(i)}),\ \ \ \ i\in\propagator, \label{eq:Tist}\end{equation}
where $\eps_*(i)$ is the parity of the interchange. We need one more definition. Each interaction point $I$ on the graph can be assigned a nonnegative integer according to how many links in the graph are required to connect $*$ to~$I$. This will be called the {\it link number}. Link number defines an order-preserving map from the vertices of the Feynman graph into natural numbers---that is, if $I$ occurs before $J$, the link number of $I$ will be less than the link number of $J$. The subset of interaction points of link number $\ell$ will be written as $\llink{\ell}$. Now let us come back to the integration density \eq{B26}. We write the differentials of the interaction times as a telescoping sum 
\begin{equation}
d(\tau_I - \tau_*) = d(\tau_I - \tau_{I_{\ell-1}(I)})+ ... + d(\tau_{I_{2}(I)} - \tau_{I_1(I)})+d(\tau_{I_1(I)} - \tau_{*}),
\end{equation}
where
\begin{equation}
I_0(I)=*,\ \ I_1(I),\ \ ..., \ \  I_{\ell-1}(I),\ \ I_\ell(I) = I,
\end{equation}
label the sequence of interaction points appearing on the path connecting $*$ to $I$. Each term in the telescoping sum is given as the difference between interaction times bounding a single strip domain. The strip domain between $I_q(I)$ and $I_{q-1}(I)$ will be labeled $i_q(I)$. In this way we can write
\begin{equation}
\tau_{I_q(I)} - \tau_{I_{q-1}(I)} = \tau_{s_*(i_q(I))} - \tau_{p_*(i_q(I))} = (-1)^{\eps_*(i_q(I))}T_{i_q(I)},
\end{equation}
and the telescoping sum is expressed using propagator widths:
\begin{equation}
d(\tau_I - \tau_*)= (-1)^{\eps_*(i_\ell(I))}dT_{i_\ell(I)}+...+ (-1)^{\eps_*(i_2(I))}dT_{i_2(I)}+(-1)^{\eps_*(i_1(I))}dT_{i_1(I)}.
\label{eq:telT}\end{equation}
The product over $d(\tau_I-\tau_*)$ can be broken into components of definite link number 
\begin{equation}
\left|\prod_{I\in\cubic\cup \quartic,\neq *}d\big(\tau_I-\tau_*\big)\right| 
= \left|\prod_{I\in \llink{1}}d\big(\tau_I-\tau_*\big)\prod_{I\in \llink{2}} d\big(\tau_I-\tau_*\big)\, ...\,\prod_{I\in \llink{N}}d\big(\tau_I-\tau_*\big)\right|,
\label{eq:B27}
\end{equation}
where $N$ links are needed to cover the whole diagram. Substituting \eq{telT} into each component gives
\begin{eqnarray}
\left|\prod_{I\in\cubic\cup \quartic,\neq *}d\big(\tau_I-\tau_*\big)\right| 
\lineup = \left|\prod_{I\in \llink{1} }(-1)^{\eps_*(i_1(I))}dT_{i_1(I)} \right.\nonumber\\
\lineup \ \ \ \prod_{I\in \llink{2}}\Big[(-1)^{\eps_*(i_1(I))}dT_{i_1(I)}+(-1)^{\eps_*(i_2(I))}dT_{i_2(I)}\Big]\nonumber\\
\lineup\ \ \ \ \ \ \ \ \ \ \ \ \ \ \vdots\phantom{\bigg]}\nonumber\\
\lineup\ \ \ \left.\prod_{I\in \llink{N}}\left[\sum_{\ell=1}^N (-1)^{\eps_*(i_\ell(I))}dT_{i_\ell (I)}\right]\right|.
\end{eqnarray}
In the product over each subset, only the final differential along the path $dT_{i_\ell(I)}$ survives, since the other differentials multiply to zero against the differentials generated from subsets with fewer links. Since we consider the integration density we can drop the signs to obtain
\begin{equation}
\left|\prod_{I\in \cubic\cup\quartic,\neq 1}d\big(\tau_I-\tau_*\big)\right|
= \left|\prod_{I\in \llink{1}}dT_{i_1(I)}\prod_{I\in\llink{2}}dT_{i_2(I)}... \prod_{I\in\llink{N}}dT_{i_N(I)} \right|.
\end{equation}
The differential $dT_i$ is produced for each propagator exactly once. Together with \eq{B26}, this implies \eq{B20}.

Now we multiply out the differentials on the right hand side of \eq{B20} to obtain
\begin{equation}
\left|\prod_{i\in \propagator} dT_i \prod_{I\in\quartic} d\theta_I\,\right| = \left|
\sum_{I=1}^{n-2}(-1)^{I+1}\big(d \rho(U_{n-2})\big) ... \widehat{\big(d\rho(U_I)\big)}... \big(d\rho(U_1)\big)\right|,\label{eq:B39}
\end{equation}
where the hat indicates omission. The next stage in the proof is learning how to compute products of the differentials $d\rho(U_I)$. Starting with a single differential,
\begin{eqnarray}
d\rho(U_I) = \sum_{i=2}^{n-2}du_i \frac{\alpha_i}{u_i-U_I}. \label{eq:B41}
\end{eqnarray}
The product of two differentials is
\begin{eqnarray}
d\rho(U_I) d\rho(U_J)\lineup = \sum_{i,j=2}^{n-2}du_i du_j \frac{\alpha_i\alpha_j}{(u_i-U_I)(u_j-U_J)} \nonumber\\
\lineup = \sum_{n-2\geq i>j\geq 2}du_i du_j \left(\frac{\alpha_i\alpha_j}{(u_i-U_I)(u_j-U_J)} -\frac{\alpha_i\alpha_j}{(u_i-U_J)(u_j-U_I)} \right)\nonumber\\
\lineup = -\sum_{n-2\geq i>j\geq 2}du_i du_j \frac{\alpha_i\alpha_j(u_i-u_j)(U_I-U_J)}{(u_i-U_I)(u_i-U_J)(u_j-U_I)(u_j-U_J)}.
\label{eq:B42}
\end{eqnarray}
In the second step we reorganized the double sum so that each term is a linearly independent basis 2-form, and in the third step we brought everything over a common denominator. The generalization to a product of $N$ differentials is
\begin{eqnarray}
\lineup \!\!\!\!\!\!\!\!\!\!\!\! d\rho(U_{I_N})...d\rho(U_{I_2})d\rho(U_{I_1})\nonumber\\
\lineup\!\!\!\!\!\! = (-1)^{\frac{N(N-1)}{2}}\!\!\!\!\!\!\sum_{n-2\geq i_N>...>i_2>i_1\geq 2}\!\!\!\!\!\!
du_{i_N}...du_{i_2}du_{i_1}\frac{\alpha_{i_N}...\alpha_{i_2}\alpha_{i_1}\prod_{N\geq q>r\geq 1}(u_{i_q}-u_{i_r})\prod_{N\geq q>r\geq 1}(U_{I_q}-U_{I_r})}{\prod_{q=1}^N \prod_{r=1}^N (u_{i_q}-U_{I_r})}.\nonumber\\
\label{eq:B43}
\end{eqnarray}
This can be proven by induction. We write 
\begin{equation}\Big(d\rho(U_{I_{N+1}})...d\rho(U_{I_3})d\rho(U_{I_2})\Big)\Big(d\rho(U_{I_1})\Big),\end{equation}
and substitute \eq{B43} in for the first factor and \eq{B41} in for the second. We label the dummy indices $i_{N+1},...i_3,i_2$ in the first factor and $i_1$ in the second. Multiplying out, reorganizing the sum so that each term is linearly independent, and bringing everything over a common denominator produces a complicated polynomial in the numerator,
\begin{eqnarray}
\sum_{s=1}^{N+1}(-1)^{s+1}\prod_{\smallpile{N+1\geq q>r\geq 1}{ q,r\neq s}}(u_{i_q}-u_{i_r})\prod_{\smallpile{q=1}{q\neq s}}^{N+1}(u_{i_q}-U_{I_1})\prod_{r=2}^{N+1}(u_{i_s}-U_{I_r}),\label{eq:B45}
\end{eqnarray}
which is supposed to be equal to
\begin{equation}(-1)^N \prod_{N+1\geq q>r\geq 1}(u_{i_q}-u_{i_r})\prod_{q=2}^{N+1}(U_{I_q}-U_{I_1}).\label{eq:B46}\end{equation}
To prove this, first note that \eq{B45} is an $N$th degree polynomial in each $u_{i_q}$. Moreover, one can check that the $N$ roots correspond to $u_{i_q}=u_{i_r}$ for $r\neq q$. The fundamental theorem of algebra then implies that \eq{B45} must be of the form 
\begin{equation}\mathrm{constant} \times \prod_{N+1\geq q>r\geq 1}(u_{i_q}-u_{i_r}),\end{equation}
where the constant of proportionality is independent of each $u_{i_q}$. Setting $u_{i_q}=U_{I_q}$ in \eq{B45} directly determines the constant in agreement with \eq{B46}. This proves \eq{B43}, which upon substitution into \eq{B39} gives
\begin{eqnarray}
\lineup \!\!\!\!\!\!\!\!\!\!   \left|\prod_{i\in \propagator} dT_i \prod_{I\in\quartic} d\theta_I\,\right| \nonumber\\
\lineup \!\!\!\! \!\!\!\! = |du_{2}...du_{n-2}| |\alpha_{2}...\alpha_{n-2}|\left(\prod_{2\leq i<j \leq n-2}|u_i-u_j|\right)
\left|\sum_{I=1}^{n-2}(-1)^{I+1}\frac{\prod_{\smallpile{n-2\geq J>K\geq 1}{ J,K\neq I}}(U_{J}-U_{K})}{\prod_{i=2}^{n-2} \prod_{J=1,J\neq I}^{n-2} (u_i-U_J)}\right|.\nonumber\\
\end{eqnarray}
Again we need to bring terms over a common denominator:
\begin{eqnarray}
\lineup \!\!\!\!\!\!\!\!\!\! \left|\prod_{i\in \propagator} dT_i \prod_{I\in\quartic} d\theta_I\,\right| \nonumber\\
\lineup  \!\!\!\!\!\!\!\!  = |du_{2}...du_{n-2}||\alpha_{2}...\alpha_{n-2}| \frac{ \prod_{2\leq i<j \leq n-2}|u_i-u_j|}{\prod_{i=2}^{n-2} \prod_{I=1}^{n-2} |u_i-U_I|}
\left|\sum_{I=1}^{n-2}(-1)^{I+1}\!\!\!\!\prod_{\smallpile{n-2\geq J>K\geq 1}{J,K\neq I}}(U_{J}-U_{K}) \prod_{i=2}^{n-2}(u_i-U_I)\right|.\nonumber\\
\label{eq:B49}
\end{eqnarray}
The above sum is supposed to be equal to 
\begin{equation}(-1)^{n+1}\prod_{n-2\geq I>J\geq 1}(U_I-U_J).\end{equation}
To prove this, first note that the sum is an $(n-3)$rd degree polynomial in each $U_I$. Moreover, one can check that the $n-3$ roots correspond to $U_I=U_J$ for $I\neq J$. The fundamental theorem of algebra then implies that the sum must be of the form 
\begin{equation}\mathrm{constant} \times \prod_{n-2\geq I>J\geq 1}(U_I-U_J),\label{eq:B51}\end{equation}
where the constant of proportionality is independent of each $U_I$. The constant can be fixed by taking for example $U_1$ to be very large, and comparing the leading order contribution from the sum to the leading order contribution from \eq{B51}. Then  \eq{B49} simplifies to 
\begin{equation}
\left|\prod_{i\in\propagator} \! dT_i \prod_{I\in\quartic} \! d\theta_I\,\right|= |du_{2}...du_{n-2}||\alpha_2... \alpha_{n-2}|\frac{\prod_{2\leq i<j \leq n-2}|u_i-u_j|\prod_{1\leq I<J\leq n-2}|U_I-U_J|}{\prod_{i=2}^{n-2}\prod_{I=1}^{n-2} |u_i-U_I|}. \label{eq:B52}
\end{equation}
Here we finally have a relation between the integration density on the Mandelstam diagram and the integration density on the upper half plane. However it is not of the form \eq{Jacobian}.

The final stage of the calculation is to remedy this. This requires some well-known identities concerning the zeros $U_I$.  The derivative of the Mandelstam mapping~is
\begin{equation}\d\rho(u) = \sum_{i=1}^n \frac{\alpha_i}{u-u_i}.\end{equation}
Bringing all terms in the sum over a common denominator gives 
\begin{equation}
\d\rho(u)=\frac{1}{\prod_{i=1}^n(u-u_i)}\sum_{i=1}^n \left(\alpha_i \prod_{j=1,\neq i}^n(u-u_i)\right).
\end{equation}
The numerator is an $(n-2)$nd degree polynomial whose roots are the $U_I$s. Therefore it may be expressed in the form
\begin{equation}\mathrm{constant}\times\prod_{I=1}^{n-2}(u-U_I). \label{eq:poly}\end{equation}
To determine the constant, we consider the expansion of the numerator for large $u$:
\begin{equation}
\sum_{i=1}^n \left(\alpha_i \prod_{j=1,\neq i}^n(u-u_j)\right) = \left(\sum_{i=1}^n \alpha_i\right)u^{n-1} -  \left(\sum_{i=1}^n \left(\alpha_i\sum_{j=1,\neq i}^n u_j\right)\right)u^{n-2} + \text{lower orders}.
\end{equation}
The $(n-1)$st order term vanishes due to momentum conservation. The coefficient of the $(n-2)$nd order term can be simplified as
\begin{equation}
-\sum_{i=1}^n \left(\alpha_i\sum_{j=1,\neq i}^n u_j\right) = -\sum_{i=1}^n\alpha_i \sum_{j=1}^n u_j + \sum_{i=1}^n \alpha_i u_i.
\end{equation}
The first term vanishes by momentum conservation, and the second fixes the coefficient in \eq{poly}. Therefore the derivative of the Mandelstam mapping can be expressed as
\begin{equation}
\d\rho(u)=\left(\sum_{i=1}^n \alpha_i u_i\right)\frac{\prod_{I=1}^{n-2}(u-U_I)}{\prod_{i=1}^n(u-u_i)}.
\end{equation}
Expanding this around a root $u=U_I$ we learn that
\begin{equation}
\d^2\rho(U_I) = \left(\sum_{i=1}^n \alpha_i u_i\right)\frac{\prod_{J=1,\neq I}^{n-2}(U_I-U_J)}{\prod_{i=1}^n(U_I-u_i)}.\label{eq:d2rho}
\end{equation}
Meanwhile, we know that the residue of the pole at $u=u_i$ should be $\alpha_i$. This implies
\begin{equation}
\alpha_i =  \left(\sum_{i=1}^n \alpha_i u_i\right)\frac{\prod_{I=1}^{n-2}(u_i-U_I)}{\prod_{j=1,\neq i}^n(u_i-u_j)}.\label{eq:ai}
\end{equation}
These are the two main identities we will need. Returning to \eq{B52}, we wish to extend the product over $\alpha_i$s
\begin{equation}
|\alpha_2 \alpha_3\, ...\, \alpha_{n-2}| = |\alpha_1\alpha_2\, ...\, \alpha_n|\frac{1}{|\alpha_1\alpha_{n-1}\alpha_n|},
\end{equation}
and rewrite the second factor using \eq{ai}
\begin{equation}
|\alpha_2 ...\alpha_{n-2}| = |\alpha_1 ...\alpha_n|\left|\sum_{i=1}^n \alpha_i u_i\right|^{-3}\frac{\prod_{i=1,\neq 1}^n|u_1-u_i|}{\prod_{I=1}^{n-2}|u_1-U_I|}\frac{\prod_{i=1,\neq n-1}^n|u_i-u_{n-1}|}{\prod_{I=1}^{n-2}|u_{n-1}-U_I|}\frac{\prod_{i=1,\neq n}^n|u_i-u_n|}{\prod_{I=1}^{n-2}|u_n-U_I|}.
\end{equation}
The denominators multiply with the denominator of \eq{B52} as follows: 
\begin{equation}\left(\prod_{I=1}^{n-2}|u_1-U_I|\right)\left(\prod_{I=1}^{n-2}|u_{n-1}-U_I|\right)\left(\prod_{I=1}^{n-2}|u_n-U_I|\right)\left(\prod_{i=2}^{n-2}\prod_{I=1}^{n-2} |u_i-U_I|\right) = \prod_{i=1}^{n}\prod_{I=1}^{n-2} |u_i-U_I|. \end{equation}
Meanwhile the numerators multiply with the numerator of \eq{B52} as follows: 
\begin{eqnarray}
\lineup \left(\prod_{i=1,\neq 1}^n|u_1-u_i|\right)\left(\prod_{i=1,\neq n-1}^n|u_i-u_{n-1}|\right)\left(\prod_{i=1,\neq n}^n|u_i-u_n|\right) \left(\prod_{2\leq i<j \leq n-2}|u_i-u_j|\right)\nonumber\\
\lineup \ \ \ \ \ =\left(|u_1-u_{n-1}||u_1-u_n|\prod_{i=2}^{n-2}|u_1-u_i|\right)\left(|u_{n-1}-u_{n}|\prod_{i=1}^{n-2}|u_i-u_{n-1}|\right)\nonumber\\
\lineup\ \ \ \ \ \ \ \ \ \ \ \times \left(\prod_{i=1}^{n-1}|u_i-u_n|\right) \left(\prod_{2\leq i<j \leq n-2}|u_i-u_j|\right)\nonumber\\
\lineup \ \ \ \ \ = |u_1-u_{n-1}||u_1-u_n||u_{n-1}-u_{n}|\prod_{i=1}^{n-2}|u_i-u_{n-1}|\prod_{i=1}^{n-1}|u_i-u_n|\prod_{1\leq i<j \leq n-2}|u_i-u_j|\nonumber\\
\lineup\ \ \ \ \ = |u_1-u_{n-1}||u_1-u_n||u_{n-1}-u_n|\prod_{1\leq i<j\leq n}|u_i-u_j|.
\end{eqnarray}
With this \eq{B52} simplifies to 
\begin{eqnarray}
\left|\prod_{i\in\propagator} dT_i \prod_{I\in\quartic} d\theta_I\,\right|\lineup = |du_{2}...du_{n-2}||u_1-u_{n-1}||u_1-u_n||u_{n-1}-u_n||\alpha_1... \alpha_{n}|\nonumber\\
\lineup\ \ \ \ \ \ \ \ \times \left|\sum_{i=1}^n \alpha_i u_i\right|^{-3}\frac{\prod_{1\leq i<j \leq n}|u_i-u_j|\prod_{1\leq I<J\leq n-2}|U_I-U_J|}{\prod_{i=1}^{n}\prod_{I=1}^{n-2} |u_i-U_I|}.\ \ \ \ \ \ \ \label{eq:B61}
\end{eqnarray}
Next notice that \eq{d2rho} implies that
\begin{equation}
\prod_{I=1}^{n-2}\sqrt{|\d^2\rho(U_I)|} = \left|\sum_{i=1}^n \alpha_i u_i\right|^{\frac{n-2}{2}}\frac{\prod_{1\leq I<J\leq n-2}|U_I-U_J|}{\prod_{i=1}^n\prod_{I=1}^{n-2}\sqrt{|U_I-u_i|}},
\end{equation}
while \eq{ai} implies
\begin{equation}
\prod_{i=1}^n \sqrt{|\alpha_i|} = \left|\sum_{i=1}^n \alpha_i u_i\right|^{\frac{n}{2}}\frac{\prod_{i=1}^n\prod_{I=1}^{n-2}\sqrt{|U_I-u_i|}}{\prod_{1\leq i<j\leq n}|u_i-u_j|}.
\end{equation}
Taking the ratio,
\begin{equation}\frac{\prod_{I=1}^{n-2}\sqrt{|\d^2\rho(U_I)|}}{\prod_{i=1}^n \sqrt{|\alpha_i|}}= \left|\sum_{i=1}^n \alpha_i u_i\right|^{-1}\frac{\prod_{1\leq i<j\leq n}|u_i-u_j|\prod_{1\leq I<J\leq n-2}|U_I-U_J|}{\prod_{i=1}^n\prod_{I=1}^{n-2}|U_I-u_i|},\end{equation}
and comparing to \eq{B61}, we obtain \eq{Jacobian} as desired.

\subsection{Schiffer form to Kugo-Zwiebach form}
\label{subapp:covtoKZ}

Next we demonstrate the equivalence of the Schiffer and Kugo-Zwiebach forms of the covariantized lightcone measure, \eq{covariantized_Schiffer} and \eq{covariantized_KZ}. This equivalence is an instance of the more general claim that two $b$-ghost insertions which represent the same tangent on the covariant fiber bundle define the same covariant measure. This  is a statement of uniqueness of the covariant measure, and is almost certainly true, though we have found relatively little discussion of it in the literature. A sketched argument is that two energy-momentum tensor insertions which represent the same tangent are equal inside the measure because they both compute the derivative of the surface state along the tangent. The corresponding $b$-ghost insertions should then also be equal because the $b$-ghost satisfies the same conservation laws inside a surface state as the energy-momentum tensor. Presently however we would like to develop a more satisfying argument which shows by direct contour deformation that the Schiffer and Kugo-Zwiebach $b$-ghosts are the same.\footnote{The author learned from S. Konopka that contour deformation can establish uniqueness of the covariant measure in general by making use of sheaf cohomology and Serre duality. The details of this argument remain unpublished.}  We start with the Schiffer form of the $b$-ghost \eq{sigmalcBn} and use the explicit form of the lightcone local coordinate maps \eq{flcinv} to simplify the Schiffer vector field as
\begin{equation}
V_i^\mathrm{lc}(u) = \frac{d(\flc_i)^{-1}(u)}{\d(\flc_i)^{-1}(u)} = \frac{ d\rho(u)}{\d\rho(u)}-\frac{d\tau_{s(i)}}{\d \rho(u)}.
\end{equation}
The first term is independent of the puncture. This means that, for this contribution, the sum of contours around each puncture can be joined into a single contour which surrounds all punctures. Shrinking the contour to infinity picks off residues from the poles in the integrand. The poles appear where $\d \rho(u)$ vanishes, which precisely coincides with the interaction points $U_I$. In this way the Schiffer $b$-ghost is expressed 
\begin{eqnarray}
(\sigmalc)^* \mathscr{B}\lineup = - \sum_{i\in\puncture} d\tau_{s(i)} \oint_{u_i}\frac{du}{2\pi i} \frac{1}{\d\rho(u)}b(u) \nonumber\\
\lineup\ \ \ -\sum_{I=1}^{n-2}\frac{d\rho(U_I)}{\d^2\rho(U_I)}b(U_I)\nonumber\\
\lineup\ \ \ + \sum_{i\in\puncture} d\lambda_i \big(\flc_i\circ e^{-\lambda_i}\circ b_0\big).
\end{eqnarray}
The differentials $d\rho(U_I)$ can be rewritten in terms of interaction times $\tau_I$ and vertical displacements $\theta_I$ through \eq{drho1}-\eq{drho3}:
\begin{eqnarray}
(\sigmalc)^* \mathscr{B}\lineup = -\sum_{i\in\puncture } d\tau_{s(i)} \oint_{u_i}\frac{du}{2\pi i} \frac{1}{\d\rho(u)}b(u) \nonumber\\
\lineup\ \ \ -\sum_{I\in\cubic}d\tau_I \frac{b(U_I)}{\d^2\rho(U_I)} -\!\!\!\! \sum_{I\in \quartic}d\tau_I \left(\frac{b(U_I)}{\d^2\rho(U_I)}+\frac{b(U_I^*)}{\d^2\rho(U_I^*)}\right) + \!\!\!\!\sum_{I\in \quartic}d\theta_I \mathrm{Im}\left(\frac{b(U_I)}{\d^2\rho(U_I)}\right)\nonumber\\
\lineup \ \ \   + \sum_{i\in\puncture } d\lambda_i \big(\flc_i\circ e^{-\lambda_i}\circ b_0\big).
\end{eqnarray}
The last two terms already agree with the Kugo-Zwiebach form of the covariantized measure. To get the rest to work out, we must replace the differentials of the interaction times $\tau_I$ with the differentials of the propagator widths $T_i$. Noting \eq{telT} the first step is to introduce a reference interaction point and write the measure using differentials $d(\tau_I-\tau_*)$. This can be achieved by subtracting zero in the form 
\begin{equation}
d\tau_*\left[\sum_{i\in\puncture } \oint_{u_i}\frac{du}{2\pi i} \frac{1}{\d\rho(u)}b(u)+\sum_{I\in\cubic}\frac{b(U_I)}{\d^2\rho(U_I)} +\sum_{I\in \quartic}\left(\frac{b(U_I)}{\d^2\rho(U_I)}+\frac{b(U_I^*)}{\d^2\rho(U_I^*)}\right)\right] =0.
\end{equation}
Again, the first term is equivalent to a contour integral which surrounds all punctures.  Shrinking the contour and picking off the residues at the interaction points exactly cancels the second two terms. Therefore the $b$-ghost is expressed 
\begin{subequations}
\begin{align}
(\sigmalc)^* \mathscr{B}= & -\sum_{i\in\puncture } d(\tau_{s(i)}-\tau_*) \oint_{u_i}\frac{du}{2\pi i} \frac{1}{\d\rho(u)}b(u)\label{eq:Bnsimp1}\\
&-\sum_{I\in\cubic}d(\tau_I-\tau_*) \frac{b(U_I)}{\d^2\rho(U_I)} -\sum_{I\in \quartic}d(\tau_I-\tau_*) \left(\frac{b(U_I)}{\d^2\rho(U_I)}+\frac{b(U_I^*)}{\d^2\rho(U_I^*)}\right) \label{eq:Bnsimp2}\\
& + \sum_{I\in \quartic}d\theta_I \mathrm{Im}\left(\frac{b(U_I)}{\d^2\rho(U_I)}\right) + \sum_{i\in\puncture } d\lambda_i \big(\flc_i\circ e^{-\lambda_i}\circ b_0\big).
\end{align}
\label{eq:Bnsimp}
\end{subequations}

\noindent Let us summarize the strategy. The contribution on the first line \eq{Bnsimp1} represents a sum of $b$-ghost contours around the punctures. The idea is to deform these contours into the propagators. In the process we pick up residues from crossing poles at the interaction points. These residues are supposed to cancel the terms on the second line \eq{Bnsimp2}. What is left should be identified as insertions of $b_0$ inside the propagators. 

\begin{figure}[t]
\begin{center}
\resizebox{4.5in}{5in}{\includegraphics{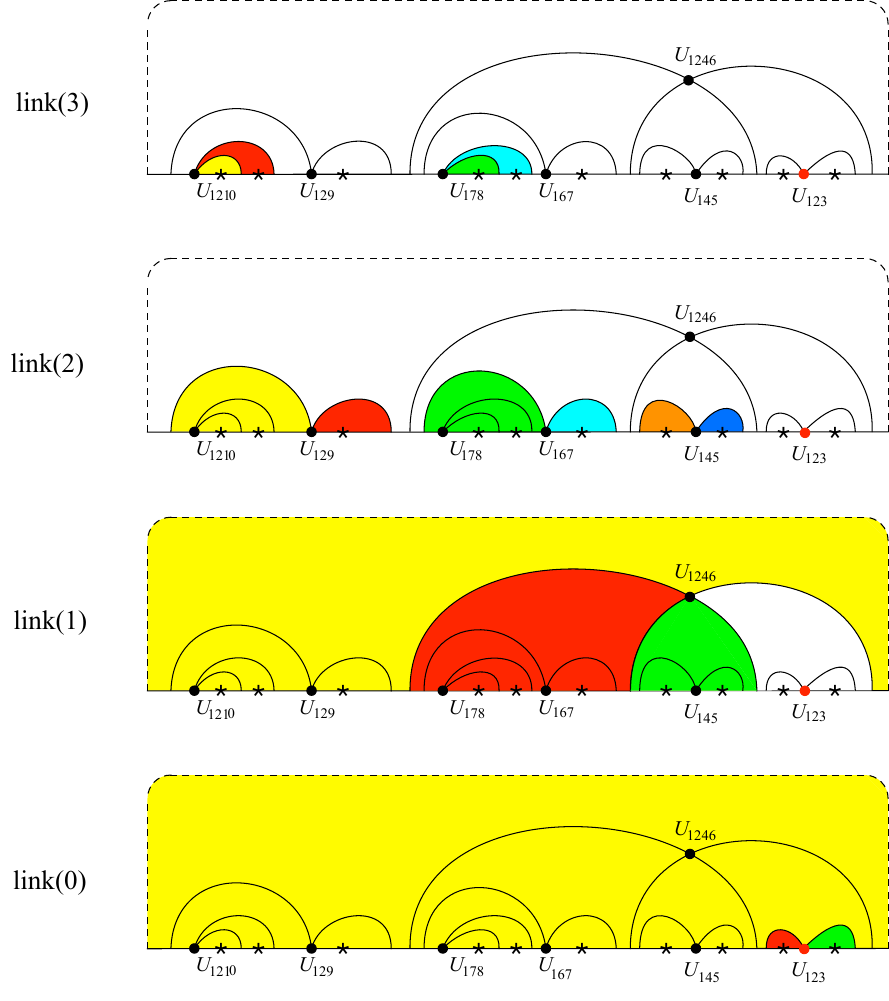}}
\end{center}
\caption{\label{fig:lightcone_gauge11} Illustration of the disks $D_i$ corresponding to the Mandelstam diagram in figure \ref{fig:lightcone_gauge5}, partitioned into classes according to the link number of $p_*(i)$. Within each class, each disk is shaded solid with a distinct color.}
\end{figure}

We will need to be precise about the definition of contours. Consider the preimage of the strip domains $\rho_i$ on the upper half plane, which we write as
\begin{eqnarray}
D_i\lineup =\rho^{-1}\circ\rho_i,\ \ \ i\in\puncture,\nonumber\\
A_i\lineup = \rho^{-1}\circ\rho_i,\ \ \ i\in\propagator.
\end{eqnarray}
Using the doubling trick, the preimage of the strip domain of a puncture will have the topology of a disk $D_i$, while the preimage of the strip domain of a propagator will have the topology of an annulus $A_i$. We want to fill the hole in the annulus with preimages of strip domains which come ``after,'' in the sense defined by the partial ordering associated to $*$. See figure \ref{fig:lightcone_gauge11}. The result is a collection of disks $D_i$ for each propagator defined recursively by 
\begin{equation}D_i = A_i \cup \left(\bigcup_{\smallpile{j\in\puncture\cup \propagator}{p_*(j)=s_*(i)}}D_j \right),\ \ \ \ i\in \propagator.\label{eq:Di} \end{equation}
The second component of the union fills the hole inside the annulus $A_i$. The purpose of filling the holes is that the boundary operation 
\begin{equation}\d D_i\end{equation} 
now gives a canonical orientation to a closed contour inside the preimage of every strip domain. Moreover, the contours satisfy a sum rule
\begin{equation}\d D_i = \sum_{\smallpile{j\in\puncture\cup \propagator}{ p_*(j)=s_*(i)}}\d D_j,\ \ \ \ i\in\propagator,\label{eq:sumrule}\end{equation}
which will be important when we deform the $b$-ghost contours into the propagators.

With this definition we will write the counterclockwise contour around the puncture $u_i$ explicitly as $\d D_i$. The sum of contour integrals in \eq{Bnsimp1} is then written as 
\begin{equation}
\text{\eq{Bnsimp1}}  = -\sum_{i\in\puncture} d(\tau_{p_*(i)}-\tau_*) \oint_{\d D_i}\frac{du}{2\pi i} \frac{b(u)}{\d\rho(u)}.\label{eq:B71}
\end{equation}
We have also substituted $s(i)=p_*(i)$ because we use the partial ordering associated to $*$ in order to set up a recursion. We want to express this as a sum over propagators rather than as a sum over punctures. For this we note that the partial ordering associated to $*$ has the property that every interaction point is the successor of exactly one propagator. Therefore we can write
\begin{equation}\sum_{i\in\puncture} = \sum_{i\in\puncture}\sum_{j\in\propagator} \delta_{p_*(i) = s_*(j)},\label{eq:sumreorg}\end{equation}
which implies 
\begin{equation}
\text{\eq{Bnsimp1}}  = -\sum_{i\in \propagator}  d(\tau_{s_*(i)}-\tau_*) \sum_{\smallpile{j\in \puncture}{ p_*(j)=s_*(i)}}\oint_{\d D_j}\frac{du}{2\pi i} \frac{b(u)}{\d\rho(u)}.\label{eq:B73}
\end{equation}
Next observe that the sum inside the sum almost takes the form of \eq{sumrule}, except that propagators are missing. If they had been there, we could use \eq{sumrule} to deform the $b$-ghost contours from the punctures into the propagators. As it happens this is still possible, but we have to work our way down in steps, starting with the subset of interaction points which have the highest link number. We decompose \eq{B73} into a sum over link $\ell$ subsets:
\begin{equation}
\text{\eq{Bnsimp1}}   = -\sum_{\ell=1}^N \sum_{\smallpile{i\in \propagator}{ s_*(i)\in \llink{\ell}}}  d(\tau_{s_*(i)}-\tau_*) \sum_{\smallpile{j\in \puncture}{ p_*(j)=s_*(i)}}\oint_{\d D_j}\frac{du}{2\pi i} \frac{b(u)}{\d\rho(u)},
\end{equation}
and pull out the contribution with highest link number: 
\begin{eqnarray}
\text{\eq{Bnsimp1}} \lineup  = - \sum_{\smallpile{i\in \propagator}{s_*(i)\in \llink{N}}}  d(\tau_{s_*(i)}-\tau_*) \sum_{\smallpile{j\in\puncture\cup\propagator}{ p_*(j)=s_*(i)}}\oint_{\d D_j}\frac{du}{2\pi i} \frac{b(u)}{\d\rho(u)}\nonumber\\
\lineup\ \ \ -\sum_{\ell=1}^{N-1} \sum_{\smallpile{i\in \propagator}{ s_*(i)\in \llink{\ell}}}  d(\tau_{s_*(i)}-\tau_*) \sum_{\smallpile{j\in \puncture}{ p_*(j)=s_*(i)}}\oint_{\d D_j}\frac{du}{2\pi i} \frac{b(u)}{\d\rho(u)}.
\end{eqnarray}
The important circumstance at link number $N$ is that the sum inside the sum includes both punctures and propagators. This is true for the trivial reason that propagators are absent; an interaction point $s_*(i)\in\{N \text{ links}\}$ cannot be the predecessor of a propagator, since otherwise there would be interaction points with higher link number. Therefore the contours can be joined using the sum rule \eq{sumrule}. In the process we pick up residues from crossing poles at the interaction points. In this way we find
\begin{subequations}
\begin{align}
\text{\eq{Bnsimp1}}   = & - \sum_{\smallpile{i\in \propagator}{ s_*(i)\in \llink{N}}}  d(\tau_{s_*(i)}-\tau_*) \oint_{\d D_i}\frac{du}{2\pi i} \frac{b(u)}{\d\rho(u)}\label{eq:B741}\\
&+\sum_{I\in\cubic\cap\llink{N}}d(\tau_I-\tau_*) \frac{b(U_I)}{\d^2\rho(U_I)}\label{eq:B742}\\
&+\sum_{I\in\quartic\cap\llink{N}}d(\tau_I-\tau_*)  \left(\frac{b(U_I)}{\d^2\rho(U_I)}+\frac{b(U_I^*)}{\d^2\rho(U_I^*)}\right)\label{eq:B743}\\
&-\sum_{\ell=1}^{N-1} \sum_{\smallpile{i\in \propagator}{s_*(i)\in \llink{\ell}}}  d(\tau_{s_*(i)}-\tau_*) \sum_{\smallpile{j\in \puncture}{ p_*(j)=s_*(i)}}\oint_{\d D_j}\frac{du}{2\pi i} \frac{b(u)}{\d\rho(u)}.
\end{align}
\label{eq:B74}
\end{subequations}

\noindent The second and third lines \eq{B742} and \eq{B743} represent the contributions from crossing the poles. As anticipated earlier, these partially cancel the sum over interaction points in \eq{Bnsimp2}. Now in \eq{B741} we want to extract a  differential of the propagator width $dT_i$. This can be done noting from \eq{Tist} that
\begin{equation}(\tau_{s_*(i)}-\tau_*)= (-1)^{\eps_*(i)}T_i + (\tau_{p_*(i)}-\tau_*).\end{equation}
Then 
\begin{subequations}
\begin{align}
\text{\eq{Bnsimp1}}   = &- \sum_{\smallpile{i\in \propagator}{ s_*(i)\in \llink{N}}}  (-1)^{\eps(i)}dT_i \oint_{\d D_i}\frac{du}{2\pi i} \frac{b(u)}{\d\rho(u)}\label{eq:B751}\\
& - \sum_{\smallpile{i\in \propagator}{ s_*(i)\in \llink{N}}}  d(\tau_{p_*(i)}-\tau_*) \oint_{\d D_i}\frac{du}{2\pi i} \frac{b(u)}{\d\rho(u)}\label{eq:B752}\\
&+\sum_{I\in\cubic\cap\llink{N}}d(\tau_I-\tau_*) \frac{b(U_I)}{\d^2\rho(U_I)}\label{eq:B753}\\
&+\sum_{I\in\quartic\cap\llink{N}}d(\tau_I-\tau_*)  \left(\frac{b(U_I)}{\d^2\rho(U_I)}+\frac{b(U_I^*)}{\d^2\rho(U_I^*)}\right)\label{eq:B754}\\
&-\sum_{\ell=1}^{N-1} \sum_{\smallpile{i\in \propagator }{ s_*(i)\in \llink{\ell}}}  d(\tau_{s_*(i)}-\tau_*) \sum_{\smallpile{j\in \puncture }{ p_*(j)=s_*(i)}}\oint_{\d D_j}\frac{du}{2\pi i} \frac{b(u)}{\d\rho(u)}.\label{eq:B755}
\end{align}
\label{eq:B75}
\end{subequations}

\noindent The $dT_i$ contributions \eq{B751} represent insertions of $b_0$ inside the lightcone gauge propagators. For now we will take this as given and prove it later. On the second line \eq{B752} we reindex the sum following \eq{sumreorg} so that the interaction times are given as successors: 
\begin{equation}
- \!\!\!\!\!\! \sum_{\smallpile{i\in \propagator }{ s_*(i)\in \llink{N}}}  d(\tau_{p_*(i)}-\tau_*) \oint_{\d D_i}\frac{du}{2\pi i} \frac{b(u)}{\d\rho(u)}
\, =\, - \!\!\!\!\!\!\!\!\!\sum_{\smallpile{i\in \propagator}{ s_*(i)\in\llink{N-1}}}d (\tau_{s_*(i)}-\tau_*) \sum_{\smallpile{j\in\propagator}{ s_*(i)=p_*(j)}}
 \oint_{\d D_j}\frac{du}{2\pi i} \frac{b(u)}{\d\rho(u)}.
\end{equation}
This achieves something important. Extracting the link number $N-1$ contribution from \eq{B755}, the sum inside the sum is missing propagators. But the above contribution supplies propagators to the sum. Therefore we can write
\begin{subequations}
\begin{align}
\text{\eq{Bnsimp1}}   = & \sum_{\smallpile{i\in \propagator }{ s_*(i)\in \{N \text{ links}\}}}  dT_i \left(\frac{1}{\alpha_i}\flc_i\circ b_0\right)\\
&+\sum_{I\in\cubic\cap\llink{N}}d(\tau_I-\tau_*) \frac{b(U_I)}{\d^2\rho(U_I)}\\
&+\sum_{I\in\quartic\cap\llink{N}}d(\tau_I-\tau_*)  \left(\frac{b(U_I)}{\d^2\rho(U_I)}+\frac{b(U_I^*)}{\d^2\rho(U_I^*)}\right)\\
&-\sum_{\smallpile{i\in \propagator }{ s_*(i)\in \llink{N-1}}}  d(\tau_{s_*(i)}-\tau_*) \sum_{\smallpile{j\in \puncture \cup\propagator }{ p_*(j)=s_*(i)}}\oint_{\d D_j}\frac{du}{2\pi i} \frac{b(u)}{\d\rho(u)}\label{eq:B764}\\
&-\sum_{\ell=1}^{N-2} \sum_{\smallpile{i\in \propagator }{ s_*(i)\in \llink{\ell}}}  d(\tau_{s_*(i)}-\tau_*) \sum_{\smallpile{j\in \puncture }{ p_*(j)=s_*(i)}}\oint_{\d D_j}\frac{du}{2\pi i} \frac{b(u)}{\d\rho(u)}.
\end{align}
\end{subequations}

\noindent Now that the sum inside the sum of \eq{B764} includes both punctures and propagators, we can use \eq{sumrule} to proceed in exactly the same way as for link number $N$. This produces $b_0$ insertions inside the propagators at link number $(N-1)$. It cancels the contribution to \eq{Bnsimp2} from interaction points at link number $(N-1)$. And finally, it provides propagators for the sum inside the sum at link number $(N-2)$. Executing the recursion to the end we obtain
\begin{subequations}
\begin{align}
\text{\eq{Bnsimp1}}   = & \sum_{\smallpile{i\in \propagator }{ s_*(i)\in \{N \text{ links}\}} } dT_i \left(\frac{1}{\alpha_i}\flc_i\circ b_0\right)\\
&+\sum_{I\in\cubic}d(\tau_I-\tau_*) \frac{b(U_I)}{\d^2\rho(U_I)}\\
&+\sum_{I\in\quartic}d(\tau_I-\tau_*)  \left(\frac{b(U_I)}{\d^2\rho(U_I)}+\frac{b(U_I^*)}{\d^2\rho(U_I^*)}\right).
\end{align}
\end{subequations}

\noindent Plugging this back into \eq{Bnsimp} we obtain the $b$-ghost insertion defining the Kugo-Zwiebach form of the covariantized lightcone measure.

Now we return to the last piece of the argument, which is to show that
\begin{equation}
\frac{1}{\alpha_i}\flc_i\circ b_0 = -(-1)^{\eps_*(i)}\oint_{\d D_i}\frac{du}{2\pi i}\frac{b(u)}{\d\rho(u)},\ \ \ \ i\in\propagator.
\end{equation}
The tricky part is establishing the correct orientation of the contour. If $C$ is the canonical counterclockwise closed contour on the unit disk, we need to show that the corresponding closed contour in the upper half plane is given by
\begin{equation}\flc_i\circ C = -(-1)^{\eps_*(i)}\d D_i,\ \ \ \ i\in\propagator. \label{eq:fCdD}\end{equation}
The first step is to determine the orientation of $\flc_i\circ C$. This can be done with the following observations:
\begin{itemize}
\item The region $A_i$ has two disjoint open string boundary segments on the real axis (not using the doubling trick). With the standard counterclockwise orientation on the boundary of $A_i$, both segments are oriented towards increasing $\mathrm{Re}(u)$.
\item The propagator strip domain $\rho_i$ has open string boundary segments on the top and bottom edges. With the standard counterclockwise orientation on the boundary of $\rho_i$, the segment on the bottom is oriented towards increasing $\mathrm{Re}(\rho)$ while the segment on top towards decreasing $\mathrm{Re}(\rho)$.
\item It follows that one open string boundary segment of $A_i$ has $\d\rho(u)>0$, and this is the preimage of the bottom edge of $\rho_i$. The other has $\d\rho(u)<0$, and is the preimage of the top edge of $\rho_i$. 
\item The contour $C$ maps onto the propagator strip $\rho_i$ with \eq{rhoi}. Since $\alpha_i>0$, the image of $C$ is a contour which passes from the bottom edge to the top edge of $\rho_i$. 
\item It follows that the image of $C$ in the upper half plane passes from the open string boundary of $A_i$ satisfying $\d\rho(u)>0$ to the open string boundary satisfying $\d\rho(u)<0$. 
\end{itemize}
Next we determine the orientation of $\d D_i$:
\begin{itemize}
\item The contour $\d D_i$ is homotopic to the level set of the predecessor interaction time,
\begin{equation}\mathrm{Re}[\rho(u)]=\tau_{p_*(i)},\label{eq:levelpr}\end{equation}
with an orientation determined by the standard counterclockwise orientation of the boundary of $A_i$. 
\item In the direction of increasing $\mathrm{Re}(u)$, the two open string boundary segments of $A_i$ are distinguished by whether they begin or terminate on the level set \eq{levelpr}. 
\item It follows that the contour $\d D_i$ passes through the upper half plane from the open string boundary segment which terminates on \eq{levelpr} to the open string boundary segment which begins on \eq{levelpr}.
\end{itemize}
Now we can compare $\d D_i$ to $\flc_i\circ C$. Consider the open string boundary segment which terminates on \eq{levelpr}. As we increase $u$ on this boundary segment, $\mathrm{Re}(\rho(u))$ will change from the interaction time of the successor $\tau_{s_*(i)}$ to the interaction time of the predecessor $\tau_{p_*(i)}$. If the first is less than the second, we conclude that $\d\rho(u)$ will be greater than zero. Therefore $\d D_i$ and $\flc_i\circ C$ will be the same.  If, on the other hand, the successor interaction time is greater than that of the predecessor, we conclude that $\d\rho(u)$ must be negative. Then $\d D_i$ and $\flc_i\circ C$ will have opposite orientation. Noting that $\eps_*(i)=1$ or $0$ depending on whether $\tau_{s_*(i)}$ is greater or less than $\tau_{p_*(i)}$, \eq{fCdD} follows.

\end{appendix}

\end{document}